\documentclass[preprint,preprintnumbers,amsmath,amssymb]{revtex4}

\usepackage{bm}
\usepackage{graphicx}
\usepackage{subfig}
\usepackage{array}
\usepackage{color}
\usepackage{float}
\allowdisplaybreaks[1]

\graphicspath{{figures/}}

\makeatletter

\newcommand{\Rmnum}[1]{\expandafter\@slowromancap\romannumeral #1@}
\makeatother

\begin{document}

\title{$B_{(s)}$ to Light Axial Vector Meson Form Factors via LCSR in HQEFT with Applications to Semileptonic Decays}

\author{Ya-Bing Zuo}\email{zuoyabing@lnnu.edu.cn}\quad
\author{Ming-Ge Li}\quad
\author{Shi-Yu Liang} \quad
\author{Wan-Ting Liu} \quad
\author{Xin-Su Liu} 

\affiliation{School of Physics and Electronic Technology, Liaoning
Normal University, Dalian 116029, P.R.China }

\begin{abstract}

In the present work, the form factors of $B_{(s)}$ to
light P-wave axial vector mesons are calculated via the light cone sum rules
(LCSR) in the framework of heavy quark effective field theory
(HQEFT). Firstly, the expressions of form factors in
terms of the light cone distribution amplitudes (DAs) of axial vector
mesons are derived via the LCSR at the leading order of heavy quark
expansion. It is found that the penguin type form factors can be obtained directly from the corresponding
semileptonic ones, which is similar to the case of S-wave mesons.
Considering the light axial vector meson DAs to twist-3, we give the
numerical results of form factors systematically. As applications, we investigate the branching ratios,
longitudinal polarization fractions and forward-backward asymmetries of relevant semileptonic decays induced by charged current. Our results may be
tested by more precise experiments in the future.

\end{abstract}

\maketitle

\section{Introduction}

$B_{(s)}$ exclusive decays are of great importance
in testing the standard model (SM) and probing new physics beyond
it. Up to now, the decays with only S-wave mesons (including pseudoscalar and
vector mesons) in the final states have been analyzed extensively
both in theory and in experiment. Yet, the studies on decays containing P-wave mesons
(including scalar, axial vector and tensor mesons) are relatively fewer. In the
last decades, a large amount of such decays have been established experimentally\cite{PDG}, promoting
corresponding theoretical investigations on these decays.

The axial-vector meson is regarded as one of
the most significant excited-state light mesons so that the exploration of the semileptonic
$B \rightarrow A$ ($A$ denotes light axial-vector mesons) transitions is of considerable necessity and
significance.
In quark model, there are two types of P-wave axial-vector mesons, namely, $1^3P_1$ and
$1^1P_1$ states due to different spin couplings of the two quarks corresponding to $J^{P C} = 1^{++}$
and $1^{+-}$, respectively. In this study, we focus on the $1^{++}$ mesons $a_1(1260)$, $f_1(1285)$,
$f_1(1420)$ and $K_{1A}$, and the $1^{+-}$ mesons $b_1(1235)$, $h_1(1170)$, $h_1(1415)$ and $K_{1B}$.
The physical mass eigenstates $K_1(1270)$, $K_1(1400)$ are obtained from the mixture of $K_{1A}$ and $K_{1B}$ due to the
mass difference of the strange and non-strange light quarks. The $1^3P_1$ states $f_1(1285)$, $f_1(1420)$ are mixtures of the light flavor $SU(3)$ singlet $f_1$ and octet $f_8$. And likewise, the $1^1P_1$ states $h_1(1170)$, $h_1(1415)$ are mixtures of $h_1$ and $h_8$.

Theoretical calculations of exclusive decays  require the knowledge of non-perturbative QCD, which
are generally parameterized in terms of decay constants, form factors and some non-factorization contributions.
The primary objective of probing the semileptonic decays of $B_{(s)}$ mesons into axial vector
mesons is to calculate the $B_{(s)} \rightarrow A$ form factors. There are a number of previous
researches on evaluating the $B_{(s)} \rightarrow A$ form factors with light cone sum rules (LCSR) based on the axial vector meson distribution amplitudes (DAs)\cite{LCSRA1,LCSRA2,LCSRA3,LCSRA4,LCSRA5,LCSRA6} or the $B_{(s)}$ meson DAs\cite{LCSRB1,LCSRB2,LCSRB3}. Another available methodology for exploring the $B_{(s)} \rightarrow A$
form factors is the perturbative QCD (pQCD) approach, which is based on transverse momentum-dependent (TMD) QCD factorization for hard processes\cite{pQCD1,pQCD2}. In addition, the $B_{(s)} \rightarrow A$ form factors have been calculated by using the QCD sum rules\cite{QCDSR1,QCDSR2}, light front quark model (LFQM)\cite{LFQM1,LFQM2,LFQM3} and AdS/QCD correspondence\cite{AdSQCD}.

For heavy hadrons containing one heavy quark, it is useful to adopt the heavy quark
effective field theory (HQEFT), in which a heavy quark
expansion is performed and the calculation of nonperturbative QCD effects can be greatly simplified\cite{HQEFT1,HQEFT2}. Some interesting
relations between the penguin type form factors and corresponding
semileptonic ones in the whole region of momentum transfered have
been found in the heavy to light S-wave meson decays at the leading
order of heavy quark expansion\cite{RELATION1,RELATION2}. Therefore,
it is worthwhile to calculate the heavy to light axial vector
meson form factors in HQEFT and investigate
if similar relations still hold in this case. In the present paper,
we intend to calculate the $B_{(s)} \rightarrow a_1(1260), b_1(1235), K_{1A}, K_{1B}, f_1,
f_8, h_1, h_8$ form factors
systematically by using the LCSR in HQEFT and, as applications, investigate the relevant semileptonic decays.

The remaining part of this paper is organized as follows. In section
II, we give the definitions of form factors and derive
their expressions in terms of the DAs of axial vector mesons by using the
LCSR in HQEFT. Numerical analysis and results of the form factors are presented in section III. Then, based on the form
factors given here, we investigate the branching ratios, longitudinal
polarization fractions and forward-backward asymmetries of all the relevant semileptonic decays in
section IV. Section V is our summary.

\section{ Definitions of form factors and LCSR in HQEFT}

\subsection{Definitions of form factors}

The form factors for $B_{(s)}$  to light axial vector meson decays
are defined via the corresponding hadronic matrix elements as
follows\cite{LCSRA1,LCSRA2},
\begin{eqnarray}
& & \langle A (P, \lambda)| \bar{q} \gamma^\mu \gamma^5 b | H (p_H) \rangle =
i \frac{2}{m_H +m_A} \varepsilon^{\mu \nu \alpha \beta} \epsilon^{(\lambda)*}
_\nu p_{H\alpha} P_\beta A (q^2)\label{DFF1} \\
& & \langle A (P, \lambda)| \bar{q} \gamma^\mu  b | H(p_H)
\rangle = -(m_H +m_A) \left [  \epsilon^{(\lambda)* \mu} -\frac{\epsilon^*_{(\lambda)} \cdot p_H}{q^2} q^\mu \right ] V_1 (q^2)\nonumber \\
& & \hspace{5cm} +  \frac{\epsilon^{(\lambda)*}
\cdot p_H}{m_H +m_A } \left [ (p_H + P)^\mu - \frac{m^2_H -m^2_A}{q^2} q^\mu \right ] V_2 (q^2) \nonumber \\
& & \hspace{5cm} -2 m_A \frac{\epsilon^{(\lambda)*} \cdot p_H}{q^2} q^\mu V_0 (q^2) \\
& & \langle A (P, \lambda)| \bar{q} \sigma^{\mu \nu} \gamma^5 q_\nu b
| H(p_H) \rangle = 2 T_1 (q^2) \varepsilon^{\mu \nu \alpha \beta}
\epsilon^{(\lambda)*}_\nu p_{H
\alpha} P_\beta \label{DFF3}\\
& & \langle A (P, \lambda)| \bar{q} \sigma^{\mu \nu}  q_\nu
b | H(p_H) \rangle = -i T_2 (q^2) \left [ ( m^2_H - m^2_A ) \epsilon^{(\lambda)* \mu
} - ( \epsilon^{(\lambda)*} \cdot p_H ) ( p_H + P )^\mu \right
] \nonumber \\
& & \hspace{5.5cm} - i T_3 (q^2)( \epsilon^{(\lambda)*} \cdot p_H ) \left
[ q^\mu - \frac{q^2}{ m^2_H - m^2_A} ( p_H + P )^\mu \right ]\label{DFF4}
\end{eqnarray}
where $q_\mu = (p_H-P)_\mu$ is the momentum transferred and $\epsilon^{(\lambda)\mu}$ is the polarization vector of the axial vector meson in the final state. $H$ represents $B_{(s)}$.
$\varepsilon^{\mu \nu \alpha \beta}$ is the total antisymmetric
Levi-Civita symbol, with $\varepsilon^{0123}=-1$. $V_1$, $V_2$,
$V_0$, $A$ and $T_1$, $T_2$, $T_3$ are the semileptonic and penguin type form factors, respectively. The latter are
only relevant for exclusive decays induced by FCNC.

To the leading order of heavy quark expansion in HQEFT\cite{HQEFT1,HQEFT2}, the hadronic
matrix elements can be written as
\begin{eqnarray}\label{HMEEP}
\langle A( P, \lambda)| \bar{q} \Gamma b | H ( p_H ) \rangle =
\sqrt{\frac{m_H}{\bar{\Lambda}_H}} \left [ \langle A( P, \lambda)|
\bar{q} \Gamma Q^{(+)}_v | M_v \rangle + {\cal O}( 1/m_Q) \right ]
\end{eqnarray}
where $\bar{\Lambda}_H = m_H -m_Q$. $Q^{(+)}_v$ and $|M_v \rangle$
are the effective heavy quark field and heavy meson state in HQEFT,
respectively. From heavy quark spin-flavor symmetry, the leading
order hadronic matrix elements
\begin{eqnarray}\label{HMELO}
\langle A( P, \lambda)| \bar{q} \Gamma Q^{(+)}_v | M_v \rangle = -
{\rm Tr} \left [ \rho(v,P) \Gamma {\cal M}_v \right ]
\end{eqnarray}
with
\begin{eqnarray}
& & \rho(v, P) = \left \{ L_1 (v \cdot P)   \epsilon^{(\lambda)*} \hspace{-0.6cm} \slash \hspace{0.3cm} + L_2 ( v
\cdot P) ( v \cdot \epsilon^{(\lambda)*} ) + \left[ L_3 (v \cdot P)\epsilon^{(\lambda)*} \hspace{-0.6cm} \slash \hspace{0.3cm}
+ L_4 ( v
\cdot P) ( v \cdot \epsilon^{(\lambda)*} ) \right ] \hat{P} \hspace{-0.25cm} \slash \right \} \nonumber \\
& & {\cal M}_v = - \sqrt{\bar{\Lambda}} \frac{1+ v \hspace{-0.2cm} \slash}{2} \gamma^5, \hspace{0.5cm} \hat{P}^\mu = \frac{P^\mu}{v \cdot P} \nonumber
\end{eqnarray}
Here, $\bar{\Lambda}=\lim_{m_Q \rightarrow \infty} \bar{\Lambda}_B$. $L_1$, $L_2$, $L_3$, $L_4$ are the leading order wave functions describing the heavy to light meson transition matrix elements in HQEFT.
${\cal M}_v$ are the heavy pseudoscalar meson wave function. According to Eqs.(\ref{DFF1})-(\ref{HMELO}), we have
\begin{eqnarray}
& & V_1 (q^2) = \frac{2}{m_H +m_A} \sqrt{\frac{m_H \bar{\Lambda}}{\bar{\Lambda}_H}} \left [ L_1 ( v \cdot P) + L_3 ( v \cdot P) \right ] \label{a1}\\
& & V_2 (q^2) =  ( m_H + m_A) \sqrt{\frac{m_H \bar{\Lambda}}{\bar{\Lambda}_H}} \left [ \frac{L_2 ( v \cdot P)}{ m^2_H}
+ \frac{L_3 ( v \cdot P) - L_4 ( v \cdot P)}{ m_H ( v \cdot P)} \right ] \label{a2}\\
& &  V_0 (q^2) = \frac{1}{m_A} \sqrt{\frac{m_H \bar{\Lambda}}{\bar{\Lambda}_H}} \left [ L_1( v \cdot P) - L_2 (v \cdot P) + \frac{v \cdot P}{m_H} L_2( v \cdot P) \right. \nonumber \\
& & \hspace{1.6cm} \left. + \frac{m^2_A}{ m_H v \cdot P} \left ( L_3 ( v \cdot P) - L_4 ( v \cdot P) \right ) + L_4 ( v \cdot P) \right ] \\
& & A ( q^2) = \sqrt{\frac{m_H \bar{\Lambda}}{\bar{\Lambda}_H}} \frac{m_H + m_A}{m_H (v \cdot P)} L_3 ( v \cdot P) \label{v}\\
& & T_1 ( q^2) = \sqrt{\frac{m_H \bar{\Lambda}}{\bar{\Lambda}_H}} \left [ \frac{1}{m_H} L^\prime_1 ( v \cdot P) + \frac{1}{v \cdot P } L^\prime_3 ( v \cdot P) \right ] \label{t1}\\
& & T_2 ( q^2) = 2 \sqrt{\frac{m_H \bar{\Lambda}}{\bar{\Lambda}_H}} \frac{1}{ m^2_H - m^2_A} \left [ ( m_H - v \cdot P ) L^\prime_1 ( v \cdot P )
+ \frac{ m_H v \cdot P - m^2_A}{ v \cdot P } L^\prime_3 ( v \cdot P ) \right ] \label{t2}\\
& & T_3 ( q^2 ) = \sqrt{\frac{m_H \bar{\Lambda}}{\bar{\Lambda}_H}} \left [ - \frac{1}{ m_H} L^\prime_1 ( v \cdot P ) + \frac{1}{ v \cdot P }  L^\prime_3 ( v \cdot P)
- \frac{ m^2_H - m^2_A}{ m^2_H v \cdot P }  L^\prime_4 ( v \cdot P) \right ] \label{t3}
\end{eqnarray}
where $v \cdot P = \frac{m^2_H +m^2_A -q^2 }{2 m_H}$ is the energy of light axial vector meson in the final state. $L^\prime_i (i=1, 3, 4)$ are in general different from
corresponding $L_i$ as they arise from different matrix elements.

\subsection{LCSR in HQEFT}

In the following, we intend to calculate the leading order wave
functions of the transition matrix elements $L_1$, $L^\prime_1$,
$L_2$, $L_3$, $L^\prime_3$, $L_4$, $L^\prime_4$ by using the LCSR in
HQEFT. The procedure is similar to the case of $B_{(s)}$ decaying to light vector mesons\cite{btov}. Firstly, we define the following vacuum-axial vector meson
correlation functions,
\begin{eqnarray}
& & C^\mu_V ( P, q) = i \int d^4 x e^{i q \cdot x} \langle A ( P ,
\lambda ) | {\cal T} \left \{ \bar{q}_1 \gamma^\mu b(x), \bar{b} (0) i
\gamma^5 q_2 (0) \right \} | 0 \rangle \label{CorreF1}\\
& & C^\mu_A ( P, q) = i \int d^4 x e^{i q \cdot x} \langle A ( P ,
\lambda ) | {\cal T} \left \{ \bar{q}_1 \gamma^\mu \gamma^5 b(x), \bar{b}
(0) i \gamma^5 q_2 (0) \right \} | 0 \rangle \label{CorreF2}\\
& & C^\mu_T ( P, q) = i \int d^4 x e^{i q \cdot x} \langle A ( P ,
\lambda ) | {\cal T} \left \{ \bar{q}_1 \sigma^{\mu \nu} q_\nu b(x),
\bar{b} (0) i \gamma^5 q_2 (0) \right \} | 0 \rangle \label{CorreF3}\\
& & C^{\prime \mu}_T ( P, q) = i \int d^4 x e^{i q \cdot x} \langle
A ( P , \lambda ) | {\cal T} \left \{ \bar{q}_1 \sigma^{\mu \nu} \gamma^5
q_\nu b(x), \bar{b} (0) i \gamma^5 q_2 (0) \right \} | 0 \rangle
\label{CorreF4}
\end{eqnarray}

On one hand, inserting a complete set of states with $B_{(s)}$ meson
quantum numbers between the currents in Eqs.
(\ref{CorreF1})-(\ref{CorreF4}) phenomenologically, we obtain
\begin{eqnarray}
& & C^\mu_V ( P, q) = \frac{ \langle A ( P, \lambda ) | \bar{q}_1
\gamma^\mu b | H \rangle \langle H | \bar{b} i \gamma^5 q_2 |0
\rangle }{ m^2_H - ( P +q )^2 } \nonumber \\
& & \hspace{2.0cm} + \sum_{H^\prime} \frac{ \langle A ( P, \lambda )
| \bar{q}_1 \gamma^\mu b | H^\prime \rangle \langle H^\prime |
\bar{b} i \gamma^5 q_2 |0 \rangle }{ m^2_{H^\prime} - ( P +q )^2 } \label{CorreFP1}\\
& & C^\mu_A ( P, q) = \frac{ \langle A ( P, \lambda ) | \bar{q}_1
\gamma^\mu \gamma^5 b | H \rangle \langle H | \bar{b} i \gamma^5 q_2
|0
\rangle }{ m^2_H - ( P +q )^2 } \nonumber \\
& & \hspace{2.0cm} + \sum_{H^\prime} \frac{ \langle A ( P, \lambda )
| \bar{q}_1 \gamma^\mu \gamma^5 b | H^\prime \rangle \langle
H^\prime | \bar{b} i \gamma^5 q_2 |0 \rangle }{ m^2_{H^\prime} - ( P
+q )^2 }\label{CorreFP2}
\\
& & C^\mu_T ( P, q) = \frac{ \langle A ( P, \lambda ) | \bar{q}_1
\sigma^{\mu \nu} q_\nu b | H \rangle \langle H | \bar{b} i \gamma^5
q_2 |0
\rangle }{ m^2_H - ( P +q )^2 } \nonumber \\
& & \hspace{2.0cm} + \sum_{H^\prime} \frac{ \langle A ( P, \lambda )
| \bar{q}_1 \sigma^{\mu \nu} q_\nu b | H^\prime \rangle \langle
H^\prime | \bar{b} i \gamma^5 q_2 |0 \rangle }{ m^2_{H^\prime} - ( P
+q )^2 }\label{CorreFP3}
\\
& & C^{\prime \mu}_T ( P, q) = \frac{ \langle A ( P, \lambda ) | \bar{q}_1
\sigma^{\mu \nu} \gamma^5 q_\nu b | H \rangle \langle H | \bar{b} i
\gamma^5 q_2 |0
\rangle }{ m^2_H - ( P +q )^2 } \nonumber \\
& & \hspace{2.0cm} + \sum_{H^\prime} \frac{ \langle A ( P, \lambda )
| \bar{q}_1 \sigma^{\mu \nu} \gamma^5 q_\nu b | H^\prime \rangle
\langle H^\prime | \bar{b} i \gamma^5 q_2 |0 \rangle }{
m^2_{H^\prime} - ( P +q )^2 }\label{CorreFP4}
\end{eqnarray}
The effective decay
constant of heavy meson in HQEFT are defined as\cite{HQEFT1, HQEFT2}
\begin{eqnarray}\label{dcHQEFT}
\langle 0 | \bar{q} \Gamma Q^{(+)}_v | {\cal M}_v \rangle =
\frac{i}{2} F {\rm Tr} \left [ \Gamma {\cal M}_v \right ]
\end{eqnarray}
Substituting Eqs.(\ref{HMEEP}), (\ref{HMELO}), (\ref{dcHQEFT}) into
(\ref{CorreFP1})-(\ref{CorreFP4}), we obtain
\begin{eqnarray}
& & C^\mu_V ( P, q) = \frac{-2  F}{ 2 \bar{\Lambda}_H - 2 v \cdot k
} \left [ ( L_1( v \cdot P ) + L_3( v \cdot P ) ) \epsilon^{(\lambda)* \mu}
- L_2( v \cdot P ) v^\mu ( \epsilon^{(\lambda) *} \cdot v ) \right.
\nonumber
\\ & & \hspace{2cm} \left. - (L_3( v \cdot P ) - L_4( v \cdot P ) ) P^\mu \frac{\epsilon^{(\lambda ) *} \cdot v}{v \cdot P}
\right ]+ \int^\infty_{s_0} ds \frac{\rho_{AV} ( v \cdot P, s)}{ s -
2 v \cdot k } + {\rm subtr.} \\
& & C^\mu_A ( P, q) = \frac{-2 i F}{ 2 \bar{\Lambda}_H - 2 v \cdot k }
\frac{1}{ v \cdot P} L_3 ( v \cdot P ) \varepsilon^{\mu \nu \alpha
\beta} \epsilon^{(\lambda) *}_\nu P_\alpha v_\beta \nonumber \\
& & \hspace{2cm} + \int^\infty_{s_0} ds \frac{\rho_{AA} ( v \cdot P,
s)}{ s - 2 v \cdot k } + {\rm subtr.} \\
& & C^ \mu_T ( P,q ) = \frac{-2 i F}{ 2 \bar{\Lambda}_H - 2 v
\cdot k } \left [ \left ( (m_H - v \cdot P) L_1^\prime ( v \cdot P)
+ \frac{m_H v \cdot P - m^2_A}{ v \cdot P} L_3^\prime ( v \cdot P)
\right )
\epsilon^{(\lambda) * \mu} \right. \nonumber \\
& & \hspace{2cm} + \epsilon^{(\lambda)*} \cdot v \left ( -m_H L_1^\prime (
v \cdot P) + \frac{m^2_A - m_H v \cdot P }{ v \cdot P } L_4^\prime (
v \cdot P)
\right ) v^\mu  \nonumber \\
& & \hspace{2cm} \left. + \epsilon^{(\lambda)*} \cdot v  \left ( -\frac{m_H
L_3^\prime ( v \cdot P)}{v \cdot P} + \frac{m_H - v \cdot P}{ v
\cdot P}
L_4^\prime ( v \cdot P) \right ) P^\mu \right ] \nonumber \\
& & \hspace{2cm}+\int^\infty_{s_0} ds \frac{\rho_{AT} ( v
\cdot P, s)}{ s - 2 v \cdot k } + {\rm subtr.}\\
& & C^{\prime \mu}_T ( P,q ) = \frac{-2 F}{ 2 \bar{\Lambda}_H - 2 v \cdot k
} \left [ L_1^\prime ( v \cdot P) + \frac{m_H}{v \cdot P} L_3^\prime
( v \cdot P ) \right ] \varepsilon^{ \mu \nu \alpha \beta}
\epsilon^{(\lambda)*}_\nu
P_\alpha v_\beta \nonumber \\
& & \hspace{2cm} +\int^\infty_{s_0} ds \frac{\rho^\prime_{AT} ( v \cdot P,
s)}{ s - 2 v \cdot k } + {\rm subtr.}
\end{eqnarray}
where $k^\mu= (P+q)^\mu - m_Q v^\mu$, denoting the residual momentum
in $B_{(s)}$ meson. The integral  terms represent contributions coming
from the higher resonances. The subtraction terms \lq \lq subtr.\rq
\rq are introduced to ensure the convergence of integral terms,
which vanish automatically after Borel transformation and therefore
do not affect the physics results.

On the other hand, we can calculate the vacuum-axial vector meson
correlation functions theoretically, {\it i.e.}
\begin{eqnarray}
& & C^\mu_V ( P, q) = \int^\infty_0 d s \frac{\rho^{{\rm th}}_{AV} (
v
\cdot P, s)}{ s - 2 v \cdot k} + {\rm subtr.} \label{CorreFth1}\\
& & C^\mu_A ( P, q) = \int^\infty_0 d s \frac{\rho^{{\rm th}}_{AA} (
v \cdot P, s)}{ s - 2 v \cdot k} + {\rm subtr.} \label{CorreFth2}\\
& & C^\mu_T ( P, q) = \int^\infty_0 d s \frac{\rho^{{\rm th}}_{AT} (
v \cdot P, s)}{ s - 2 v \cdot k} + {\rm subtr.} \label{CorreFth3}\\
& & C^{\prime \mu}_T ( P, q) = \int^\infty_0 d s \frac{\rho^{\prime
{\rm th}}_{AT} ( v \cdot P, s)}{ s - 2 v \cdot k} + {\rm
subtr.}\label{CorreFth4}
\end{eqnarray}

According to the assumption of quark hadron duality, the theoretical
spectral densities $\rho^{{\rm th}}_{AV} ( v \cdot P, s )$,
$\rho^{{\rm th}}_{AA} ( v \cdot P, s )$, $\rho^{{\rm th}}_{AT} ( v
\cdot P, s )$, $\rho^{\prime {\rm th}}_{AT} ( v \cdot P, s )$ are
equal to their physical counterparts $\rho_{AV} ( v \cdot P, s )$,
$\rho_{AA} ( v \cdot P, s )$, $\rho_{AT} ( v \cdot P, s )$,
$\rho^\prime_{AT} ( v \cdot P, s )$, respectively. Therefore, we
have
\begin{eqnarray}
& & \frac{-2 F}{2 \bar{\Lambda}_H - 2 v \cdot k } \left [ \left (
L_1 ( v \cdot P ) + L_3 ( v \cdot P ) \right ) \epsilon^{(\lambda)* \mu } -
L_2 ( v \cdot P ) v^\mu \epsilon^{(\lambda ) * } \cdot v \right. \nonumber
\\
& & \hspace{1cm} \left. - \left ( L_3 ( v \cdot P ) - L_4 ( v \cdot
P ) \right ) P^\mu \frac{\epsilon^{(\lambda)*} \cdot v }{ v \cdot P} \right
] =\int^{s_0}_0 d s \frac{ \rho_{AV} ( v \cdot P, s )}{s - 2 v \cdot
k} + {\rm subtr. } \label{qhd2}\\
& & \frac{-2 i F}{2 \bar{\Lambda}_H - 2 v \cdot k } \frac{1}{ v \cdot
P} L_3 ( v \cdot P) \varepsilon^{\mu \nu \alpha \beta}
\epsilon^{(\lambda)*} P_\alpha v_\beta = \int^{s_0}_0 d s \frac{
\rho_{AA} ( v \cdot P, s )}{s - 2 v \cdot k} + {\rm subtr. } \label{qhd1}\\
& & \frac{-2 i F}{ 2 \bar{\Lambda}_H - 2 v \cdot k } \left [ \left (
(m_H - v \cdot P) L_1^\prime ( v \cdot P) + \frac{m_H v \cdot P -
m^2_T}{ v \cdot P} L_3^\prime ( v \cdot P) \right )
\epsilon^{(\lambda) *} \right. \nonumber \\
& & \hspace{2cm} + \epsilon^{(\lambda)*} \cdot v \left ( -m_H L_1^\prime (
v \cdot P) + \frac{m^2_T - m_H v \cdot P }{ v \cdot P } L_4^\prime (
v \cdot P)
\right ) v^\mu  \nonumber \\
& & \hspace{2cm} \left. + \epsilon^{(\lambda)*} \cdot v  \left ( -\frac{m_H
L_3^\prime ( v \cdot P)}{v \cdot P} + \frac{m_H - v \cdot P}{ v
\cdot P}
L_4^\prime ( v \cdot P) \right ) P^\mu \right ] \nonumber \\
& & \hspace{3cm} =\int^{s_0}_0 d s \frac{ \rho_{AT} ( v \cdot
P, s )}{s - 2 v \cdot k} + {\rm subtr. } \label{qhd4}\\
& &\frac{-2 F}{ 2 \bar{\Lambda}_H - 2 v \cdot k } \left [
L_1^\prime ( v \cdot P) + \frac{m_H}{v \cdot P} L_3^\prime ( v \cdot
P ) \right ] \varepsilon^{ \mu \nu \alpha \beta} \epsilon^{(\lambda)*}_\nu
P_\alpha
v_\beta \nonumber \\
& & \hspace{3cm} =\int^{s_0}_0 d s \frac{ \rho^\prime_{AT} ( v \cdot P, s
)}{s - 2 v \cdot k} + {\rm subtr. } \label{qhd3}
\end{eqnarray}

Defining $y = v \cdot P$, $\omega = 2 v \cdot k$ and performing two
sequential Borel transformation on
Eq.(\ref{CorreFth1})-(\ref{CorreFth4}), we have
\begin{eqnarray}
& & \rho^{{\rm th}}_{AV}( y, s )= \hat{B}^{(-1/T)}_{1/s}
\hat{B}^{(\omega)}_T C^\mu_V ( y, \omega ) \label{SDth1}\\
& & \rho^{{\rm th}}_{AA}( y, s )= \hat{B}^{(-1/T)}_{1/s}
\hat{B}^{(\omega)}_T C^\mu_A ( y, \omega ) \label{SDth2}\\
& & \rho^{{\rm th}}_{AT}( y, s )= \hat{B}^{(-1/T)}_{1/s}
\hat{B}^{(\omega)}_T C^\mu_T ( y, \omega ) \label{SDth3}\\
& & \rho^{\prime {\rm th}}_{AT}( y, s )= \hat{B}^{(-1/T)}_{1/s}
\hat{B}^{(\omega)}_T C^{\prime \mu}_T ( y, \omega )\label{SDth4}
\end{eqnarray}
Calculating the
vacuum-axial vector meson correlation functions to the leading order of
heavy quark expansion in HQEFT and substituting into
Eqs.(\ref{SDth1})-(\ref{SDth4}), we obtain
\begin{eqnarray}
& & \rho^{{\rm th}}_{AV} ( y, s) = \frac{1}{2y} \left \{ \epsilon^{(\lambda
) * \mu} \left [ -f_A m_A g^{(a)}_\perp (u) + f^\perp_A y
\phi_\perp (u) + \frac{f^\perp_A m^2_A}{2 y} \left ( h_3 (u)
-\phi_\perp ( u )
\right ) \right ] \right. \nonumber \\
& & \hspace{2cm} - v^\mu \epsilon^{(\lambda)*} \cdot v \left [ - \frac{1}{2
y^2} f_A m^3_A \left ( g_3 (u) + \phi_\parallel (u) - 2 g^{(a)}_\perp (u)
\right ) + \frac{f^\perp_A m^2_A}{2 y} \frac{\partial}{\partial u}
h^{(p)}_\parallel (u) \right. \nonumber \\
& & \hspace{2cm} \left. + \frac{ f^\perp_A m^2_A}{  y} \left ( h^{(t)}_\parallel
(u) -\phi_\perp (u) \right ) \right ] - P^\mu \frac{\epsilon^{(\lambda)*}
\cdot v}{y} \left [  f_A m_A \left (
\phi_\parallel (u) - g^{(a)}_\perp (u) \right ) \right. \nonumber \\
& & \hspace{2cm} \left. \left. + f^\perp_A y \phi_\perp (u) -
f^\perp_A \frac{m^2_A}{y} \left ( h^{(t)}_\parallel (u) - \frac{1}{2} \phi_\perp
(u) - \frac{1}{2} h_3 (u) \right ) \right ] \right \}_{u = 1-
\frac{s}{2y}} \label{spectrald2}\\
& & \rho^{{\rm th}}_{AA}( y, s ) = \frac{i}{2y} \left [ f_A m_A
\frac{1}{4 y} \frac{\partial}{\partial u} g^{(v)}_\perp (u) + f^\perp_A \phi_\perp (u) \right ]_{u= 1- \frac{s}{2y}}
\varepsilon^{\mu \nu \alpha \beta} \epsilon^{(\lambda)*}_\nu P_\alpha v_\beta \label{spectrald1}\\
& & \rho^{ {\rm th}}_{AT} ( y, s) = \frac{i}{2y} \left \{
\epsilon^{(\lambda)* \mu} \left [
 f^\perp_A  \left ( m_H y - m^2_A \right ) \phi_\perp (u)  - \frac{1}{2y} f^\perp_A m^2_A ( m_H -
y) \left ( \phi_\perp (u)-h_3 (u)\right ) \right. \right. \nonumber \\
& & \hspace{2cm} \left. - f_A m_A  (m_H -y) g^{(a)}_\perp (u) - \frac{1}{4 y} f_A m_A ( m^2_A - y^2)
\frac{\partial}{ \partial u} g^{(v)}_\perp (u) \right ]  \nonumber \\
& & \hspace{2cm} + \epsilon^{(\lambda)*} \cdot v P^\mu \left [ - f^\perp_A
m_H \phi_\perp (u)  + f^\perp_A m^2_A \frac{1}{y^2}(m_H -y) \left ( h^{(t)}_\parallel
(u) -\frac{1}{2} \phi_\perp (u) \right. \right. \nonumber \\
& & \hspace{2cm}\left. - \frac{1}{2} h_3 (u) \right )- f_A m_A \frac{1}{y} (m_H -y)
\phi_\parallel (u) \nonumber \\
& & \hspace{2cm} \left. + \frac{1}{y} f_A m_A ( m_H -y) g^{(a)}_\perp (u)
- \frac{1}{4} f_A m_A \frac{\partial}{ \partial u} g^{(v)}_\perp (u) \right
] \nonumber \\
& & \hspace{2cm} + \epsilon^{(\lambda)*} \cdot v v^\mu \left [ -\frac{1}{y^2}
f^\perp_A m^2_A (m_H y -m^2_A) \left ( h^{(t)}_\parallel (u) - \frac{1}{2} \phi_\perp (u) - \frac{1}{2} h_3 (u) \right
) \right. \nonumber \\
& & \hspace{2cm} +  f^\perp_A \frac{m^2_A m_H}{2 y} \left ( \phi_\perp (u)- h_3 (u) \right ) + \frac{1}{2y^2} f_A m^3_A \left( \phi_\parallel (u) - 2 g^{(a)}_\perp (u) + g_3 (u) \right )  \nonumber \\
& & \hspace{2cm} + f_A  m_A (m_H y - m^2_A ) \phi_\parallel (u) + \frac{1}{y} f_A m_A g^{(a)}_\perp (u)  \nonumber \\
& & \hspace{2cm}  - f_A m^3_A \frac{1}{2y^2} (m_H -y) \left ( \phi_\parallel (u) - 2 g^{(a)}_\perp (u) + g_3 (u) \right ) \nonumber \\
& & \hspace{2cm} \left. \left. +
\frac{1}{4 y } f_A m^3_T \frac{\partial }{\partial u} g^{(v)}_\perp (u)
\right ] \right \}_{u=1-\frac{s}{2y}} \label{spectrald4}\\
& & \rho^{\prime{\rm th}}_{AT} ( y, s) =  \frac{1}{2y} \left [  f^\perp_A m_H  \phi_\perp (u) + \frac{f^\perp_A m^2_A}{2 y}
\left ( h_3 (u) - \phi_\perp (u) \right ) \right. \nonumber \\
& & \hspace{2cm} \left. + \frac{f_A m_A}{ 4 y } ( m_H -y)
\frac{\partial }{ \partial u} g^{(v)}_\perp ( u) - f_A m_A g^{(a)}_\perp (u)
\right ]_{ u = 1 - \frac{s}{2y} } \label{spectrald3}
\end{eqnarray}
Here $\phi_\parallel (u)$, $\phi_\perp (u)$, $g^{(v)}_\perp (u)$, $g^{(a)}_\perp (u)$, $h^{(t)}_\parallel (u)$, $h^{(p)}_\parallel (u)$, $g_3 (u)$, $h_3 (u)$ are the light cone DAs of axial vector mesons,
which are defined as\cite{LCSRA2,DAs}
\begin{eqnarray}
& & \langle A (P, \lambda ) | \bar{q}_1 (x) \gamma_\mu \gamma_5 q_2 (0) | 0 \rangle = i f_A m_A \int_0^1 d u e^{i u P \cdot x} \left \{ P_\mu \frac{\epsilon^{(\lambda)*} \cdot x }{ P \cdot x } \phi_\parallel (u) + \left ( \epsilon^{(\lambda)*}_\mu \right.  \right. \nonumber \\
& & \hspace{2cm} \left. \left. - P_\mu
\frac{\epsilon^{(\lambda)*} \cdot x }{
P \cdot x } \right ) g^{(a)}_\perp (u) - \frac{1}{2} x_\mu
\frac{\epsilon^{(\lambda)*} \cdot x }{ (
P \cdot x )^2} m^2_A \left ( g_3 (u) + \phi_\parallel (u) - 2 g^{(a)}_\perp (u)
\right ) \right \} \\
& & \langle A (P, \lambda ) | \bar{q}_1 (x) \gamma_\mu  q_2
(0) | 0 \rangle = -\frac{i}{4} f_A m_A  \varepsilon_{\mu \nu \alpha \beta} \epsilon^{(\lambda)*\nu} P^\alpha x^\beta \int_0^1 du e^{i u P \cdot
x} g^{(v)}_\perp (u) \\
& &  \langle A (P, \lambda ) | \bar{q}_1 (x) \sigma_{\mu \nu} \gamma_5 q_2
(0) | 0 \rangle = f^\perp_A \int_0^1 d u e^{i u P \cdot x}
\left \{ \left [ \epsilon^{(\lambda)*}_\mu
P_\nu - \epsilon^{(\lambda)*}_\nu P_\mu \right ]
\phi_\perp (u) \right.
\nonumber \\
& & \hspace{2cm} + (
P_\mu x_\nu - P_\nu x_\mu ) \frac{m^2_A \epsilon^{(\lambda)*} \cdot x }{ (P \cdot x)^2} \left ( h^{(t)}_\parallel (u) -
\frac{1}{2} \phi_\perp (u)  - \frac{1}{2} h_3 (u) \right ) \nonumber \\
& & \hspace{2cm} \left. +
\frac{1}{2} \left ( \epsilon^{(\lambda) *}_\mu
x_\nu - \epsilon^{(\lambda) *}_\nu x_\mu \right )
\frac{m^2_A}{ P \cdot x} (h_3 (u) - \phi_\perp (u)) \right \} \\
& & \langle A (P, \lambda ) | \bar{q}_1 (x)\gamma_5 q_2 (0) | 0 \rangle =  \frac{1}{2} f^\perp_A m^2_A \epsilon^{(\lambda)*} \cdot x \int^1_0 d u
e^{i u P \cdot x }  h^{(p)}_\parallel (u)
\end{eqnarray}
where \{$\phi_\parallel (u)$, $\phi_\perp (u)$\}, \{$g^{(v)}_\perp (u)$, $g^{(a)}_\perp (u)$, $h^{(t)}_\parallel (u)$, $h^{(p)}_\parallel (u)$\} and \{$g_3 (u)$, $h_3 (u)$\} are of twist-2, twist-3 and twist-4, respectively.
Substituting Eqs.(\ref{spectrald1})-(\ref{spectrald4}) into
(\ref{qhd1})-(\ref{qhd4}) and performing the Borel transformation
$\hat{B}^{(\omega)}_T$ on both sides, we obtain the leading order wave
functions describing heavy to light meson transition matrix elements
in HQEFT,
\begin{eqnarray}
& & L_1 (y) = \frac{1}{4F} e^{2 \bar{\Lambda}_H/t} \int^{s_0}_0 d s
e^{-s/T}  \left [ \frac{1}{y} f_A m_A g^{(a)}_\perp (u) + \frac{1}{4y} f_A
m_A \frac{\partial}{ \partial u} g^{(v)}_\perp (u) \right. \nonumber \\
& & \hspace{1.5cm} \left. - \frac{1}{2y^2} f^\perp_A m^2_A (h_3 (u)
-\phi_\perp (u)) \right ]_{u = 1 - \frac{s}{2y}} \label{l1} \\
& & L_2 (y) = \frac{1}{4F} e^{2 \bar{\Lambda}_H/t} \int^{s_0}_0 d s
e^{-s/T} \left [ \frac{1}{y^2} f^\perp_A m^2_A \left ( \phi_\perp
(u) - h^{(t)}_\parallel (u)  - \frac{1}{2} \frac{\partial}{\partial u} h^{(p)}_\parallel (u)
\right ) \right. \nonumber \\
& & \hspace{1.5cm} \left. +  \frac{1}{2y^3} f_A m^3_A ( g_3 (u) +
\phi_\parallel (u) - 2 g^{(a)}_\perp (u) ) \right ]_{u= 1- \frac{s}{2y}} \label{l2}\\
& & L_3 (y) = \frac{1}{4F} e^{2 \bar{\Lambda}_H/t} \int^{s_0}_0 d s
e^{-s/T} \left [ - \frac{1}{4y} f_A m_A \frac{\partial}{\partial
u} g^{(v)}_\perp (u) -  f^\perp_A \phi_\perp (u) \right ]_{u =
1- \frac{s}{2y}} \label{l3}\\
& & L_4 (y) = \frac{1}{4F} e^{2 \bar{\Lambda}_H/t} \int^{s_0}_0 d s
e^{-s/T}  \left [ \frac{1}{y} f_A m_A \left ( \phi_\parallel (u)
- g^{(a)}_\perp (u) - \frac{1}{4} \frac{\partial}{\partial u} g^{(v)}_\perp (u) \right )
\right.   \nonumber \\
& & \hspace{1.5cm} \left. - f^\perp_A \frac{m^2_A}{y^2} \left ( h^{(t)}_\parallel
(u) - \frac{1}{2} \phi_\perp (u) - \frac{1}{2} h_3 (u) \right )
\right ]_{u = 1- \frac{s}{2y}} \label{l4}
\end{eqnarray}
It is found that the wave functions $L^\prime_1$, $L^\prime_3$ and $L^\prime_4$ are exactly the same as $L_1$, $L_2$ and $L_3$, respectively. With this consideration,
from Eqs.(\ref{a1}) to
(\ref{t3}), we obtain the following relations between the semileptonic and penguin type form factors,
\begin{eqnarray}
& & T_1 (q^2) = \frac{m^2_H - m^2_A +q^2}{ 2 m_H} \frac{A(q^2)}{m_H + m_A} + \frac{m_H + m_A}{2 m_H} V_1( q^2) \label{t1new}\\
& & T_2 (q^2) = \frac{2}{m^2_H - m^2_A} \left [ \frac{(m_H -y)(m_H + m_A)}{2} V_1 (q^2) + \frac{m_H ( y^2- m^2_A)}{m_H + m_A} A(q^2) \right ]\label{t2new} \\
& & T_3(q^2) = - \frac{m_H +m_A}{2 m_H} V_1 (q^2) + \frac{m_H -m_A}{2 m_H} [ V_2 (q^2) - V_3 (q^2) ]\nonumber \\
& & \hspace{1.5cm}  + \frac{m^2_H +3 m^2_A -q^2}{ 2m_H ( m_H + m_A)} A(q^2) \label{t3new}
\end{eqnarray}
with
\begin{eqnarray}
V_3(q^2) = \frac{2 m_H(m_H+ m_A)}{q^2}\left [ \frac{m_H + m_A}{2m_A } V_1(q^2) -  \frac{m_H - m_A}{2m_A } V_2(q^2) -V_0 (q^2) \right ]\label{t4new}
\end{eqnarray}
Therefore, the penguin type form factors can be obtained directly from the corresponding semileptonic ones, which is similar to the
case of S-wave mesons\cite{HQEFT1,HQEFT2}.

\section{Numerical analysis and results of the form factors}

With Eqs.(\ref{a1})-(\ref{v}), (\ref{l1})-(\ref{l4}) and
(\ref{t1new})-(\ref{t4new}) given above, we are now in a position to
calculate the $B_{(s)}$ $\rightarrow$ $a_1(1260)$, $b_1(1235)$, $K_{1A}$, $K_{1B}$, $f_1$,
$f_8$, $h_1$, $h_8$ form factors. For this
purpose, it needs to know the light cone DAs of axial vector mesons, which
have been studied via QCD sum rules in Ref.\cite{DAs}. In the
present work, we shall neglect the contributions of $g_3 (u)$ and $h_3 (u)$ and
consider the light cone DAs to twist-3.

The twist-2 light cone DAs for the $1^3P_1$ mesons  have the following forms\cite{LCSRA1,DAs}
\begin{eqnarray}
& & \phi_\parallel (u) = 6 u (1-u) \left [ 1+ 3 a^\parallel_1 (2u-1) + a^\parallel_2 \frac{3}{2} ( 5 (2u-1)^2-1) \right ] \\
& & \phi_\perp (u) =6 u (1-u) \left [ a^\perp_0 + 3 a^\perp_1 (2u-1) + a^\perp_2 \frac{3}{2} ( 5 (2u-1)^2-1) \right ] \\
\end{eqnarray}
For the $1^1P_1$ mesons,
\begin{eqnarray}
& & \phi_\parallel (u) = 6 u (1-u) \left [ a^\parallel_0+ 3 a^\parallel_1 (2u-1) + a^\parallel_2 \frac{3}{2} ( 5 (2u-1)^2-1) \right ] \\
& & \phi_\perp (u) =6 u (1-u) \left [ 1 + 3 a^\perp_1 (2u-1) + a^\perp_2 \frac{3}{2} ( 5 (2u-1)^2-1) \right ] \\
\end{eqnarray}
The explicit expressions for the twist-3 light cone DAs are given in APPENDIX~\ref{twist3DAs}. The parameters appearing in the light cone DAs depend on the factorization scale $\mu_f$ and we choose $\mu_f = \sqrt{m^2_H - m^2_b} \simeq 2.2 {\rm GeV}$ for $B_{(s)}$ decays.

The parameters relevant to specific light mesons are collected in TABLE~\ref{tab:3P1para} and \ref{tab:1P1para} for the $1^3P_1$ and $1^1P_1$ axial vector mesons, respectively. Considering the $SU(3)$ breaking correction, the sign of the $G$-parity violation parameters are opposite for $K_{1A(B)}$ and $\bar{K}_{1A(B)}$. Here we adopt the convention that the former contain a strange quark , while the later contain an antistrange quark.
\begin{table}[H]
\centering
\begin{tabular}{|c|c|c|c|c|c|c|}
\hline
State & $a_1(1260)$ & $f_1(1^3P_1)$ & $f_8(1^3P_1)$ & $K_{1A}$ & $\bar{K}_{1A}$ \\
\hline
$m_A$ & $1.23 \pm 0.06$ & $1.28 \pm 0.06$ & $1.29 \pm 0.05$ & $1.31 \pm 0.06$ & $1.31 \pm 0.06$ \\
\hline
$f_A$ & $0.238 \pm 0.010$ & $0.245 \pm 0.013$  & $0.239 \pm 0.013$ & $0.250 \pm 0.013$ & $0.250 \pm 0.013$ \\
\hline
$a^\parallel_1$ & --- & ---  & --- & $-0.25^{+0.00}_{-0.17}$ & $0.25^{+0.17}_{-0.00}$\\
\hline
$a^\parallel_2$ & $-0.01 \pm 0.01 $ & $-0.03 \pm 0.02$  & $-0.05 \pm 0.03$ & $-0.04 \pm 0.02$ & $-0.04 \pm 0.02$\\
\hline
$a^\perp_0$ & --- & --- & --- & $0.25^{+0.13}_{-0.16}$ & $-0.25^{+0.16}_{-0.13}$  \\
\hline
$a^\perp_1$ & $-0.85 \pm 0.28$ & $-0.86 \pm 0.29$ & $-0.90 \pm 0.25$ & $-0.88 \pm 0.39$ & $-0.88 \pm 0.39$\\
\hline
$a^\perp_2$ & --- & --- & --- & $0.01 \pm 0.15$ & $-0.01 \pm 0.15$\\
\hline
$f^V_{3,^3P_1}$ & $0.0036 \pm 0.0018$ & $0.0036 \pm 0.0018$ & $0.0035 \pm 0.0018$ & $0.0034 \pm 0.0018$ & $0.0034 \pm 0.0018$\\
\hline
$\omega^V_{^3P_1}$ & $-2.9 \pm 0.9$ & $-2.8 \pm 0.9$ & $-3.0 \pm 1.0$ & $-3.1 \pm 1.1$ & $-3.1 \pm 1.1$ \\
\hline
$\sigma^V_{^3P_1}$ & --- & --- & --- & $0.01 \pm 0.04$ & $-0.01 \pm 0.04$ \\
\hline
$f^A_{3,^3P_1}$ & $0.0012 \pm 0.0005$ & $0.0012 \pm 0.0005$ & $0.0015 \pm 0.0005$ & $0.0014 \pm 0.0007$ & $0.0014 \pm 0.0007$ \\
\hline
$\lambda^A_{^3P_1}$ & --- & --- & --- & $-0.12 \pm 0.22$ & $0.12 \pm 0.22$ \\
\hline
$\sigma^A_{^3P_1}$ & --- & --- & --- & $-1.9 \pm 1.1$ & $1.9 \pm 1.1$ \\
\hline
$f^\perp_{3,^3P_1}$ & $-0.009 \pm 0.002$ & $-0.009 \pm 0.002$ & $-0.009 \pm 0.002$ & $-0.009 \pm 0.002$ & $-0.009 \pm 0.002$ \\
\hline
$\omega^\perp_{^3P_1}$ & $-2.9 \pm 0.3$ & $-2.6 \pm 0.3$ & $-2.4 \pm 0.4$ & $-2.6 \pm 0.4$ & $-2.6 \pm 0.4$ \\
\hline
$\sigma^\perp_{^3P_1}$ & --- & --- & --- & $0.05 \pm 0.15$ & $-0.05 \pm 0.15$ \\
\hline
\end{tabular}
\caption{Parameters relevant to specific $1^3P_1$ axial vector mesons\cite{LCSRA1,DAs}.} \label{tab:3P1para}
\end{table}

\begin{table}[H]
\centering
\begin{tabular}{|c|c|c|c|c|c|c|}
\hline
State & $b_1(1235)$ & $h_1(1^1P_1)$ & $h_8(1^1P_1)$ & $K_{1B}$ & $\bar{K}_{1B}$ \\
\hline
$m_A$ & $1.21 \pm 0.07$ & $1.23 \pm 0.07$ & $1.37 \pm 0.07$ & $1.34 \pm 0.08$ & $1.34 \pm 0.08$ \\
\hline
$f_A$ & $0.180 \pm 0.008$ & $0.180 \pm 0.012$  & $0.190 \pm 0.010$ & $0.190 \pm 0.010$ & $0.190 \pm 0.010$ \\
\hline
$a^\parallel_0$ & --- & --- & --- & $-0.19 \pm 0.07$ & $0.19 \pm 0.07$ \\
\hline
$a^\parallel_1$ & $-1.61 \pm 0.29$ & $-1.65 \pm 0.29$  & $-1.61 \pm 0.29$ & $-1.57 \pm 0.37$ & $-1.57 \pm 0.37$\\
\hline
$a^\parallel_2$ & --- & ---  & --- & $0.07^{+0.11}_{-0.14}$ & $-0.07^{+0.14}_{-0.11}$ \\
\hline
$a^\perp_1$ & --- & --- & --- & $0.24^{+0.00}_{-0.27}$ & $-0.24^{+0.27}_{-0.00}$\\
\hline
$a^\perp_2$ & $0.02 \pm 0.15$ & $0.14 \pm 0.17$ & $0.11 \pm 0.17$ & $-0.02 \pm 0.17$ & $-0.02 \pm 0.17$\\
\hline
$f^V_{3,^1P_1}$ & $0.0030 \pm 0.0011$ & $0.0027 \pm 0.0012$ & $0.0027 \pm 0.0012$ & $0.0029 \pm 0.0012$ & $0.0029 \pm 0.0012$\\
\hline
$\lambda^V_{^1P_1}$ & --- & --- & --- & $-0.23 \pm 0.18$ & $0.23 \pm 0.18$ \\
\hline
$\sigma^V_{^1P_1}$ & --- & --- & --- & $1.3 \pm 0.8$ & $-1.3 \pm 0.8$ \\
\hline
$f^A_{3,^1P_1}$ & $-0.0036 \pm 0.0014$ & $-0.0033 \pm 0.0014$ & $-0.0035 \pm 0.0014$ & $-0.0041 \pm 0.0018$ & $-0.0041 \pm 0.0018$ \\
\hline
$\omega^A_{^1P_1}$ & $-1.4 \pm 0.3$ & $-1.7 \pm 0.4$ & $-2.9 \pm 0.8$ & $-1.7 \pm 0.4$ & $-1.7 \pm 0.4$ \\
\hline
$\sigma^A_{^1P_1}$ & --- & --- & --- & $0.03 \pm 0.03$ & $-0.03 \pm 0.03$ \\
\hline
$f^\perp_{3,^1P_1}$ & $0.006 \pm 0.003$ & $0.006 \pm 0.003$ & $0.006 \pm 0.003$ & $0.006 \pm 0.003$ & $0.006 \pm 0.003$ \\
\hline
$\lambda^\perp_{^1P_1}$ & --- & --- & --- & $0.25 \pm 0.25$ & $-0.25 \pm 0.25$ \\
\hline
$\sigma^\perp_{^1P_1}$ & --- & --- & --- & $-0.76 \pm 0.56$ & $0.76 \pm 0.56$ \\
\hline
\end{tabular}
\caption{Parameters relevant to specific $1^1P_1$ axial vector mesons\cite{LCSRA1,DAs}.} \label{tab:1P1para}
\end{table}

Other input parameters involved are as
follows\cite{HQEFT1,HQEFT2,PDG},
\begin{eqnarray}
& & m_B = 5.28 {\rm GeV}, \hspace{0.5cm} m_{B_s} = 5.37 {\rm GeV},
\hspace{0.5cm} \bar{\Lambda}_B = 0.53 {\rm GeV} \nonumber
\\
& & \bar{\Lambda}_{B_s} =0.62 {\rm GeV}, \hspace{0.5cm}
\bar{\Lambda} =0.53 {\rm GeV}, \hspace{0.5cm } F = 0.3 {\rm GeV}^{3/2}
\nonumber
\end{eqnarray}

\begin{figure}
\centering
\includegraphics[width=3in]{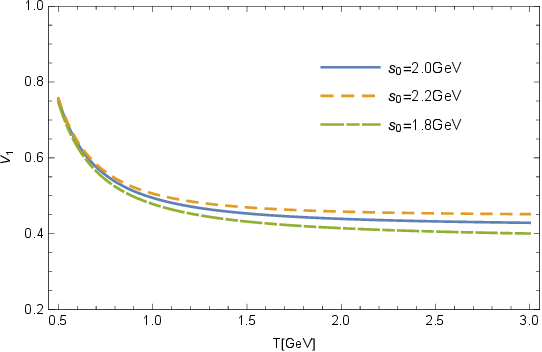} \hfill
\includegraphics[width=3in]{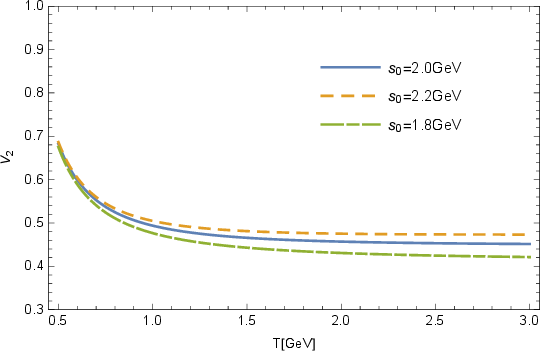} \\
\includegraphics[width=3in]{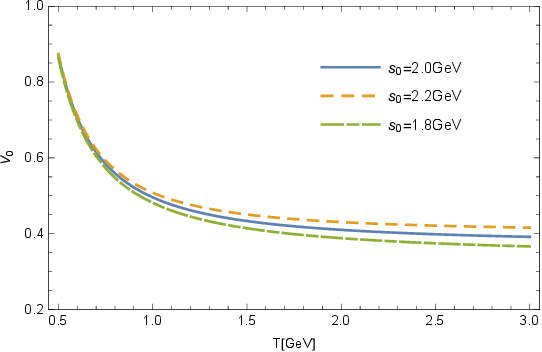}\hfill
\includegraphics[width=3in]{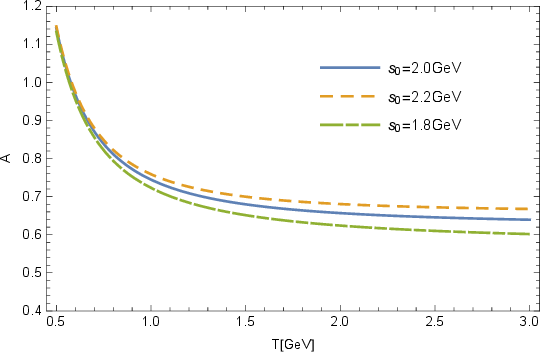}
\caption{$B \rightarrow a_1(1260)$ form factors as
functions of $T$ for different $s_0$ at $q^2=0$. } \label{Fig:Ba1T}
\end{figure}

\begin{figure}
\centering
\includegraphics[width=3in]{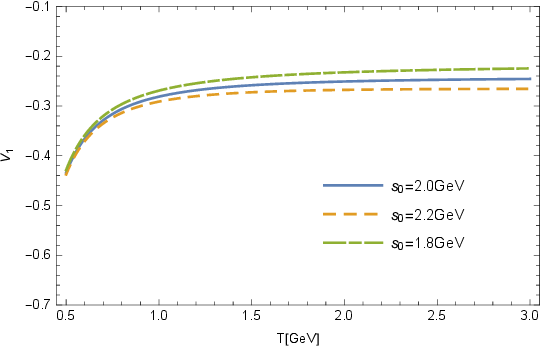} \hfill
\includegraphics[width=3in]{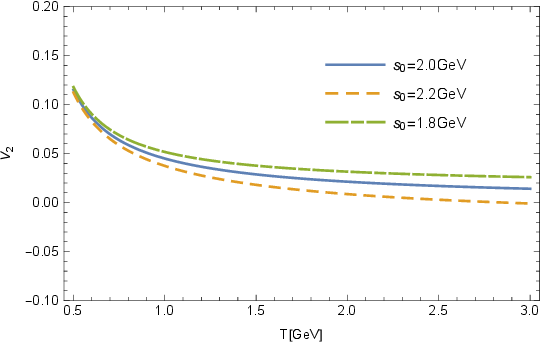} \\
\includegraphics[width=3in]{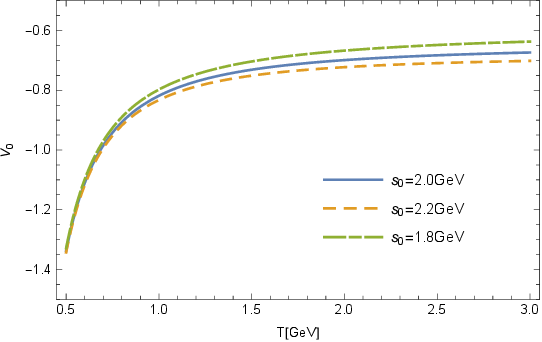}\hfill
\includegraphics[width=3in]{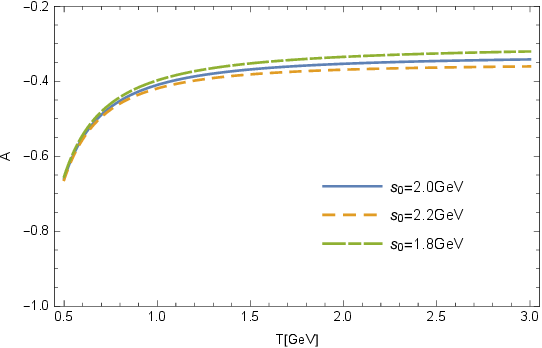}
\caption{$B \rightarrow b_1(1235)$ form factors as
functions of $T$ for different $s_0$ at $q^2=0$. } \label{Fig:Bb1T}
\end{figure}

As shown in Eqs.(\ref{l1})-(\ref{l4}), two free parameters, {\it
i.e.} $s_0$, $T$ are involved in our calculations. $s_0$ is related
to the threshold energy of initial heavy meson and $T$ is the Borel
parameter. According to the requirements of LCSR, we choose the
region of these two parameters so that the curves of form
factors become most stable. In the evaluation, we adjust $s_0$ and
$T$ for all relevant $B \rightarrow A$ decays consistently and the
same procedure is also performed for $B_s \rightarrow A$ decays. The
variation of $B \rightarrow a_1 (1260), b_1(1235)$
 form factors as functions of $T$ for different $s_0$ at
the maximally recoiling point ($q^2=0$) are shown in FIG.\ref{Fig:Ba1T},
FIG.\ref{Fig:Bb1T}, respectively.
The corresponding figures for other decay modes are collected in APPENDIX~\ref{FFfigT}. Note that we have only shown the curves of the semileptonic type
form factors here in that the penguin type form factors can be
obtained directly from the corresponding semileptonic ones as
mentioned above.
Based on the curves given here, we can see that choosing $s_0 = 2.0
\pm 0.2 {\rm GeV}, T=2.0 \pm 0.5 {\rm GeV} $ for $B \rightarrow T$
decays is reasonable. Similarly, we determine the region of
$s_0$ and $T$ for $B_s \rightarrow T$ decays. It is found that the
region for $B_s \rightarrow T$ can be chosen as $s_0 =2.2 \pm 0.2 {\rm GeV}, T=2.0 \pm 0.5 {\rm GeV} $.

As well known, the LCSR may be broken down at large momentum transfer, {\it i.e.} $q^2 \sim m^2_b - 2 m_b \bar{\Lambda}$. Therefore, it is
needed to extrapolate our LCSR results to the entire physical region via certain parametrization. Here, we shall adopt
the Bourrely Caprini Lellouch (BCL) version of the z-series
expansion\cite{BCLP1,BCLP2}
\begin{eqnarray}
F (q^2) = \frac{F(0)}{1- q^2/m^2_{F, pole}} \left \{ 1+ b_1 \left [ z(q^2) -z(0) + \frac{1}{2} \left ( z(q^2)^2 - z(0)^2 \right ) \right ]\right \}, \label{param}
\end{eqnarray}
where $F$ can be any of the form factors $A_1$, $A_2$, $A_0$, $V$, and
\begin{eqnarray}
z(q^2)=\frac{\sqrt{t_+ - q^2} - \sqrt{t_+ - t_0}}{\sqrt{t_+ - q^2} + \sqrt{t_+ - t_0}},
\end{eqnarray}
with $t_+ = (m_H-m_A)^2$, $t_0= (m_H + m_A) (\sqrt{m_H}-\sqrt{m_A})^2$. For the resonance masses $m_{F, pole}$, we take the values from PDG\cite{PDG} and heavy meson chiral perturbation theory\cite{HMChPT}, as listed in TABLE~\ref{tab:RMass}.
\begin{table}[H]
\centering
\begin{tabular}{|c|c|c|c|}
\hline
~~~~$F$~~~~ & ~~~$J^P$ ~~~ & ~~~$m^{b \rightarrow u(d)}_{F, pole}$ ~~~ & ~~~$m^{b \rightarrow s}_{F, pole}$ ~~~ \\
\hline
$V_0$ & $0^+$ & $5.681$ & $5.711$  \\
\hline
$A$ & $1^+$ & $5.726$  & $5.829$  \\
\hline
$V_1, V_2$ &$1^-$ & 5.325 & 5.415  \\
\hline
\end{tabular}
\caption{The resonance masses entering the z-series expansion of form
factors\cite{PDG,HMChPT}, in units of GeV.} \label{tab:RMass}
\end{table}

\begin{figure}[H]
\centering
\includegraphics[width=3in]{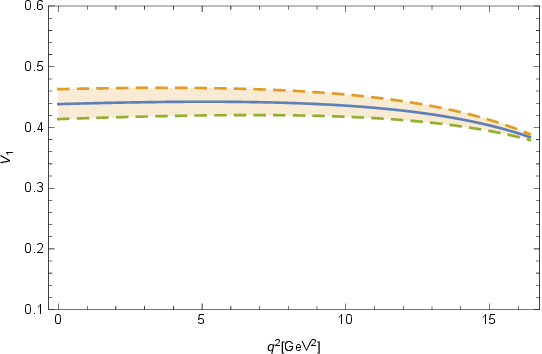} \hfill
\includegraphics[width=3in]{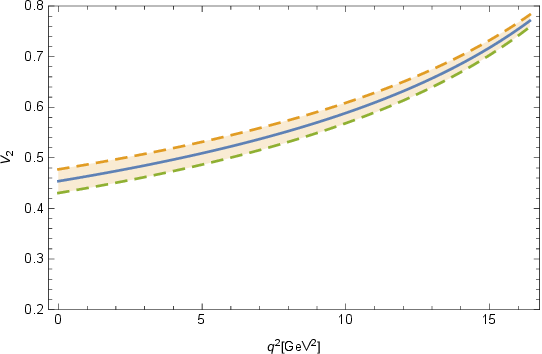} \\
\includegraphics[width=3in]{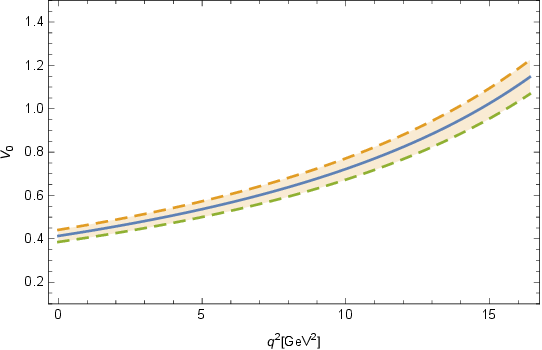}\hfill
\includegraphics[width=3in]{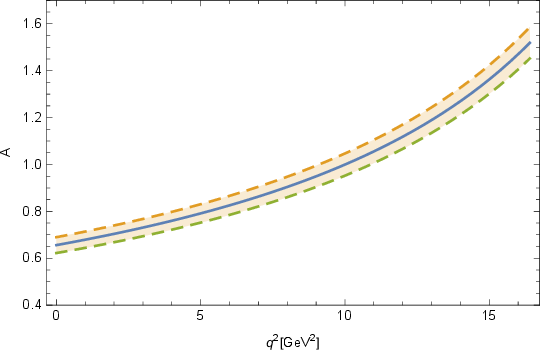}
\caption{$B \rightarrow a_1(1260)$ form factors as
functions of $q^2$. } \label{Fig:Ba1q}
\end{figure}

\begin{figure}[H]
\centering
\includegraphics[width=3in]{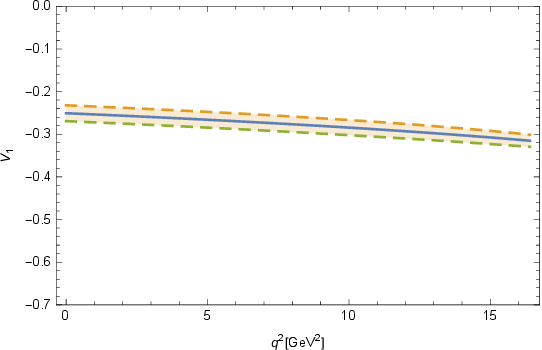} \hfill
\includegraphics[width=3in]{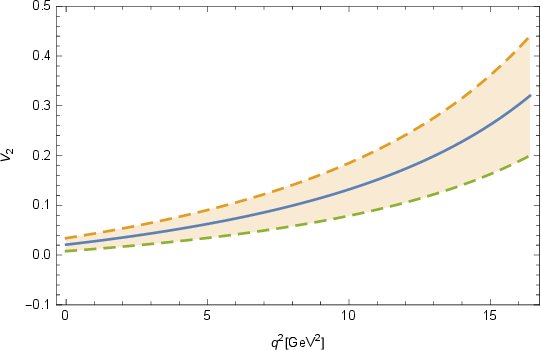} \\
\includegraphics[width=3in]{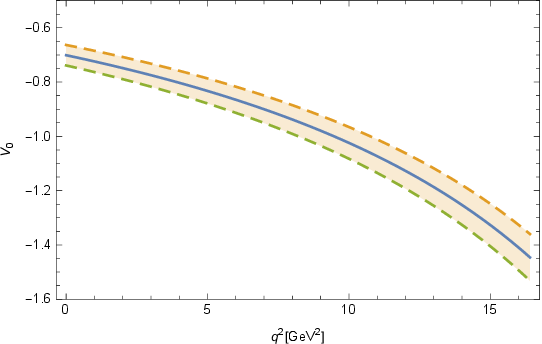}\hfill
\includegraphics[width=3in]{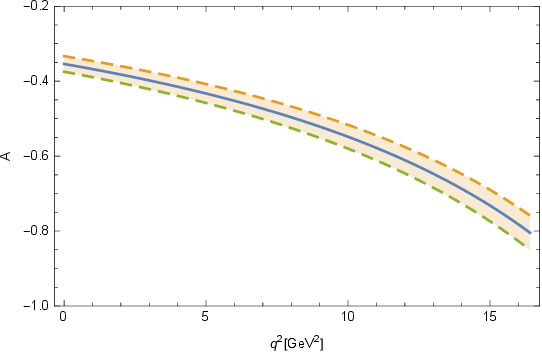}
\caption{$B \rightarrow b_1(1235)$ form factors as
functions of $q^2$. } \label{Fig:Bb1q}
\end{figure}

Using the LCSR predictions in low $q^2$ region (specifically, we choose $0 \leq q^2 \leq 10{\rm GeV}^2$) to fit the parameter
$b_1$ in Eq.(\ref{param}), we obtain the behavior of form factors
in the whole physical region, as shown in FIG.\ref{Fig:Ba1q} and FIG.\ref{Fig:Bb1q} for $B \rightarrow a_1(1260)$ and $B \rightarrow b_1(1235)$ decays, respectively. The corresponding figures for other decay modes are collected in APPENDIX~\ref{FFfigq}. Generally speaking, for the case of $1^3P_1$ axial vector meson in the final states, the form factors $V_1, V_2, V_0, A$ are positive and the $V_1$ decreases with $q^2$, while the behaviors of $V_2$, $V_0$, $A$ are opposite. For the decays with $1^1P_1$ axial vector meson in the final states, the form factors $V_1, V_0, A$ are negative and decrease with $q^2$, while the $V_2$ are positive and increases with $q^2$. Note that the signs and behaviors with $q^2$ of form factors are all opposite with the descriptions above for the $B_s \rightarrow f_8, h_8$ decays due to the fact that the flavor contents of these two mesons are $1/\sqrt{6} (u \bar{u} +d \bar{d} - 2 s \bar{s})$.
Numerical results are presented in TABLE~\ref{tab:FFBto3P1}-\ref{tab:FFBsto1P1}. The uncertainties of form factors come form the two free parameters
$s_0$ and $T$. It is found that the uncertainties of form factors are about (5-8)\% at $q^2=0$, except the $V_2$ for the decays with a $1^1P_1$ axial vector meson in the final states. In these cases, the uncertainties of $V_2$ are apparently large due to the tininess of its central values at $q^2=0$.
For the $B \rightarrow a_1(1260), b_1(1235)$ modes, the form factors given here correspond to $a^\pm_1(1260), b^\pm_1(1235)$. In the case of
$a^0_1(1260), b^0_1(1235)$ in the final states, the form factors can be obtained by multiplying $1/\sqrt{2}$ in light of the fact that the
flavor contents of $a^0_1(1260)$ and $b^0_1(1235)$ are $1/\sqrt{2} (d\bar{d} - u\bar{u})$.
The comparison of form factors at $q^2=0$ given in this work with
other groups are shown in TABLE \ref{tab:FFSLCBto3P1}-\ref{tab:FFSLCBsto1P1}. The penguin type form
factors are also listed in TABLE \ref{tab:FFPCBto3P1}-\ref{tab:FFPCBsto1P1} for a complete
analysis. We can find that large differences
exist between the results of different approaches. For $B_{(s)} \rightarrow A(1^1P_1)$ decays, the form factors $V_1, V_0, A$ predicted by light cone sum rules (this work, LCSR\cite{LCSRA2}) all have
negative signs, which is opposite from those obtained via pQCD\cite{pQCD2} and LFQM\cite{LFQM2}. This is due to the fact that the decay constants, $f_{^3P_1}$ and $f^\perp_{^1P_1}$, are of the same sign\cite{LCSRA2}. The form factors obtained in this work are in general
some larger, yet still compatible with those results of
LCSR\cite{LCSRA1} as a whole.

\begin{table}[H]
\centering
\begin{tabular}{|c|c|c|c|}
\hline Decays & ~~~$F$~~~ & ~~~~~~$F(0)$ ~~~~~~& ~~~~~~$b_1$ ~~~~~~\\
\hline $B \rightarrow a_1 (1260)$ & $V_1$ & $0.438 \pm 0.025$ & $5.16, 5.29, 5.02$  \\
\cline{2-4} & $V_2$ & $0.454 \pm 0.023 $ & $2.32, 2.52, 2.10$   \\
\cline{2-4}  & $V_0$ & $0.413 \pm 0.028$ & $-2.98, -2.98, -2.97$   \\
\cline{2-4}  & $A$ & $0.656 \pm 0.034$ & $0.20, 0.25, 0.13$   \\
\hline
$B \rightarrow K_{1A} $ & $V_1$ & $0.389 \pm 0.021$ &  $5.36, 5.47, 5.24$ \\
\cline{2-4} & $V_2$ & $0.469 \pm 0.021 $ & $1.83, 1.75, 1.93$  \\
\cline{2-4}  & $V_0$ & $0.267 \pm 0.021 $ & $-4.95, -5.02, -4.88$  \\
\cline{2-4}  & $A$ & $0.622 \pm 0.031 $ & $-0.92, -0.94, -0.91$   \\
\hline
$B \rightarrow \bar{K}_{1A}$ & $V_1$ & $0.536 \pm 0.032$ &  $4.92, 5.10, 4.70$ \\
\cline{2-4} & $V_2$ & $0.499 \pm 0.030 $ & $2.63, 3.17, 2.03$  \\
\cline{2-4}  & $V_0$ & $0.592 \pm 0.037 $ & $-2.76, -2.68, -2.84$  \\
\cline{2-4}  & $A$ & $0.819 \pm 0.044 $ & $-1.44, -1.23, -1.68$   \\
\hline
$B \rightarrow f_1 (1^3P_1)$ & $V_1$ & $0.262 \pm 0.015$ & $5.26, 5.39, 5.11$  \\
\cline{2-4} & $V_2$ & $0.272 \pm 0.014 $ & $2.43, 2.65, 2.18$   \\
\cline{2-4}  & $V_0$ & $0.246 \pm 0.017$ & $-3.19, -3.19, -3.19$   \\
\cline{2-4}  & $A$ & $0.399 \pm 0.021$ & $0.14, 0.21, 0.06$   \\
\hline
$B \rightarrow f_8 (1^3P_1)$ & $V_1$ & $0.186 \pm 0.011$ & $5.32, 5.46, 5.17$  \\
\cline{2-4} & $V_2$ & $0.196 \pm 0.010 $ & $2.41, 2.66, 2.14$   \\
\cline{2-4}  & $V_0$ & $0.170 \pm 0.012$ & $-3.30, -3.30, -3.30$   \\
\cline{2-4}  & $A$ & $0.285 \pm 0.015$ & $0.18, 0.26, 0.09$   \\
\hline
\end{tabular}
\caption{The form factors of $B \rightarrow A(1^3P_1)$ decays, where the first, second and third results of $b_1$ correspond to the central, maximum and minimum values of form factors, respectively.} \label{tab:FFBto3P1}
\end{table}

\begin{table}[H]
\centering
\begin{tabular}{|c|c|c|c|}
\hline Decays & ~~~$F$~~~ & ~~~~~~$F(0)$ ~~~~~~& ~~~~~~$b_1$ ~~~~~~\\
\hline
$B \rightarrow b_1 (1235)$ & $V_1$ & $-0.251 \pm 0.019$ &  $3.84, 3.71, 3.96$ \\
\cline{2-4} & $V_2$ & $0.021 \pm 0.013 $ & $-44.60, -36.54, -79.29$  \\
\cline{2-4}  & $V_0$ & $-0.701 \pm 0.038 $ & $-0.12, -0.07, -0.16$  \\
\cline{2-4}  & $A$ & $-0.354 \pm 0.021 $ & $-1.10, -1.12, -1.09$ \\
\hline
$B \rightarrow K_{1B}$ & $V_1$ & $-0.337 \pm 0.024$ &  $4.29, 4.12, 4.43$ \\
\cline{2-4} & $V_2$ & $-0.031 \pm 0.017 $ & $45.59, 125.90, 23.54$   \\
\cline{2-4}  & $V_0$ & $-0.792 \pm 0.044 $ & $-0.64, -0.60, -0.68$  \\
\cline{2-4}  & $A$ & $-0.510 \pm 0.030 $ & $-1.62, -1.75, -1.50$   \\
\hline
$B \rightarrow \bar{K}_{1B}$ & $V_1$ & $-0.157 \pm 0.014$ &  $2.42, 2.30, 2.51$ \\
\cline{2-4} & $V_2$ & $0.179 \pm 0.010 $ & $-5.32, -7.75, -2.61$  \\
\cline{2-4}  & $V_0$ & $-0.650 \pm 0.033 $ & $0.12, 0.09, 0.13$  \\
\cline{2-4}  & $A$ & $-0.213 \pm 0.015 $ & $-2.32, -2.23, -2.39$   \\
\hline
$B \rightarrow h_1 (1^1P_1)$ & $V_1$ & $-0.160 \pm 0.010$ & $4.63, 4.56, 4.68$  \\
\cline{2-4} & $V_2$ & $-0.006 \pm 0.005 $ & $127.34, 1331.26, 54.13$   \\
\cline{2-4}  & $V_0$ & $-0.414 \pm 0.022$ & $0.00, 0.08, -0.06$   \\
\cline{2-4}  & $A$ & $-0.227 \pm 0.012$ & $0.88, 1.02, 0.76$   \\
\hline
$B \rightarrow h_8 (1^1P_1)$ & $V_1$ & $-0.116 \pm 0.008$ & $4.89, 4.83, 4.94$  \\
\cline{2-4} & $V_2$ & $0.016 \pm 0.004 $ & $-43.70, -42.27, -46.06$   \\
\cline{2-4}  & $V_0$ & $-0.305 \pm 0.017$ & $-0.02, 0.06, -0.09$   \\
\cline{2-4}  & $A$ & $-0.170 \pm 0.009$ & $0.80, 0.93, 0.68$   \\
\hline
\end{tabular}
\caption{The form factors of $B \rightarrow A(1^1P_1)$ decays, where the first, second and third results of $b_1$ correspond to the central, maximum and minimum values of form factors, respectively.} \label{tab:FFBto1P1}
\end{table}

\begin{table}[H]
\centering
\begin{tabular}{|c|c|c|c|}
\hline Decays & ~~~$F$~~~ & ~~~~~~$F(0)$ ~~~~~~& ~~~~~~$b_1$ ~~~~~~\\
\hline $B_s \rightarrow \bar{K}_{1A}$ & $V_1$ & $0.569 \pm 0.032$ & $5.45, 5.47, 5.43$  \\
\cline{2-4} & $V_2$ & $0.532 \pm 0.026 $ & $3.43, 3.57, 3.28$   \\
\cline{2-4}  & $V_0$ & $0.625 \pm 0.042$ & $-2.17, -2.25, -2.08$   \\
\cline{2-4}  & $A$ & $0.860 \pm 0.046$ & $-0.47, -0.60, -0.32$   \\
\hline
$B_s \rightarrow K_{1A} $ & $V_1$ & $0.410 \pm 0.023$ &  $5.97, 5.89, 6.06$ \\
\cline{2-4} & $V_2$ & $0.493 \pm 0.021 $ & $3.17, 2.58, 3.81$  \\
\cline{2-4}  & $V_0$ & $0.283 \pm 0.025 $ & $-4.58, -4.66, -4.49$  \\
\cline{2-4}  & $A$ & $0.649 \pm 0.034 $ & $0.23, -0.16, 0.65$   \\
\hline
$B_s \rightarrow f_1 (1^3P_1)$ & $V_1$ & $0.270 \pm 0.015$ & $5.36, 5.35, 5.37$  \\
\cline{2-4} & $V_2$ & $0.282 \pm 0.012 $ & $2.97, 2.76, 3.20$   \\
\cline{2-4}  & $V_0$ & $0.250 \pm 0.018$ & $-3.05, -3.10, -2.99$   \\
\cline{2-4}  & $A$ & $0.416 \pm 0.021$ & $0.49, 0.27, 0.74$   \\
\hline
$B_s \rightarrow f_8 (1^3P_1)$ & $V_1$ & $-0.382 \pm 0.021$ & $5.43, 5.43, 5.42$  \\
\cline{2-4} & $V_2$ & $-0.403 \pm 0.018 $ & $3.04, 3.24, 2.86$   \\
\cline{2-4}  & $V_0$ & $-0.348 \pm 0.026$ & $-3.09, -3.04, -3.13$   \\
\cline{2-4}  & $A$ & $-0.594 \pm 0.030$ & $0.53, 0.77, 0.31$   \\
\hline
\end{tabular}
\caption{The form factors of $B_s \rightarrow A(1^3P_1)$ decays, where the first, second and third results of $b_1$ correspond to the central, maximum and minimum values of form factors, respectively.} \label{tab:FFBsto3P1}
\end{table}

\begin{table}[H]
\centering
\begin{tabular}{|c|c|c|c|}
\hline Decays & ~~~$F$~~~ & ~~~~~~$F(0)$ ~~~~~~& ~~~~~~$b_1$ ~~~~~~\\
\hline
$B_s \rightarrow \bar{K}_{1B}$ & $V_1$ & $-0.170 \pm 0.015$ &  $2.83, 2.72, 2.92$ \\
\cline{2-4} & $V_2$ & $0.169 \pm 0.016 $ & $-2.98, -5.61, 0.20$  \\
\cline{2-4}  & $V_0$ & $-0.684 \pm 0.037 $ & $0.87, 1.18, 0.60$  \\
\cline{2-4}  & $A$ & $-0.226 \pm 0.017 $ & $-1.92, -1.70, -2.11$ \\
\hline
$B_s \rightarrow K_{1B}$ & $V_1$ & $-0.361 \pm 0.025$ &  $4.74, 4.68, 4.80$ \\
\cline{2-4} & $V_2$ & $-0.052 \pm 0.019 $ & $23.94, 45.99, 13.58$   \\
\cline{2-4}  & $V_0$ & $-0.832 \pm 0.053 $ & $0.06, 0.35, -0.19$  \\
\cline{2-4}  & $A$ & $-0.537 \pm 0.033 $ & $-0.87, -0.76, -0.97$   \\
\hline
$B_s \rightarrow h_1 (1^1P_1)$ & $V_1$ & $-0.160 \pm 0.010$ & $4.65, 4.66, 4.64$  \\
\cline{2-4} & $V_2$ & $-0.004 \pm 0.006 $ & $168.93, -412.32, 48.24$   \\
\cline{2-4}  & $V_0$ & $-0.425 \pm 0.024$ & $0.46, 0.77, 0.19$   \\
\cline{2-4}  & $A$ & $-0.228 \pm 0.013$ & $0.88, 1.16, 0.63$   \\
\hline
$B_s \rightarrow h_8 (1^1P_1)$ & $V_1$ & $0.233 \pm 0.015$ & $4.91, 4.90, 4.92$  \\
\cline{2-4} & $V_2$ & $-0.035 \pm 0.009 $ & $-35.02, -34.80, -35.14$   \\
\cline{2-4}  & $V_0$ & $0.626 \pm 0.036$ & $0.44, 0.17, 0.74$   \\
\cline{2-4}  & $A$ & $0.343 \pm 0.021$ & $0.78, 0.54, 1.05$   \\
\hline
\end{tabular}
\caption{The form factors of $B_s \rightarrow A(1^1P_1)$ decays, where the first, second and third results of $b_1$ correspond to the central, maximum and minimum values of form factors, respectively.} \label{tab:FFBsto1P1}
\end{table}

\begin{table}[H]
\centering
\begin{tabular}{|c|c|c|c|c|c|}
\hline Decays & Reference & $V_1(0)$ & $V_2 (0)$ & $V_0 (0)$ & $A (0)$ \\
\hline
$B \rightarrow a_1 (1260)$ & This work & $0.438 \pm 0.025$ & $0.454 \pm 0.023$ & $0.413 \pm 0.028 $ &  $0.656 \pm 0.034$ \\
\cline{2-6}
& LCSR\cite{LCSRA2} & $0.37 \pm 0.07$ & $0.42 \pm 0.08$ & $0.30 \pm 0.05$ & $0.48 \pm 0.09$ \\
\cline{2-6}
& pQCD\cite{pQCD2} & $0.43^{+0.16}_{-0.15}$ & $0.13^{+0.03}_{-0.04}$ & $0.34^{+0.16}_{-0.17}$ & $0.26^{+0.09}_{-0.09}$ \\
\cline{2-6}
& LFQM\cite{LFQM2} & $0.36$ & $0.17$ & $0.14$ & $0.24$ \\
\hline
$B \rightarrow K_{1A}$ & This work & $0.389 \pm 0.021$ & $0.469 \pm 0.021$ & $0.267 \pm 0.021 $ &  $0.622 \pm 0.031 $ \\
\cline{2-6}
& LCSR\cite{LCSRA2} & $0.34 \pm 0.07$ & $0.41 \pm 0.07$ & $0.22 \pm 0.04$ & $0.45 \pm 0.09$ \\
\cline{2-6}
& pQCD\cite{pQCD2} & $0.47^{+0.13}_{-0.11}$ & $0.14^{+0.05}_{-0.06}$ & $0.35^{+0.22}_{-0.22}$ & $0.27^{+0.12}_{-0.12}$ \\
\cline{2-6}
& LFQM\cite{LFQM2} & $0.39$ & $0.17$ & $0.16$ & $0.27$ \\
\hline
$B \rightarrow f_1(1^3P_1)$ & This work & $0.262 \pm 0.015$ & $0.272 \pm 0.014$ & $0.246 \pm 0.017 $ &  $0.399 \pm 0.021 $ \\
\cline{2-6}
& LCSR\cite{LCSRA2} & $0.23 \pm 0.04$ & $0.26 \pm 0.05$ & $0.18 \pm 0.03$ & $0.30 \pm 0.05$ \\
\cline{2-6}
& pQCD\cite{pQCD2} & $0.27^{+0.09}_{-0.09}$ & $0.08^{+0.02}_{-0.02}$ & $0.21^{+0.11}_{-0.10}$ & $0.16^{+0.06}_{-0.05}$ \\
\cline{2-6}
& LFQM\cite{LFQM2} & $0.37$ & $0.17$ & $0.14$ & $0.24$\\
\hline
$B \rightarrow f_8(1^3P_1)$ & This work & $0.186 \pm 0.011$ & $0.196 \pm 0.010$ & $0.170 \pm 0.012 $ &  $0.285 \pm 0.015 $ \\
\cline{2-6}
& LCSR\cite{LCSRA2} & $0.16 \pm 0.03$ & $0.19 \pm 0.03$ & $0.12 \pm 0.02$ & $0.22 \pm 0.04$ \\
\cline{2-6}
& pQCD\cite{pQCD2} & $0.19^{+0.06}_{-0.06}$ & $0.05^{+0.01}_{-0.01}$ & $0.15^{+0.07}_{-0.07}$ & $0.11^{+0.04}_{-0.03}$ \\
\hline
\end{tabular}
\caption{Comparison of the $B \rightarrow A(1^3P_1) $ form factors at $q^2=0$ given in this
work with other groups. }\label{tab:FFSLCBto3P1}
\end{table}

\begin{table}[H]
\centering
\begin{tabular}{|c|c|c|c|c|c|}
\hline Decays & Reference & $V_1(0)$ & $V_2 (0)$ & $V_0 (0)$ & $A (0)$ \\
\hline
$B \rightarrow b_1 (1235)$ & This work & $-0.251 \pm 0.019$ & $0.021 \pm 0.013$ & $-0.701 \pm 0.038 $ &  $-0.354 \pm 0.021$ \\
\cline{2-6}
& LCSR\cite{LCSRA2} & $-0.20 \pm 0.04$ & $-0.09 \pm 0.02$ & $-0.39 \pm 0.07$ & $-0.25 \pm 0.05$ \\
\cline{2-6}
& pQCD\cite{pQCD2} & $0.33^{+0.13}_{-0.13}$ & $0.03^{+0.03}_{-0.03}$ & $0.45^{+0.15}_{-0.14}$ & $0.19^{+0.08}_{-0.08}$ \\
\cline{2-6}
& LFQM\cite{LFQM2} & $0.19$ & $-0.02$ & $0.38$ & $0.11$ \\
\hline
$B \rightarrow K_{1B}$ & This work & $-0.337 \pm 0.024$ & $-0.031 \pm 0.017$ & $-0.792 \pm 0.044 $ &  $-0.510 \pm 0.030 $ \\
\cline{2-6}
& LCSR\cite{LCSRA2} & $-0.29^{+0.08}_{-0.05}$ & $-0.17^{+0.05}_{-0.03}$ & $-0.45^{+0.12}_{-0.08}$ & $-0.37^{+0.10}_{-0.06}$ \\
\cline{2-6}
& pQCD\cite{pQCD2} & $0.36^{+0.18}_{-0.17}$ & $0.00^{+0.03}_{-0.03}$ & $0.52^{+0.20}_{-0.19}$ & $0.20^{+0.10}_{-0.10}$ \\
\cline{2-6}
& LFQM\cite{LFQM2} & $0.21$ & $-0.05$ & $0.45$ & $0.12$ \\
\hline
$B \rightarrow h_1(1^1P_1)$ & This work & $-0.160 \pm 0.010$ & $-0.006 \pm 0.005$ & $-0.414 \pm 0.022 $ &  $-0.227 \pm 0.012 $ \\
\cline{2-6}
& LCSR\cite{LCSRA2} & $-0.13 \pm 0.03$ & $-0.07 \pm 0.02$ & $-0.24 \pm 0.04$ & $-0.17 \pm 0.03$ \\
\cline{2-6}
& pQCD\cite{pQCD2} & $0.20^{+0.08}_{-0.08}$ & $0.03^{+0.02}_{-0.01}$ & $0.26^{+0.08}_{-0.08}$ & $0.12^{+0.05}_{-0.05}$ \\
\cline{2-6}
& LFQM\cite{LFQM2} & $0.19$ & $-0.02$ & $0.37$ & $0.10$\\
\hline
$B \rightarrow h_8(1^1P_1)$ & This work & $-0.116 \pm 0.008$ & $0.016 \pm 0.004$ & $-0.305 \pm 0.017 $ &  $-0.170 \pm 0.009 $ \\
\cline{2-6}
& LCSR\cite{LCSRA2} & $-0.11 \pm 0.02$ & $-0.06 \pm 0.01$ & $-0.18 \pm 0.03$ & $-0.13 \pm 0.02$ \\
\cline{2-6}
& pQCD\cite{pQCD2} & $0.16^{+0.07}_{-0.06}$ & $0.01^{+0.01}_{-0.01}$ & $0.21^{+0.07}_{-0.07}$ & $0.09^{+0.03}_{-0.03}$ \\
\hline
\end{tabular}
\caption{Comparison of the $B \rightarrow A(1^1P_1) $ form factors at $q^2=0$ given in this
work with other groups. }\label{tab:FFSLCBto1P1}
\end{table}

\begin{table}[H]
\centering
\begin{tabular}{|c|c|c|c|c|c|}
\hline Decays & Reference & $V_1(0)$ & $V_2 (0)$ & $V_0 (0)$ & $A (0)$ \\
\hline
$B_s \rightarrow \bar{K}_{1A}$ & This work & $0.569 \pm 0.032$ & $0.532 \pm 0.026$ & $0.625 \pm 0.042 $ &  $0.860 \pm 0.046 $ \\
\cline{2-6}
& LCSR\cite{LCSRA2} & $0.30 \pm 0.06$ & $0.36 \pm 0.07$ & $0.19 \pm 0.04$ & $0.40 \pm 0.08$ \\
\cline{2-6}
& pQCD\cite{pQCD2} & $0.43^{+0.19}_{-0.18}$ & $0.11^{+0.05}_{-0.05}$ & $0.36^{+0.18}_{-0.19}$ & $0.25^{+0.10}_{-0.11}$ \\
\cline{2-6}
& LFQM\cite{LFQM2} & $0.37$ & $0.17$ & $0.12$ & $0.24$ \\
\hline
$B_s \rightarrow f_1$ & This work & $0.270 \pm 0.015$ & $0.282 \pm 0.012$ & $0.250 \pm 0.018 $ &  $0.416 \pm 0.021 $ \\
\cline{2-6}
& LCSR\cite{LCSRA2} & $0.20 \pm 0.04$ & $0.23 \pm 0.04$ & $0.16 \pm 0.03$ & $0.26 \pm 0.04$ \\
\cline{2-6}
& pQCD\cite{pQCD2} & $0.23^{+0.08}_{-0.08}$ & $0.07^{+0.01}_{-0.01}$ & $0.18^{+0.08}_{-0.07}$ & $0.14^{+0.05}_{-0.05}$ \\
\cline{2-6}
& LFQM\cite{LFQM2} & $0.41$ & $0.18$ & $0.13$ & $0.28$\\
\hline
$B_s \rightarrow f_8$ & This work & $-0.382 \pm 0.021$ & $-0.403 \pm 0.018$ & $-0.348 \pm 0.026 $ &  $-0.594 \pm 0.030 $ \\
\cline{2-6}
& LCSR\cite{LCSRA2} & $-0.28 \pm 0.05$ & $-0.33 \pm 0.05$ & $-0.21 \pm 0.04$ & $-0.39 \pm 0.07$ \\
\cline{2-6}
& pQCD\cite{pQCD2} & $-0.33^{+0.11}_{-0.11}$ & $-0.10^{+0.03}_{-0.02}$ & $-0.26^{+0.11}_{-0.10}$ & $-0.19^{+0.07}_{-0.06}$ \\
\hline
\end{tabular}
\caption{Comparison of the $B_s \rightarrow A(1^3P_1) $ form factors at $q^2=0$ given in this
work with other groups. }\label{tab:FFSLCBsto3P1}
\end{table}

\begin{table}[H]
\centering
\begin{tabular}{|c|c|c|c|c|c|}
\hline Decays & Reference & $V_1(0)$ & $V_2 (0)$ & $V_0 (0)$ & $A (0)$ \\
\hline
$B_s \rightarrow \bar{K}_{1B}$ & This work & $-0.170 \pm 0.015$ & $0.169 \pm 0.016$ & $-0.684 \pm 0.037 $ &  $-0.226 \pm 0.017 $ \\
\cline{2-6}
& LCSR\cite{LCSRA2} & $-0.25^{+0.07}_{-0.04}$ & $-0.15^{+0.04}_{-0.03}$ & $-0.40^{+0.11}_{-0.07}$ & $-0.33^{+0.09}_{-0.05}$ \\
\cline{2-6}
& pQCD\cite{pQCD2} & $0.33^{+0.14}_{-0.14}$ & $0.03^{+0.03}_{-0.03}$ & $0.42^{+0.16}_{-0.15}$ & $0.18^{+0.08}_{-0.08}$ \\
\cline{2-6}
& LFQM\cite{LFQM2} & $0.15$ & $-0.06$ & $0.38$ & $0.08$ \\
\hline
$B_s \rightarrow h_1$ & This work & $-0.160 \pm 0.010$ & $-0.004 \pm 0.006$ & $-0.425 \pm 0.024 $ &  $-0.228 \pm 0.013 $ \\
\cline{2-6}
& LCSR\cite{LCSRA2} & $-0.11 \pm 0.03$ & $-0.06 \pm 0.02$ & $-0.21 \pm 0.04$ & $-0.15 \pm 0.03$ \\
\cline{2-6}
& pQCD\cite{pQCD2} & $0.18^{+0.07}_{-0.06}$ & $0.03^{+0.01}_{-0.01}$ & $0.23^{+0.07}_{-0.06}$ & $0.10^{+0.04}_{-0.04}$ \\
\cline{2-6}
& LFQM\cite{LFQM2} & $0.17$ & $-0.10$ & $0.51$ & $0.09$\\
\hline
$B_s \rightarrow h_8$ & This work & $0.233 \pm 0.015$ & $-0.035 \pm 0.009$ & $0.626 \pm 0.036 $ &  $0.343 \pm 0.021 $ \\
\cline{2-6}
& LCSR\cite{LCSRA2} & $0.19 \pm 0.04$ & $0.11 \pm 0.02$ & $0.32 \pm 0.05$ & $0.23 \pm 0.04$ \\
\cline{2-6}
& pQCD\cite{pQCD2} & $-0.28^{+0.10}_{-0.10}$ & $-0.02^{+0.01}_{-0.01}$ & $-0.36^{+0.11}_{-0.10}$ & $-0.16^{+0.05}_{-0.05}$ \\
\hline
\end{tabular}
\caption{Comparison of the $B_s \rightarrow A(1^1P_1) $ form factors at $q^2=0$ given in this
work with other groups. }\label{tab:FFSLCBsto1P1}
\end{table}

\begin{table}[H]
\centering
\begin{tabular}{|c|c|c|c|}
\hline Decays & Reference &$T_1(0) = T_2(0)$ & $T_3(0)$  \\
\hline
$B \rightarrow a_1 (1260)$ & This work & $0.522 \pm 0.028$ & $0.440 \pm 0.020$\\
\cline{2-4}
& pQCD\cite{pQCD2} & $0.34^{+0.13}_{-0.13}$ & $0.30^{+0.17}_{-0.12}$ \\
\hline
$B \rightarrow K_{1A}$ & This work & $0.479 \pm 0.024$ & $0.457 \pm 0.019$\\
\cline{2-4}
& pQCD\cite{pQCD2} & $0.37^{+0.10}_{-0.09}$ & $0.33^{+0.16}_{-0.16}$ \\
\hline
$B \rightarrow f_1(1^3P_1)$ & This work & $0.314 \pm 0.017$ & $0.267 \pm 0.012$\\
\cline{2-4}
& pQCD\cite{pQCD2} & $0.21^{+0.08}_{-0.07}$ & $0.19^{+0.07}_{-0.07}$ \\
\hline
$B \rightarrow f_8(1^3P_1)$ & This work & $0.223 \pm 0.012$ & $0.192 \pm 0.009$\\
\cline{2-4}
& pQCD\cite{pQCD2} & $0.15^{+0.05}_{-0.05}$ & $0.13^{+0.05}_{-0.05}$ \\
\hline
\end{tabular}
\caption{The penguin type form factors for the $B \rightarrow A(1^3P_1)$ decays at $q^2=0$ given via the relations of Eqs.(\ref{t1new})-(\ref{t4new}), where the values obtained via pQCD\cite{pQCD2} are also listed for comparison. }\label{tab:FFPCBto3P1}
\end{table}

\begin{table}[H]
\centering
\begin{tabular}{|c|c|c|c|}
\hline Decays & Reference &$T_1(0) = T_2(0)$ & $T_3(0)$  \\
\hline
$B \rightarrow b_1 (1260)$ & This work & $-0.290 \pm 0.019$ & $-0.032 \pm 0.007$\\
\cline{2-4}
& pQCD\cite{pQCD2} & $0.27^{+0.11}_{-0.10}$ & $0.18^{+0.08}_{-0.08}$ \\
\hline
$B \rightarrow K_{1B}$ & This work & $-0.402 \pm 0.026$ & $-0.113 \pm 0.012$\\
\cline{2-4}
& pQCD\cite{pQCD2} & $0.29^{+0.13}_{-0.13}$ & $0.20^{+0.11}_{-0.10}$ \\
\hline
$B \rightarrow h_1(1^1P_1)$ & This work & $-0.185 \pm 0.011$ & $-0.035 \pm 0.003$\\
\cline{2-4}
& pQCD\cite{pQCD2} & $0.17^{+0.06}_{-0.06}$ & $0.12^{+0.06}_{-0.05}$ \\
\hline
$B \rightarrow h_8(1^1P_1)$ & This work & $-0.136 \pm 0.008$ & $-0.015 \pm 0.002$\\
\cline{2-4}
& pQCD\cite{pQCD2} & $0.13^{+0.05}_{-0.06}$ & $0.09^{+0.03}_{-0.04}$ \\
\hline
\end{tabular}
\caption{The penguin type form factors for the $B \rightarrow A(1^1P_1)$ decays at $q^2=0$ given via the relations of Eqs.(\ref{t1new})-(\ref{t4new}), where the values obtained via pQCD\cite{pQCD2} are also listed for comparison. }\label{tab:FFPCBto1P1}
\end{table}

\begin{table}[H]
\centering
\begin{tabular}{|c|c|c|c|}
\hline Decays & Reference &$T_1(0) = T_2(0)$ & $T_3(0)$  \\
\hline
$B_s \rightarrow \bar{K}_{1A}$ & This work & $0.679 \pm 0.037$ & $0.527 \pm 0.024$\\
\cline{2-4}
& pQCD\cite{pQCD2} & $0.34^{+0.15}_{-0.14}$ & $0.30^{+0.13}_{-0.14}$ \\
\hline
$B_s \rightarrow f_1(1^3P_1)$ & This work & $0.325 \pm 0.017$ & $0.276 \pm 0.012$\\
\cline{2-4}
& pQCD\cite{pQCD2} & $0.18^{+0.07}_{-0.07}$ & $0.16^{+0.06}_{-0.06}$ \\
\hline
$B_s \rightarrow f_8(1^3P_1)$ & This work & $-0.462 \pm 0.024$ & $-0.395 \pm 0.017$\\
\cline{2-4}
& pQCD\cite{pQCD2} & $-0.26^{+0.09}_{-0.08}$ & $-0.23^{+0.08}_{-0.08}$ \\
\hline
\end{tabular}
\caption{The penguin type form factors for the $B_s \rightarrow A(1^3P_1)$ decays at $q^2=0$ given via the relations of Eqs.(\ref{t1new})-(\ref{t4new}), where the values obtained via pQCD\cite{pQCD2} are also listed for comparison. }\label{tab:FFPCBsto3P1}
\end{table}

\begin{table}[H]
\centering
\begin{tabular}{|c|c|c|c|}
\hline Decays & Reference &$T_1(0) = T_2(0)$ & $T_3(0)$  \\
\hline
$B_s \rightarrow \bar{K}_{1B}$ & This work & $-0.191 \pm 0.015$ & $0.100 \pm 0.009$\\
\cline{2-4}
& pQCD\cite{pQCD2} & $0.26^{+0.11}_{-0.12}$ & $0.17^{+0.08}_{-0.08}$ \\
\hline
$B_s \rightarrow h_1(1^1P_1)$ & This work & $-0.186 \pm 0.011$ & $-0.034 \pm 0.003$\\
\cline{2-4}
& pQCD\cite{pQCD2} & $0.15^{+0.05}_{-0.05}$ & $0.10^{+0.03}_{-0.03}$ \\
\hline
$B_s \rightarrow h_8(1^1P_1)$ & This work & $0.274 \pm 0.017$ & $0.030 \pm 0.004$\\
\cline{2-4}
& pQCD\cite{pQCD2} & $-0.23^{+0.07}_{-0.08}$ & $0.15^{+0.06}_{-0.05}$ \\
\hline
\end{tabular}
\caption{The penguin type form factors for the $B_s \rightarrow A(1^1P_1)$ decays at $q^2=0$ given via the relations of Eqs.(\ref{t1new})-(\ref{t4new}), where the values obtained via pQCD\cite{pQCD2} are also listed for comparison. }\label{tab:FFPCBsto1P1}
\end{table}

\section{$B_{(s)}$ to light axial vector meson semileptonic decays}

In this section, we shall apply the form factors obtained above to
investigate the corresponding semileptonic decays.

As mentioned in the INTRODUCTION, the physical states $K_1(1270)$ and $K_1(1400)$ are the mixtures of the $K_{1A}$ and $K_{1B}$. their relations can be written as
\begin{eqnarray}
& & |K_1(1270)\rangle = | K_{1A} \sin \theta_{K_1} + |K_{1B} \rangle \cos \theta_{K_1}, \nonumber \\
& & |K_1(1400)\rangle = | K_{1A} \cos \theta_{K_1} - |K_{1B} \rangle \sin \theta_{K_1}.
\end{eqnarray}
The sign ambiguity for $\theta_{K_1}$ is due to the fact that one can add arbitrary phases to $|K_{1A}\rangle$ and $|K_{1B}\rangle$. This sign ambiguity can be removed by fixing the signs for $f_{K_{1A}}$ and $f^\perp_{K_{1B}}$, which do not vanish in the $SU(3)$ limit. Following Ref.\cite{DAs}, we adopt the convention: $f_{K_{1A}} >0, f^\perp_{K_{1B}}>0$. Combing the analyses for the data of the decays $B \rightarrow \gamma$ and $\tau^- \rightarrow K^-_1 (1270) \nu_\tau$, the mixing angle was found to be $\theta_{K_1}= - (34 \pm 13)^\circ$\cite{LCSRA1}.

Analogous to the $\eta-\eta^\prime$ mixing in the pseudoscalar sector, the $1^3P_1$ states, $f_1(1285)$ and $f_1(1420)$, have mixing via
\begin{eqnarray}
& & |f_1(1285) \rangle = |f_1 \rangle \cos \theta_{^3P_1} + | f_8 \rangle \sin \theta_{^3P_1}, \nonumber \\
& & |f_1(1420) \rangle = -|f_1 \rangle \sin \theta_{^3P_1} + | f_8 \rangle \cos \theta_{^3P_1},
\end{eqnarray}
and likewise the $1^1P_1$ states, $h_1(1170)$ and $h_1(1415)$, can be mixed in terms of the pure octet $h_8$ and singlet $h_1$,
\begin{eqnarray}
& & |h_1(1285) \rangle = |h_1 \rangle \cos \theta_{^1P_1} + | h_8 \rangle \sin \theta_{^1P_1}, \nonumber \\
& & |h_1(1420) \rangle = -|h_1 \rangle \sin \theta_{^1P_1} + | h_8 \rangle \cos \theta_{^1P_1}.
\end{eqnarray}
Using the Gell-Mann-Okubo mass formula\cite{DAs}, the mixing angles $\theta_{^3P_1}$ and $\theta_{^1P_1}$ can be determined to be
\begin{eqnarray}
\theta_{^3P_1} = (23.6^{+17.0}_{-23.6})^\circ, \hspace{0.5cm} \theta_{^1P_1} = (28.1^{+9.8}_{-17.2})^\circ.
\end{eqnarray}

For the charged current induced semileptonic decays $B_{(s)} \rightarrow A l \bar{\nu}_l$,
the differential decay width with respect to $q^2$ can be written as\cite{LCSRB3,DecayR1,DecayR2}
\begin{eqnarray}\label{ddw}
\frac{d \Gamma}{d q^2} =\frac{d \Gamma_L}{d q^2}+\frac{d \Gamma_T}{d
q^2}
\end{eqnarray}
where $\frac{d \Gamma_L}{d q^2} $ and $\frac{d \Gamma_T}{d q^2} $
denote the longitudinal and transverse differential decay width,
respectively. Their expressions have the following forms,
\begin{eqnarray}
\frac{d \Gamma_L}{d q^2} & = & \frac{ G^2_F |V_{ub}|^2}{ 384 m^3_H
\pi^3 } \sqrt{\lambda_A} \left ( 1- \frac{m^2_l}{q^2}
\right )^2  X_L \\
\frac{d \Gamma_T}{d q^2} & = & \frac{ G^2_F |V_{ub}|^2}{ 384 m^3_H
\pi^3 } \sqrt{\lambda_A} \left ( 1- \frac{m^2_l}{q^2}
\right )^2 (X_+ + X_- )
\end{eqnarray}
with
\begin{eqnarray}
 X_L & = &  ( 2 q^2 + m^2_l )H^2_0 (q^2) + 3 m^2_l H^2_t (q^2)  \\
 X_\pm & = & ( 2 q^2 + m^2_l)H^2_\pm(q) \\
\end{eqnarray}
and
\begin{eqnarray}
H_0 (q^2) & = & \frac{1}{2 m_A \sqrt{q^2}} \left [ ( m^2_H -m^2_A - q^2) ( m_H + m_A ) V_1 (q^2) - \frac{\lambda_A}{m_H + m_A} V_2 (q^2) \right ], \\
H_t (q^2) & = & \sqrt{\frac{\lambda_A}{q^2}} V_0 (q^2), \\
H_\pm (q^2) & = & \frac{\sqrt{\lambda_A}}{m_H + m_A} A(q^2) \mp (m_H+m_A) V_1 (q^2).
\end{eqnarray}
Here $m_l$ is the mass of charged lepton and $q^2$ represents the
momentum square of lepton pair.

The branching ratio can be obtained by integrating Eq.(\ref{ddw})
over $q^2$ in the whole physical region and using the lifetimes of
$B_{(s)}$ ( denoted as $\tau_H$ in the following ) as inputs,
\begin{eqnarray}
Br = \frac{\tau_H}{\hbar} \Gamma = \frac{\tau_H}{\hbar} \int^{(m_H-
m_A)^2}_{m^2_l} d q^2 \frac{d \Gamma}{d q^2}
\end{eqnarray}
In addition, it is also meaningful to define the longitudinal
polarization fraction
\begin{eqnarray}
f_L = \frac{\Gamma_L}{\Gamma} = \frac{\int^{(m_H- m_A)^2}_{m^2_l} d
q^2 \frac{d \Gamma_L}{d q^2}}{\int^{(m_H- m_A)^2}_{m^2_l} d q^2
\frac{d \Gamma}{d q^2}}
\end{eqnarray}
and forward-backward asymmetry (FBA) of lepton
\begin{eqnarray}
\frac{d A_{FB}}{d q^2} & = & \frac{\left ( \int_0^1  - \int_{-1}^0
\right ) d \cos \theta \frac{d^2 \Gamma}{d q^2 d \cos \theta }}{d
\Gamma/ d q^2} \nonumber \\
& = & \frac{3}{2} \frac{q^2(H^2_-(q^2)-H^2_+(q^2))+2m^2_l H_t(q) H_0 (q)}{(2q^2+m^2_l)(H^2_0(q^2)+H^2_+(q^2)+H^2_-(q^2))+3m^2_lH^2_t(q^2)}.
\end{eqnarray}

For the lifetimes of $B_{(s)}$, masses of charged leptons,  CKM
matrix element $|V_{ub}|$ and Fermi coupling constant $G_F$, we use
the latest values given by the particle data group (PDG)\cite{PDG},
\begin{eqnarray}\label{inputpara}
& & \tau_{\bar{B}^0} = 1.520 \times 10^{-12}{\rm s}, \hspace{0.5cm}
\tau_{B^-} = 1.638 \times 10^{-12}{\rm s}, \hspace{0.5cm} \tau_{B_s} =
1.510 \times 10^{-12}{\rm s} \nonumber \\
& & m_e = 0.51\times 10^{-3} {\rm GeV}, \hspace{0.5cm} m_\mu = 0.106 {\rm GeV},
\hspace{0.5cm} m_\tau = 1.777 {\rm GeV} \nonumber \\
& & |V_{ub}|=(3.82 \pm 0.20) \times 10^{-3}, \hspace{0.5cm} G_F =
1.166 \times 10^{-5} {\rm GeV}^{-2}.
\end{eqnarray}

\begin{figure}
\centering
\includegraphics[width=3in]{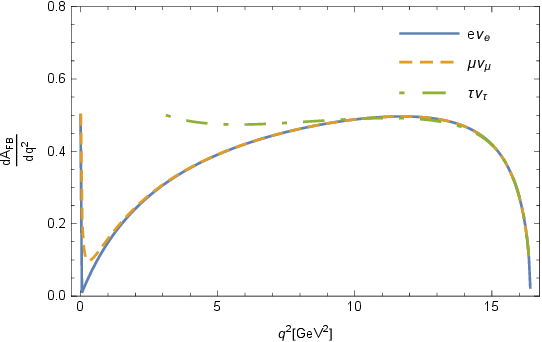}\hfill
\includegraphics[width=3in]{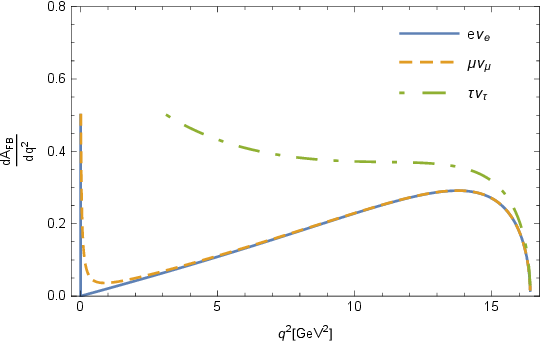}
\caption{The FBAs of lepton as functions of $q^2$ for the decays $\bar{B}^0 \rightarrow a_1^+(1260) l^- \bar{\nu}_l$ (left graph)  and $\bar{B}^0 \rightarrow b_1^+(1235) l^- \bar{\nu}_l$ (right graph), with $l=e, \mu, \tau $.}
\label{fig:dAFB}
\end{figure}

The FBAs of lepton as functions of $q^2$ for the decays $\bar{B}^0 \rightarrow a_1^+(1260) l^- \bar{\nu}_l$ and $\bar{B}^0 \rightarrow b_1^+(1235) l^- \bar{\nu}_l$ with $l=e, \mu, \tau $
are shown in FIG. \ref{fig:dAFB}.
It is easily seen that the FBAs of lepton are positive in the whole physical region. For other charged current induced semileptonic decay modes, the behaviors of FBAs are similar. The numerical results of branching ratios $Br$, longitudinal
polarization fractions $f_L$ and integrated FBAs $A_{FB}$ for all $B_{(s)} \rightarrow A l
\bar{\nu}_l$ decay modes are collected in TABLE \ref{tab:Bto3P1SLD}-\ref{tab:BstoASLD}. For comparison, we also list the values of $Br$ and $A_{FB}$ calculated via LCSR\cite{LCSRA2,LCSRB3}. We can see that our results of $Br$ are some larger, yet still compatible with the values given by Ref.\cite{LCSRA2} for $l=e$. On the whole, the $Br$ and $A_{FB}$ given in this work are significantly smaller than the corresponding results obtained in Ref.\cite{LCSRB3} for $l=\mu, \tau$.
The $Br$ are at the order of ${\cal O}(10^{-4})$ and ${\cal O}(10^{-6})$ for the $B$ decays into $a_1(1260), b_1(1235), f_1(1285), h_1(1170)$ and $f_1(1420), h_1(1415)$, respectively. For the $B_s$ decays into $K^+_1(1270)$ and $K^+_1(1400)$, the $Br$ are at the order of ${\cal O}(10^{-3})$ and ${\cal O}(10^{-4})$, respectively. In contrast, for the corresponding charge conjugate processes, the $Br$ for the $B_s$ decays into $K^-_1(1400)$ are at the order of ${\cal O}(10^{-5})$. The results are almost identical for $l=e, \mu$, while those for $l=\tau$ are significantly different. Especially, the $Br$ for $l=\tau$ are about $(25-40)\%$ of those for $l=e, \mu$. The $Br$, $f_L$ decrease with the charged lepton mass $m_l$, while the $A_{FB}$ increases with $m_l$, except for the $\bar{B}^0_s \rightarrow K^+_1(1400) l^- \bar{\nu}_l$ decays. For these modes, the behaviors of $f_L$, $A_{FB}$ are opposite. For the $B$ decays into pure $1^3P_1$ (i.e. $a_1(1260)$, $f_1(1285)$, $f_1(1420)$) and $1^1P_1$ (i.e. $b_1(1235)$, $h_1(1170)$, $h_1(1415)$) states, the $f_L$ are about $40\%$ and $80\%$, respectively. For the $\bar{B}^0_s$ decays into $K^+_1(1270)$ and $K^+_1(1400)$, the $f_L$ are approximately $70\%$ and $20\%$, respectively. For the corresponding charge conjugate processes, the $f_L$ for the decays with $K_1(1400)$ in the final states are close to those with $K_1(1270)$ in the final states. The $A_{FB}$ are more than $40\%$ for the $B$ decays into $1^3P_1$ states, while the corresponding results for the decays into $1^1P_1$ states are approximately $(10-40)\%$ in which the values for $l=e, \mu$ are much smaller. For the $B_s$ decays into $K_1(1270)$, the $A_{FB}$ are about $20\%$ and $40\%$ for $l=e, \mu$ and $l=\tau$. The $A_{FB}$ for the $\bar{B}^0_s$ decays into $K^+_1(1400)$ are more than $50\%$, and the values for corresponding charge conjugate processes are around $(30-50)\%$.

The first and second uncertainties of $Br$ come from the form factors and CKM matrix element $|V_{ub}|$, respectively. Except for the $B^0_s \rightarrow K^-_1(1400) l^+ \nu_l$ decays, the uncertainties from these two sources are at the same order of magnitude, roughly (10-15)\%. For the $B^0_s \rightarrow K^-_1(1400) l^+ \nu_l$ decays, the uncertainties from the form factors are significantly larger, almost $30\%$, due to the fact that the variations of relevant form factors are much larger at the high $q^2$ region. In contrast, the uncertainties of $f_L$ and $A_{FB}$ only come from the form factors, which are less than $10\%$ for most decay channels.

\begin{table}[H]
\centering
\begin{tabular}{|c|c|c|c|c|}
\hline Decays & Obs. & $e \nu_e$ & $\mu \nu_\mu$ & $\tau \nu_\tau$ \\
\hline
& $Br$ & $(3.96^{+0.53+0.42}_{-0.48-0.40}) \times 10^{-4}$ & $(3.94^{+0.52+0.42}_{-0.48-0.40})\times 10^{-4}$ & $(1.56^{+0.13+0.17}_{-0.12-0.16})\times 10^{-4}$ \\
\cline{3-5}
& & --- &  $(3.23 \pm 0.98 )\times 10^{-4}$\cite{LCSRB3} & $(1.08 \pm 0.32) \times 10^{-4}$\cite{LCSRB3} \\
\cline{3-5}
& & $(3.02^{+1.03}_{-1.03})\times 10^{-4}$\cite{LCSRA2} & --- & --- \\
\cline{2-5}
$\bar{B}^0 \rightarrow a^+_1  l^- \bar{\nu}_l$ & $f_L$ & $0.441^{+0.043}_{-0.047}$ & $0.441^{+0.043}_{-0.046}$ & $0.416^{+0.020}_{-0.020}$ \\
\cline{2-5}
& $A_{FB}$ & $0.401^{+0.037}_{-0.035}$ & $0.404^{+0.037}_{-0.035} $ & $0.475^{+0.013}_{-0.014} $ \\
\cline{3-5}
& & --- &  $0.452 \pm 0.013 $\cite{LCSRB3} & $0.528 \pm 0.015$\cite{LCSRB3} \\
\hline
 & $Br$ & $(1.96^{+0.26+0.21}_{-0.24-0.20})\times 10^{-4}$ & $(1.95^{+0.25+0.21}_{-0.23-0.20})\times 10^{-4}$ & $(0.76^{+0.06+0.08}_{-0.06-0.08})\times 10^{-4}$ \\
 \cline{3-5}
 & & --- &  $(5.31 \pm 1.29) \times 10^{-4}$\cite{LCSRB3} & $(1.77 \pm 0.42) \times 10^{-4}$\cite{LCSRB3} \\
 \cline{3-5}
 & & $(1.63^{+0.60+0.04}_{-0.51-0.48}) \times 10^{-4}$\cite{LCSRA2} & --- & --- \\
\cline{2-5}
$B^- \rightarrow f_1(1285)  l^- \bar{\nu}_l$ & $f_L$ & $0.439^{+0.042}_{-0.045}$ & $0.439^{+0.041}_{-0.044}$ & $0.415^{+0.019}_{-0.019}$ \\
\cline{2-5}
& $A_{FB}$ & $0.400^{+0.036}_{-0.034} $ & $0.404^{+0.036}_{-0.034} $ & $0.471^{+0.013}_{-0.014} $ \\
\cline{3-5}
& & --- & $0.452 \pm 0.010$\cite{LCSRB3} & $0.528 \pm 0.012$\cite{LCSRB3} \\
\hline
 & $Br$ & $(6.72^{+0.86+0.72}_{-0.79-0.68})\times 10^{-6}$ & $(6.69^{+0.85+0.72}_{-0.78-0.68})\times 10^{-6}$ & $(2.41^{+0.19+0.26}_{-0.18-0.25})\times 10^{-6}$ \\
 \cline{3-5}
 & & --- &  $(0.88 \pm 0.49) \times 10^{-4}$\cite{LCSRB3} & $(0.26 \pm 0.14) \times 10^{-4}$\cite{LCSRB3} \\
 \cline{3-5}
 & & $(0.05^{+0.03+0.42}_{-0.02-0.03})\times 10^{-4}$\cite{LCSRA2} & --- & --- \\
\cline{2-5}
$B^- \rightarrow f_1(1420)  l^- \bar{\nu}_l$ & $f_L$ & $0.406^{+0.041}_{-0.044}$ & $0.405^{+0.040}_{-0.043}$ & $0.382^{+0.018}_{-0.018}$ \\
\cline{2-5}
& $A_{FB}$ & $0.420^{+0.036}_{-0.034} $ & $0.423^{+0.035}_{-0.034} $ & $0.469^{+0.014}_{-0.014} $ \\
\cline{3-5}
& & --- & $0.441 \pm 0.025$\cite{LCSRB3} & $0.521 \pm 0.030$\cite{LCSRB3} \\
\hline
\end{tabular}
\caption{The $Br$, $f_L$ and $A_{FB}$ for semileptonic $B
\rightarrow A(1^3P_1) l \bar{\nu}_l$ decays, where the values of $Br$, $A_{FB}$ calculated via LCSR\cite{LCSRA2,LCSRB3} are also listed for comparison. }\label{tab:Bto3P1SLD}
\end{table}

\begin{table}[H]
\centering
\begin{tabular}{|c|c|c|c|c|}
\hline Decays & Obs. & $e \nu_e$ & $\mu \nu_\mu$ & $\tau \nu_\tau$ \\
\hline
& $Br$ & $(4.33^{+0.60+0.47}_{-0.57-0.44}) \times 10^{-4}$ & $(4.31^{+0.59+0.46}_{-0.56-0.44})\times 10^{-4}$ & $(1.48^{+0.14+0.16}_{-0.14-0.15})\times 10^{-4}$ \\
\cline{3-5}
& & --- &  $(5.63 \pm 1.69 )\times 10^{-4}$\cite{LCSRB3} & $(1.89 \pm 0.55) \times 10^{-4}$\cite{LCSRB3} \\
\cline{3-5}
& & $(1.93^{+0.84}_{-0.68})\times 10^{-4}$\cite{LCSRA2} & --- & --- \\
\cline{2-5}
$\bar{B}^0 \rightarrow b^+_1  l^- \bar{\nu}_l$ & $f_L$ & $0.806^{+0.015}_{-0.016}$ & $0.806^{+0.014}_{-0.016}$ & $0.750^{+0.014}_{-0.015}$ \\
\cline{2-5}
& $A_{FB}$ & $0.129^{+0.013}_{-0.012}$ & $0.137^{+0.014}_{-0.013} $ & $0.380^{+0.011}_{-0.011} $ \\
\cline{3-5}
& & --- & $0.457 \pm 0.011$\cite{LCSRB3} & $0.533 \pm 0.013$\cite{LCSRB3} \\
\hline
 & $Br$ & $(2.70^{+0.33+0.29}_{-0.31-0.28})\times 10^{-4}$ & $(2.69^{+0.33+0.29}_{-0.31-0.27})\times 10^{-4}$ & $(0.92^{+0.08+0.10}_{-0.08-0.10})\times 10^{-4}$ \\
 \cline{3-5}
 & & --- & $(10.49 \pm 2.56) \times 10^{-4}$\cite{LCSRB3} & $(3.84 \pm 0.92) \times 10^{-4}$\cite{LCSRB3} \\
 \cline{3-5}
& & $(1.24^{+0.57+0.06}_{-0.45-0.25})\times 10^{-4}$\cite{LCSRA2} & --- & --- \\
\cline{2-5}
$B^- \rightarrow h_1(1170)  l^- \bar{\nu}_l$ & $f_L$ & $0.823^{+0.011}_{-0.012}$ & $0.823^{+0.011}_{-0.012}$ & $0.775^{+0.013}_{-0.013}$ \\
\cline{2-5}
& $A_{FB}$ & $0.120^{+0.011}_{-0.011} $ & $0.129^{+0.011}_{-0.011} $ & $0.380^{+0.010}_{-0.010} $ \\
\cline{3-5}
& & --- & $0.462 \pm 0.011$\cite{LCSRB3} & $0.533 \pm 0.012$\cite{LCSRB3} \\
\hline
& $Br$ & $(5.72^{+0.62+0.61}_{-0.59-0.58})\times 10^{-6}$ & $(5.70^{+0.61+0.61}_{-0.58-0.58})\times 10^{-6}$ & $(1.66^{+0.14+0.18}_{-0.13-0.17})\times 10^{-6}$ \\
 \cline{3-5}
 & & --- & $(0.73 \pm 0.57) \times 10^{-4}$\cite{LCSRB3} & $(0.23 \pm 0.17) \times 10^{-4}$\cite{LCSRB3} \\
 \cline{3-5}
& & $(0.04^{+0.01+0.18}_{-0.01-0.00})\times 10^{-4}$\cite{LCSRA2} & --- & --- \\
\cline{2-5}
$B^- \rightarrow h_1(1415)  l^- \bar{\nu}_l$ & $f_L$ & $0.858^{+0.008}_{-0.008}$ & $0.858^{+0.008}_{-0.008}$ & $0.808^{+0.012}_{-0.012}$ \\
\cline{2-5}
& $A_{FB}$ & $0.095^{+0.008}_{-0.007} $ & $0.105^{+0.008}_{-0.008} $ & $0.381^{+0.008}_{-0.009} $ \\
\cline{3-5}
& & --- & $0.432 \pm 0.075$\cite{LCSRB3} & $0.507 \pm 0.086$\cite{LCSRB3} \\
\hline
\end{tabular}
\caption{The $Br$, $f_L$ and $A_{FB}$ for semileptonic $B
\rightarrow A(1^1P_1) l \bar{\nu}_l$ decays, where the values of $Br$, $A_{FB}$ calculated via LCSR\cite{LCSRA2,LCSRB3} are also listed for comparison. }\label{tab:Bto1P1SLD}
\end{table}

\begin{table}[H]
\centering
\begin{tabular}{|c|c|c|c|c|}
\hline Decays & Obs. & $e \nu_e$ & $\mu \nu_\mu$ & $\tau \nu_\tau$ \\
\hline
& $Br$ & $(1.07^{+0.17+0.12}_{-0.16-0.11}) \times 10^{-3}$ & $(1.06^{+0.15+0.11}_{-0.14-0.11})\times 10^{-3}$ & $(0.39^{+0.04+0.04}_{-0.04-0.04})\times 10^{-3}$ \\
\cline{3-5}
& & --- &  $(1.15 \pm 0.85 )\times 10^{-4}$\cite{LCSRB3} & $(0.40 \pm 0.30) \times 10^{-4}$\cite{LCSRB3} \\
\cline{3-5}
& & $(4.53^{+1.67+0.00}_{-2.00-0.44})\times 10^{-4}$\cite{LCSRA2} & --- & --- \\
\cline{2-5}
$\bar{B}^0_s \rightarrow K^+_1(1270)  l^- \bar{\nu}_l$ & $f_L$ & $0.725^{+0.019}_{-0.021}$ & $0.724^{+0.019}_{-0.021}$ & $0.669^{+0.016}_{-0.016}$ \\
\cline{2-5}
& $A_{FB}$ & $0.188^{+0.019}_{-0.018}$ & $0.196^{+0.019}_{-0.018} $ & $0.400^{+0.012}_{-0.013} $ \\
\cline{3-5}
& & --- & $0.434 \pm 0.060$\cite{LCSRB3} & $0.506 \pm 0.070$\cite{LCSRB3} \\
\hline
 & $Br$ & $(1.79^{+0.28+0.19}_{-0.24-0.18})\times 10^{-4}$ & $(1.79^{+0.27+0.19}_{-0.24-0.18})\times 10^{-4}$ & $(0.71^{+0.09+0.08}_{-0.08-0.07})\times 10^{-4}$ \\
 \cline{3-5}
 & & --- & $(9.09 \pm 2.24) \times 10^{-4}$\cite{LCSRB3} & $(2.77 \pm 0.68) \times 10^{-4}$\cite{LCSRB3} \\
 \cline{3-5}
& & $(3.86^{+1.43+0.03}_{-1.70-0.40})\times 10^{-4}$\cite{LCSRA2} & --- & --- \\
\cline{2-5}
$\bar{B}^0_s \rightarrow K^+_1(1400)  l^- \bar{\nu}_l$ & $f_L$ & $0.208^{+0.064}_{-0.063}$ & $0.208^{+0.063}_{-0.063}$ & $0.228^{+0.039}_{-0.039}$ \\
\cline{2-5}
& $A_{FB}$ & $0.572^{+0.052}_{-0.054} $ & $0.572^{+0.052}_{-0.053} $ & $0.521^{+0.027}_{-0.029} $ \\
\cline{3-5}
& & --- & $0.442 \pm 0.009$\cite{LCSRB3} & $0.522 \pm 0.010$\cite{LCSRB3} \\
\hline
& $Br$ & $(1.06^{+0.15+0.11}_{-0.14-0.11})\times 10^{-3}$ & $(1.06^{+0.15+0.11}_{-0.14-0.11})\times 10^{-3}$ & $(0.40^{+0.04+0.04}_{-0.04-0.04})\times 10^{-3}$ \\
\cline{2-5}
$B_s^0 \rightarrow K^-_1(1270)  l^+ \nu_l$ & $f_L$ & $0.656^{+0.023}_{-0.026}$ & $0.656^{+0.023}_{-0.026}$ & $0.607^{+0.017}_{-0.017}$ \\
\cline{2-5}
& $A_{FB}$ & $0.243^{+0.024}_{-0.022} $ & $0.250^{+0.024}_{-0.022} $ & $0.425^{+0.014}_{-0.015} $ \\
\hline
& $Br$ & $(3.37^{+0.95+0.36}_{-0.72-0.34})\times 10^{-5}$ & $(3.35^{+0.93+0.36}_{-0.71-0.34})\times 10^{-5}$ & $(0.88^{+0.17+0.09}_{-0.12-0.09})\times 10^{-5}$ \\
\cline{2-5}
$B_s^0 \rightarrow K^-_1(1400)  l^+ \nu_l$ & $f_L$ & $0.545^{+0.129}_{-0.149}$ & $0.545^{+0.128}_{-0.148}$ & $0.412^{+0.129}_{-0.113}$ \\
\cline{2-5}
& $A_{FB}$ & $0.341^{+0.109}_{-0.099} $ & $0.351^{+0.108}_{-0.098} $ & $0.524^{+0.013}_{-0.042} $ \\
\hline
\end{tabular}
\caption{The $Br$, $f_L$ and $A_{FB}$ for semileptonic $B_s
\rightarrow A l \bar{\nu}_l$ decays, where the values of $Br$, $A_{FB}$ calculated via LCSR\cite{LCSRA2,LCSRB3} are also listed for comparison. }\label{tab:BstoASLD}
\end{table}

\section{Summary}

By using the LCSR in the framework of HQEFT, we systematically calculated the form factors of $B_{(s)}$ decays into light P-wave axial vector mesons. We derived the expressions of form factors in terms of the light cone DAs of axial vector mesons via LCSR and found that the penguin type form factors can be obtained directly from the corresponding
semileptonic ones at the leading order of heavy quark expansion.
Considering the light axial vector meson DAs to twist-3, we give the
numerical results of form factors systematically. The uncertainties of form factors come from the two free parameters $s_0$, $T$, and are about $(5-8)\%$ at $q^2=0$, except the $V_2$ for the decays with a $1^1P_1$ axial vector meson in the final states.
Large differences exist between the results of different appoaches (i.e. this work, LCSR\cite{LCSRA2}, pQCD\cite{pQCD2}, LFQM\cite{LFQM2}). The form factors obtained in this work are in general some larger, yet still compatible with those results of
LCSR\cite{LCSRA2} as a whole.

With the form factors given here, we investigated the relevant charged current induced semileptonic decays,
predicting the branching ratios,
longitudinal polarization fractions and FBAs. It is found that the FBAs are positive in the whole physical region.
The branching ratios are at the order of ${\cal O}(10^{-4})$ and ${\cal O}(10^{-6})$ for the $B$ decays into $a_1(1260)$, $b_1(1235)$, $f_1(1285)$, $h_1(1170)$ and $f_1(1420)$, $h_1(1415)$, respectively. For the $B_s$ decays into $K_1(1270)$, $K_1(1400)$, the branching ratios are at the order of ${\cal O}(10^{-5})-{\cal O}(10^{-3})$. For the decays with $K_1(1270)$, $K_1(1400)$ in the final states, the results differ from those of the corresponding charge conjugate processes, implying the existence of CP violation. The branching ratios for $l=\tau$ are about $(25-40)\%$ of those for $l=e, \mu$. For the $B$ decays into pure $1^3P_1$ and $1^1P_1$ states, the longitudinal polarization fractions are about $40\%$ and $80\%$, respectively. The FBAs ae more than $40\%$ for the $B$ decays into $1^3P_1$ states, while the corresponding results for the decays into $1^1P_1$ states are approximately $(10-40)\%$. The uncertainties of branching ratios come from the form factors and CKM matrix element $|V_{ub}|$, which are at the same order of magnitude, {\it i.e.} roughly $(10-15)\%$, except for the $B^0_s \rightarrow K^-_1(1400) l^+ \nu_l$ decays. In contrast, the uncertainties of longitudinal polarization fractions and FBAs only come from the form factors, and less than $10\%$ for most decay channels. Our results may be tested by more precise experiments in the future.




\appendix

\section{The twist-3 light cone distribution amplitudes of axial vector meson}\label{twist3DAs}

In this appendix, we give the explicit expressions of the twist-3 light cone DAs used in this work\cite{LCSRA1,DAs},
\begin{eqnarray}
 & & g_\perp^{(a)}(u)  =   \frac{3}{4}(1+(2u-1)^2)
+ \frac{3}{2} a_1^\parallel (2u-1)^3
 + \left(\frac{3}{7}
a_2^\parallel + 5 \zeta_{3,^3P_1}^V \right) \left(3(2u-1)^2-1\right)
 \nonumber\\
& & \hspace{1.7cm} + \left( \frac{9}{112} a_2^\parallel + \frac{105}{16}
 \zeta_{3,^3P_1}^A - \frac{15}{64} \zeta_{3,^3P_1}^V \omega_{^3P_1}^V
 \right) \left( 35(2u-1)^4 - 30 (2u-1)^2 + 3\right) \nonumber\\
 & & \hspace{1.7cm}
 + 5\Bigg[ \frac{21}{4}\zeta_{3,^3P_1}^V \sigma_{^3P_1}^V
  + \zeta_{3,^3P_1}^A \bigg(\lambda_{^3P_1}^A -\frac{3}{16}
 \sigma_{^3P_1}^A\Bigg) \Bigg](2u-1)(5(2u-1)^2-3)
 \nonumber\\
& & \hspace{1.7cm} -\frac{9}{2} {a}_1^\perp
\tilde{\delta}_+ \left(\frac{3}{2}+\frac{3}{2}(2u-1)^2+\ln u
 +\ln(1-u)\right) \nonumber \\
 & & \hspace{1.7cm} - \frac{9}{2} {a}_1^\perp \tilde{\delta}_- (
3(2u-1) + \ln(1-u) - \ln u),\\
& & g_\perp^{(v)}(u)  =  6 u (1-u) \Bigg\{ 1 +
 \Bigg(a_1^\parallel + \frac{20}{3} \zeta_{3,^3P_1}^A
 \lambda_{^3P_1}^A\Bigg) (2u-1)\nonumber\\
 && \hspace{1.7cm} + \Bigg[\frac{1}{4}a_2^\parallel + \frac{5}{3}\,
 \zeta^V_{3,^3P_1} \left(1-\frac{3}{16} \omega^V_{^3P_1}\right)
 +\frac{35}{4} \zeta^A_{3,^3P_1}\Bigg] (5(2u-1)^2-1) \nonumber\\
 && \hspace{1.7cm} + \frac{35}{4}\Bigg(\zeta_{3,^3P_1}^V
 \sigma_{^3P_1}^V -\frac{1}{28}\zeta_{3,^3P_1}^A
 \sigma_{^3P_1}^A \Bigg) (2u-1)(7(2u-1)^2-3) \Bigg\}\nonumber\\
& & \hspace{1.7cm} -18 a_1^\perp \tilde{\delta}_+  (3u (1-u) +
(1-u) \ln (1-u) + u \ln u ) \nonumber \\
& & \hspace{1.7cm} - 18
a_1^\perp \tilde{\delta}_-   (u (1-u)(2u-1) + (1-u) \ln (1-u) -
u \ln u),\\
& & h_\parallel^{(t)}(u) =  3a_0^\perp (2u-1)^2+ \frac{3}{2} a_1^\perp
(2u-1) (3 (2u-1)^2-1) \nonumber \\
& & \hspace{1.7cm} + \frac{3}{2} \Bigg[a_2^\perp (2u-1) +
 \zeta^\perp_{3,^3P_1}\Bigg(5
-\frac{\omega_{^3P_1}^{\perp}}{2}\Bigg)\Bigg] (2u-1)(5(2u-1)^2-3)
\nonumber\\
 && \hspace{1.7cm} +\frac{35}{4}\zeta^\perp_{3,^3P_1} \sigma^\perp_{^3P_1}
 (35(2u-1)^4-30(2u-1)^2+3)
  \nonumber \\
  & & \hspace{1.7cm} + 18 a_2^\parallel
  \Bigg[\delta_+ (2u-1) -\frac{5}{8}\delta_- (3(2u-1)^2-1)\Bigg] -
  \frac{3}{2} \Bigg( \delta_+  (2u-1) \nonumber\\
 && \hspace{1.7cm} \times [2 +  \ln ((1-u)u)]
   +\delta_-  [ 1 + (2u-1) \ln \frac{1-u}{u} ]\Bigg)
   (1+ 6 a_2^\parallel),\\
& & h_\parallel^{(p)}(u)  =  6u(1-u) \Bigg\{ a_0^\perp +
\Bigg[a_1^\perp +5\zeta^\perp_{3,
^3P_1}\Bigg(1-\frac{1}{40}(7(2u-1)^2-3)
 \omega_{^3P_1}^{\perp} \Bigg)\Bigg] (2u-1)\nonumber\\
 &&  \hspace{1.7cm} + \Bigg( \frac{1}{4}a_2^\perp
 +
 \frac{35}{6} \zeta^\perp_{3,^3P_1} \sigma^\perp_{^3P_1} \Bigg)
 (5(2u-1)^2-1)
 -5a_2^\parallel
  \Bigg[\delta_+ (2u-1)\nonumber \\
  & & \hspace{1.7cm}  + \frac{3}{2} \delta_- (1-(1-u) u) \Bigg]\Bigg\}
  - 3[ \delta_+ ((1-u) \ln (1-u) - u \ln u)
 \nonumber \\
 & & \hspace{1.7cm} + \delta_-  ( u (1-u) + (1-u) \ln (1-u) + u \ln u)]
 (1+ 6 a_2^\parallel),
 \end{eqnarray}
for the $1^3P_1$ states, and
\begin{eqnarray}
& &  g_\perp^{(a)}(u)  =  \frac{3}{4} a_0^\parallel (1+(2u-1)^2)
+ \frac{3}{2} a_1^\parallel (2u-1)^3
 + 5\left[\frac{21}{4} \zeta_{3,^1P_1}^V
 + \zeta_{3,^1P_1}^A \Bigg(1-\frac{3}{16}\omega_{^1P_1}^A\Bigg)\right]\nonumber\\
& & \hspace{1.7cm} \times
 (2u-1)\left(5(2u-1)^2-3\right)
  + \frac{3}{16} a_2^\parallel \left(15(2u-1)^4 -6 (2u-1)^2 -1\right)\nonumber\\
& & \hspace{1.7cm}
 + 5 \zeta^V_{3,^1P_1}\lambda^V_{^1P_1}\left(3(2u-1)^2 -1\right)
 + \frac{105}{16}\left(\zeta^A_{3,^1P_1}\sigma^A_{^1P_1}
-\frac{1}{28} \zeta^V_{^1P_1}\sigma^V_{^1P_1}\right)\nonumber\\
 & & \hspace{1.7cm} \times
 \left(35(2u-1)^4 -30 (2u-1)^2 +3\right) -15 {a}_2^\perp \bigg[ \tilde{\delta}_+ (2u-1)^3 \nonumber\\
& & \hspace{1.7cm}+
 \frac{1}{2}\tilde{\delta}_-(3(2u-1)^2-1) \bigg]
  -\frac{3}{2} \bigg[\tilde{\delta}_+ ( 2 (2u-1) + \ln(1-u) -\ln u)
\nonumber \\
& & \hspace{1.7cm} + \tilde{\delta}_- (2+\ln u + \ln(1-u))\bigg](1+6a_2^\perp),\\
& & g_\perp^{(v)}(u)  =  6 u (1-u) \Bigg\{ a_0^\parallel +
a_1^\parallel (2u-1) +
 \Bigg[\frac{1}{4}a_2^\parallel
  +\frac{5}{3} \zeta^V_{3,^1P_1}
  \Bigg(\lambda^V_{^1P_1} -\frac{3}{16} \sigma^V_{^1P_1}\Bigg)
  \nonumber \\
  & & \hspace{1.7cm} +\frac{35}{4} \zeta^A_{3,^1P_1}\sigma^A_{^1P_1}\Bigg](5(2u-1)^2-1)  + \frac{20}{3}  (2u-1)
 \left[\zeta^A_{3, ^1P_1}\right. \nonumber \\
 & & \hspace{1.7cm} \left.
 + \frac{21}{16}
 \Bigg(\zeta^V_{3,^1P_1}  - \frac{1}{28}  \zeta^A_{3,^1P_1}\omega^A_{^1P_1}
  \Bigg)
 (7(2u-1)^2-3)\right]\nonumber\\
 & & \hspace{1.7cm} -5 a_2^\perp [2\tilde\delta_+ (2u-1) + \tilde\delta_- (1+(2u-1)^2)]
 \Bigg\}\nonumber\\
 & & \hspace{1.7cm} - 6 \bigg[ \tilde{\delta}_+  ((1-u) \ln(1-u) -u\ln u )
  \nonumber \\
  & & \hspace{1.7cm} + \tilde{\delta}_-  (2u (1-u) + (1-u) \ln (1-u) + u \ln u)\bigg]
  (1+6 a_2^\perp),\\
 & & h_\parallel^{(t)}(u) =  3(2u-1)^2+ \frac{3}{2} a_1^\perp (2u-1)
(3(2u-1)^2-1) + \Bigg[\frac{3}{2} a_2^\perp (2u-1)
 \nonumber \\
 & & \hspace{1.7cm} + \frac{15}{2}\zeta^\perp_{3, ^1P_1} \Bigg(\lambda^\perp_{^1P_1} - \frac{1}{10} \sigma^\perp_{^1P_1}\Bigg)
 \Bigg]
 (2u-1)(5(2u-1)^2-3) \nonumber\\
 && \hspace{1.7cm} +\frac{35}{4}\zeta^\perp_{3,^1P_1}(35(2u-1)^4-30 (2u-1)^2+3)\nonumber\\
 & & \hspace{1.7cm} +\frac{9}{2} a_1^\parallel (2u-1)
 \Bigg[\delta_+ (\ln u - \ln (1-u) -3 (2u-1))
 \nonumber \\
 & & \hspace{1.7cm} - \delta_-  \Bigg( \ln u + \ln (1-u)
  +\frac{8}{3}\Bigg)\Bigg], \\
& & h_\parallel^{(p)}(u)  =  6u (1-u) \Bigg\{ 1 + a_1^\perp (2u-1)
 +  \left(\frac{1}{4}a_2^\perp +
 \frac{35}{6} \zeta^\perp_{3,^1P_1} \right)(5 (2u-1)^2-1)
  \nonumber\\
 & & \hspace{1.7cm} +5\zeta^\perp_{3,^1P_1}
 \Bigg[\lambda^\perp_{^1P_1}-\frac{1}{40}(7(2u-1)^3-3)\sigma^\perp_{^1P_1}
 \Bigg] (2u-1)
 \Bigg\}  \nonumber\\
 & & \hspace{1.7cm} -9 a_1^\parallel \delta_+ (3 u (1-u) + (1-u) \ln
(1-u) + u \ln u)
 \nonumber \\
 & & \hspace{1.7cm} -9 a_1^\parallel \delta_-  \Bigg( \frac{2}{3} (2u-1) u (1-u)
 + (1-u) \ln (1-u) - u \ln u \Bigg),
\end{eqnarray}
for the $1^1P_1$ states, where
\begin{eqnarray}
& & \tilde{\delta}_\pm  ={f_{A}^{\perp}\over f_{A}}{m_{q_2} \pm m_{q_1}
\over m_{A}},\qquad \delta_\pm = \frac{f_A}{f^\perp_A} \frac{m_{q_2} \pm m_{q_1}}{m_A}, \\
& & \zeta_{3,A}^{V(A)} = \frac{f^{V(A)}_{3A}}{f_{A} m_{A}}, \qquad \zeta^\perp_{3,A} = \frac{f^\perp_{3,A}}{f^\perp_A m_A}.
\end{eqnarray}

\section{Variation of $B_{(s)} \rightarrow A$
 form factors as functions of $T$ for different $s_0$ at
$q^2=0$}\label{FFfigT}

In this appendix, we give the variation of form factors as functions of $T$ for different $s_0$ at maximal recoil point ($q^2=0$) for the $B_{(s)}$ decays into $K_{1A}$, $K_{1B}$, $f_1(1285)$, $f_1(1420)$, $h_1(1415)$, $h_1(1170)$.
\begin{figure}[H]
\centering
\includegraphics[width=3in]{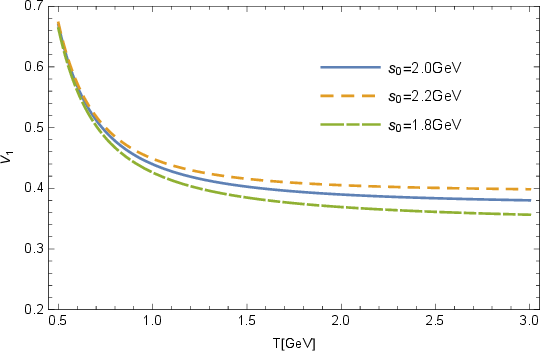} \hfill
\includegraphics[width=3in]{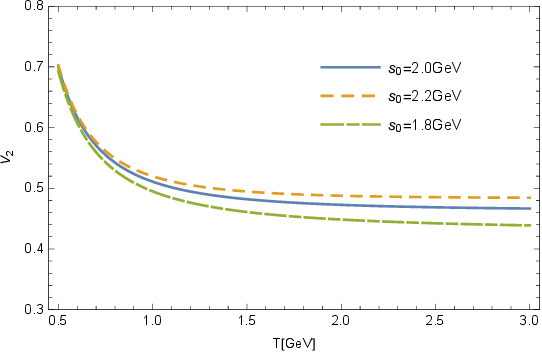} \\
\includegraphics[width=3in]{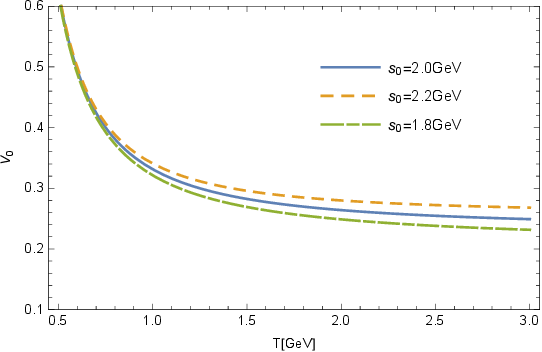}\hfill
\includegraphics[width=3in]{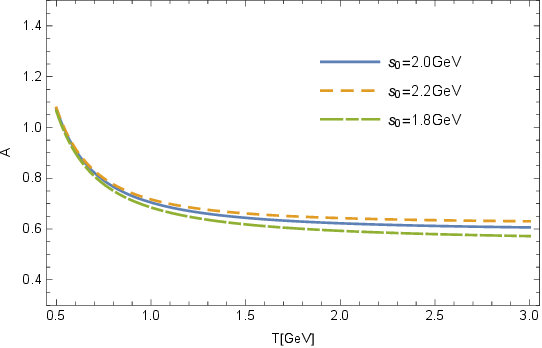}
\caption{$B \rightarrow K_{1A}$ form factors as
functions of $T$ for different $s_0$ at $q^2=0$. } \label{Fig:BK1AT}
\end{figure}
\begin{figure}[H]
\centering
\includegraphics[width=3in]{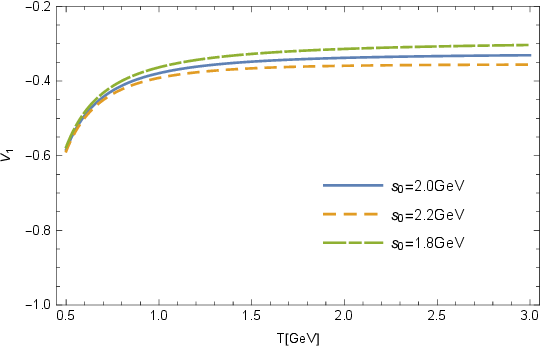} \hfill
\includegraphics[width=3in]{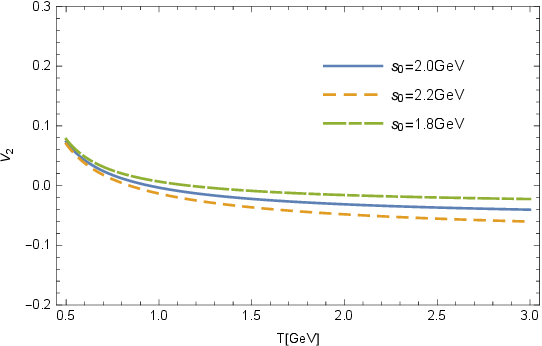} \\
\includegraphics[width=3in]{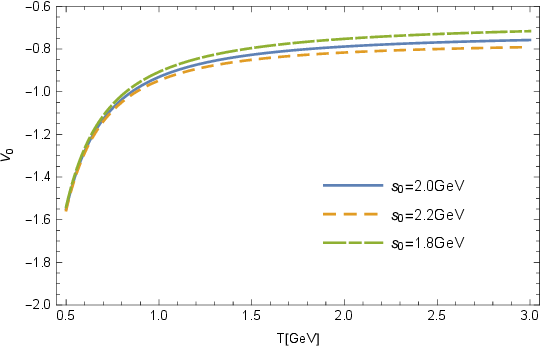}\hfill
\includegraphics[width=3in]{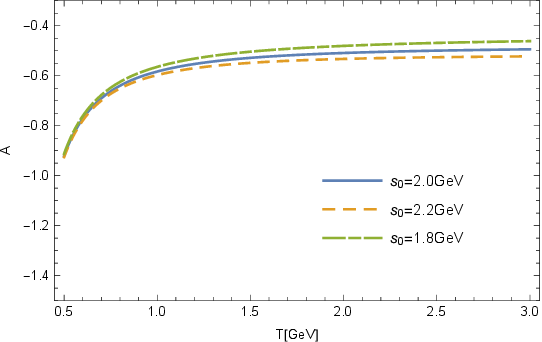}
\caption{$B \rightarrow K_{1B}$ form factors as
functions of $T$ for different $s_0$ at $q^2=0$. } \label{Fig:BK1BT}
\end{figure}

\begin{figure}[H]
\centering
\includegraphics[width=3in]{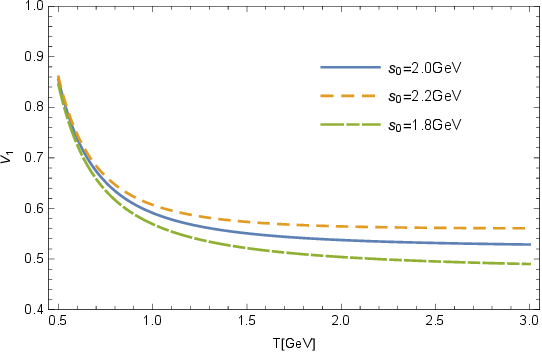} \hfill
\includegraphics[width=3in]{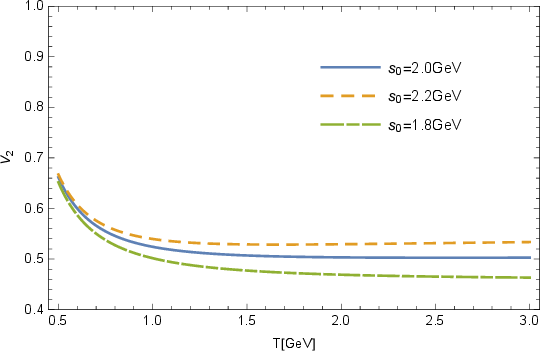} \\
\includegraphics[width=3in]{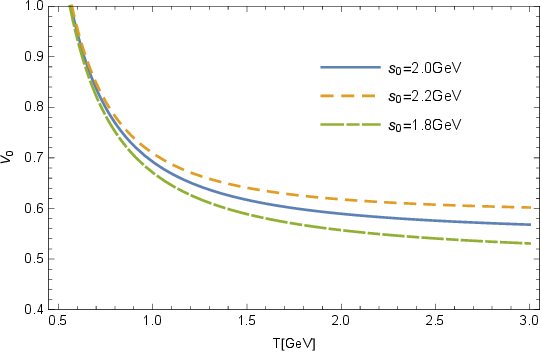}\hfill
\includegraphics[width=3in]{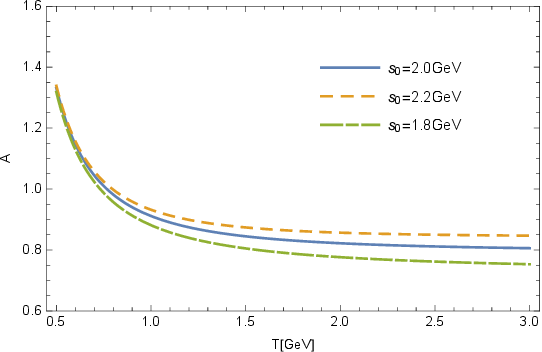}
\caption{$B \rightarrow \bar{K}_{1A}$ form factors as
functions of $T$ for different $s_0$ at $q^2=0$. } \label{Fig:BK1AbT}
\end{figure}
\begin{figure}[H]
\centering
\includegraphics[width=3in]{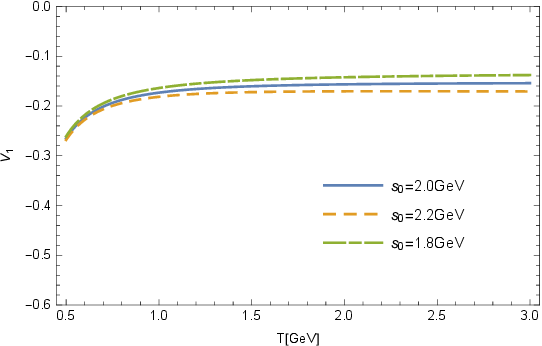} \hfill
\includegraphics[width=3in]{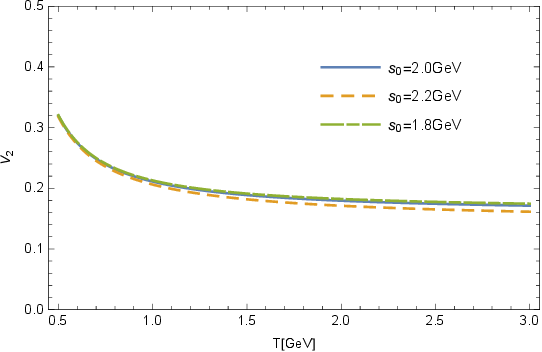} \\
\includegraphics[width=3in]{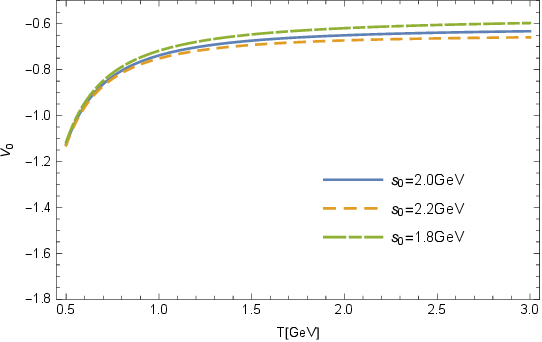}\hfill
\includegraphics[width=3in]{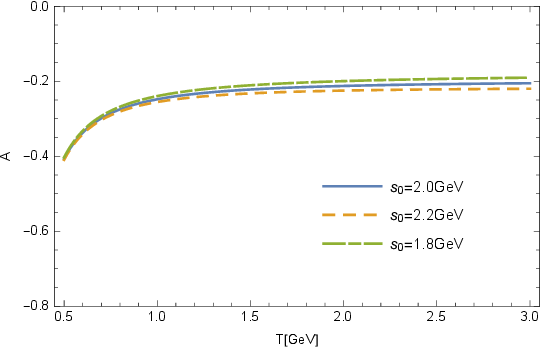}
\caption{$B \rightarrow \bar{K}_{1B}$ form factors as
functions of $T$ for different $s_0$ at $q^2=0$. } \label{Fig:BK1BbT}
\end{figure}

\begin{figure}[H]
\centering
\includegraphics[width=3in]{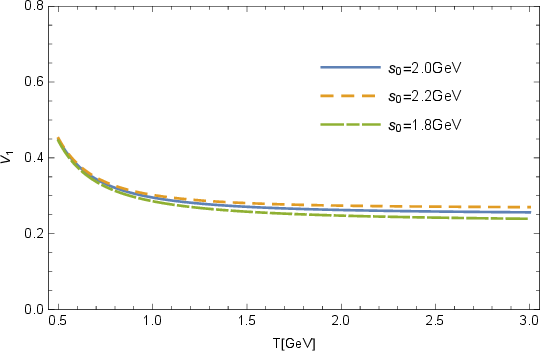} \hfill
\includegraphics[width=3in]{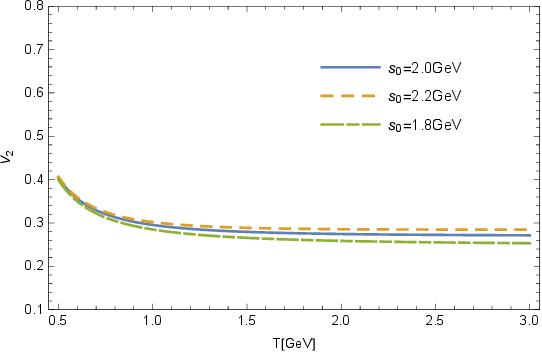} \\
\includegraphics[width=3in]{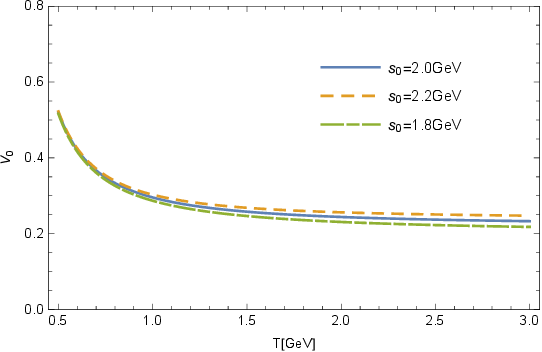}\hfill
\includegraphics[width=3in]{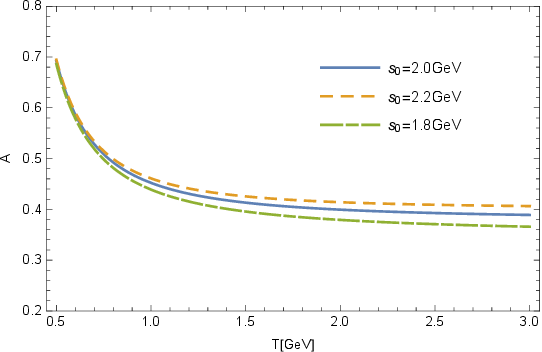}
\caption{$B \rightarrow f_1(1^3P_1)$ form factors as
functions of $T$ for different $s_0$ at $q^2=0$. } \label{Fig:Bf1T}
\end{figure}
\begin{figure}[H]
\centering
\includegraphics[width=3in]{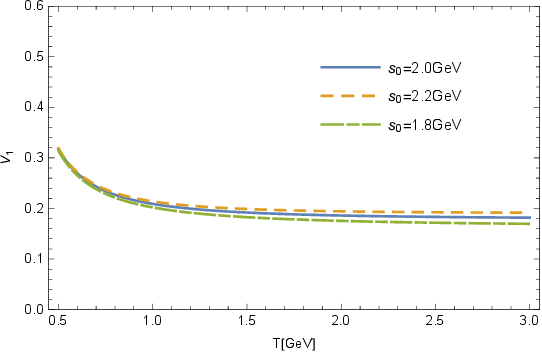} \hfill
\includegraphics[width=3in]{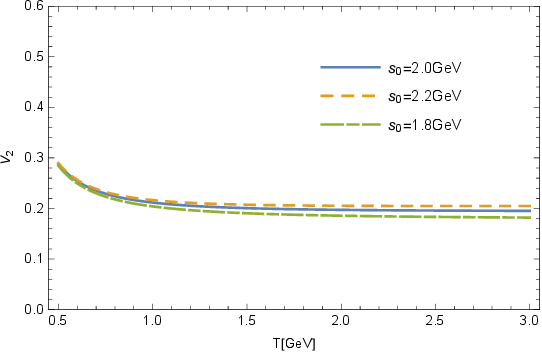} \\
\includegraphics[width=3in]{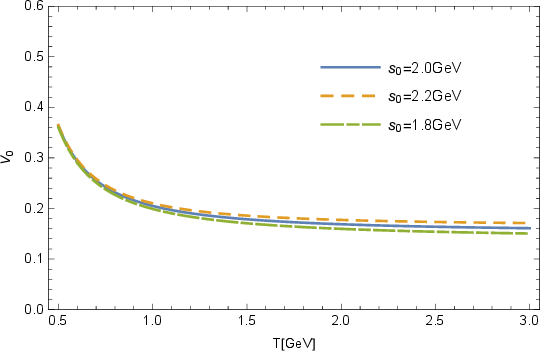}\hfill
\includegraphics[width=3in]{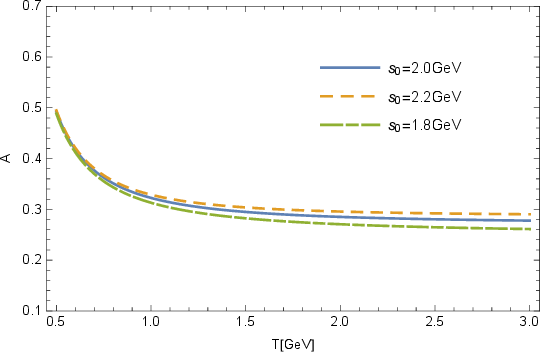}
\caption{$B \rightarrow f_8(1^3P_1)$ form factors as
functions of $T$ for different $s_0$ at $q^2=0$. } \label{Fig:Bf8T}
\end{figure}

\begin{figure}[H]
\centering
\includegraphics[width=3in]{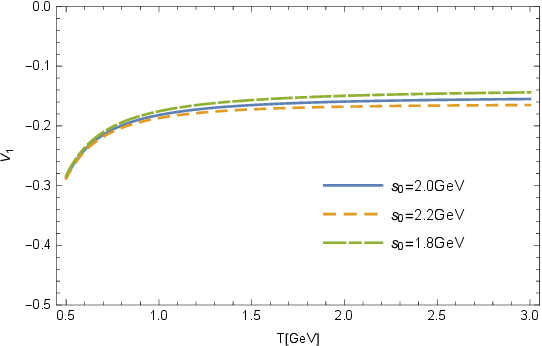} \hfill
\includegraphics[width=3in]{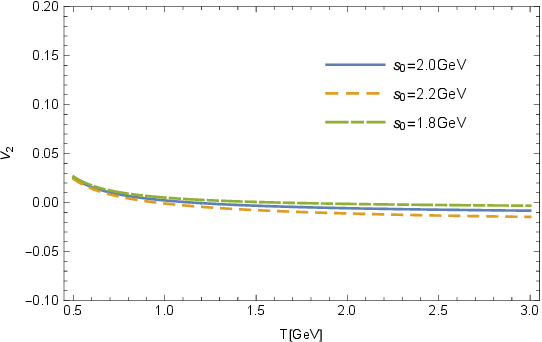} \\
\includegraphics[width=3in]{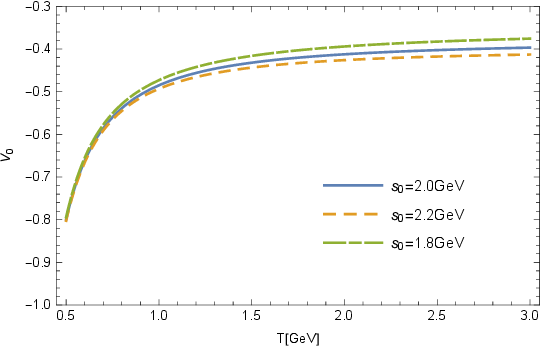}\hfill
\includegraphics[width=3in]{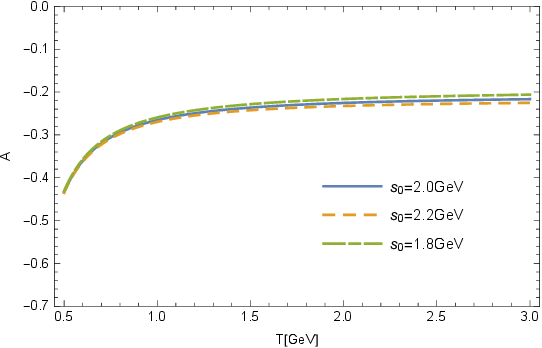}
\caption{$B \rightarrow h_1(1^1P_1)$ form factors as
functions of $T$ for different $s_0$ at $q^2=0$. } \label{Fig:Bh1T}
\end{figure}
\begin{figure}[H]
\centering
\includegraphics[width=3in]{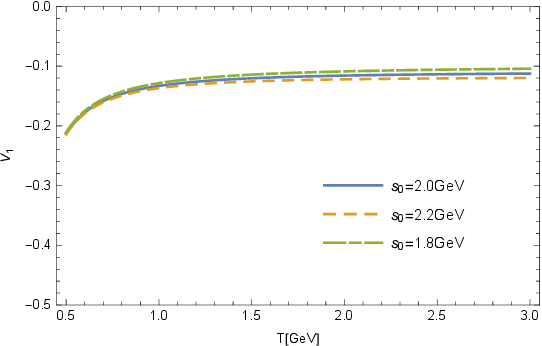} \hfill
\includegraphics[width=3in]{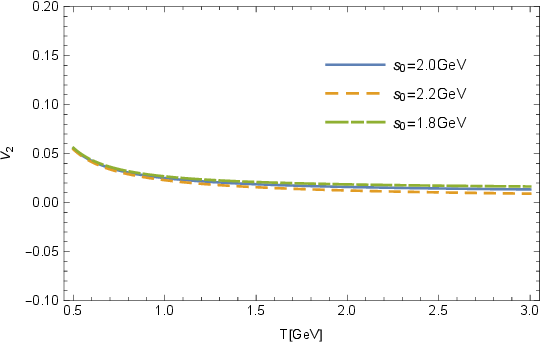} \\
\includegraphics[width=3in]{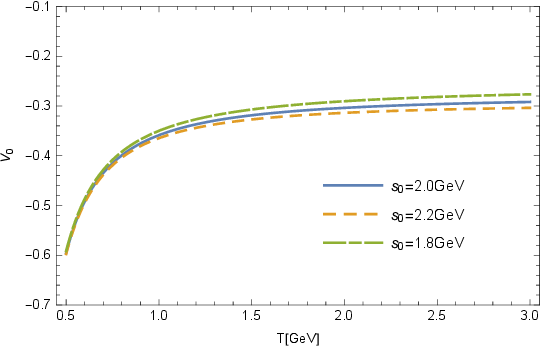}\hfill
\includegraphics[width=3in]{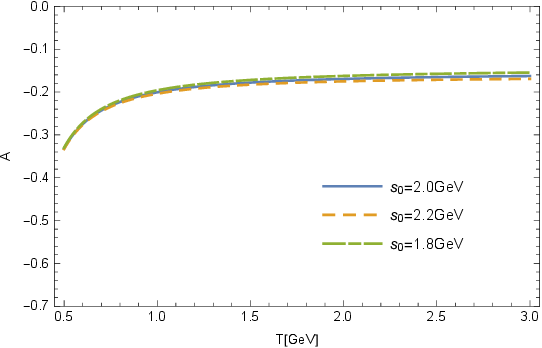}
\caption{$B \rightarrow h_8(1^1P_1)$ form factors as
functions of $T$ for different $s_0$ at $q^2=0$. } \label{Fig:Bh8T}
\end{figure}

\begin{figure}[H]
\centering
\includegraphics[width=3in]{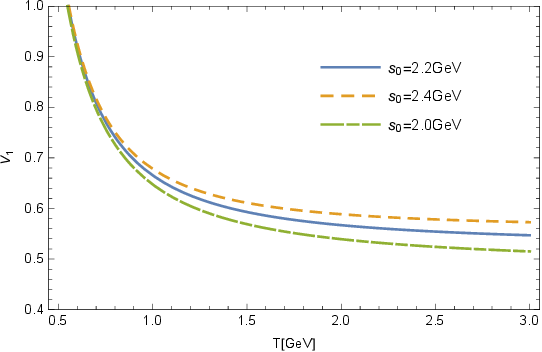} \hfill
\includegraphics[width=3in]{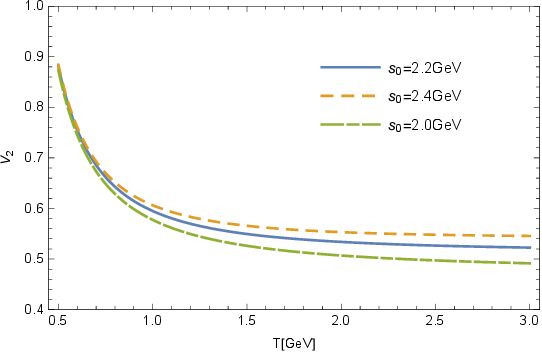} \\
\includegraphics[width=3in]{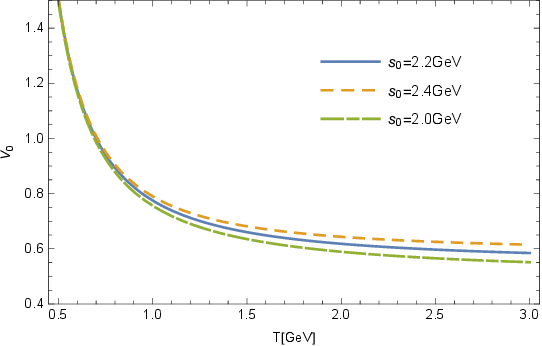}\hfill
\includegraphics[width=3in]{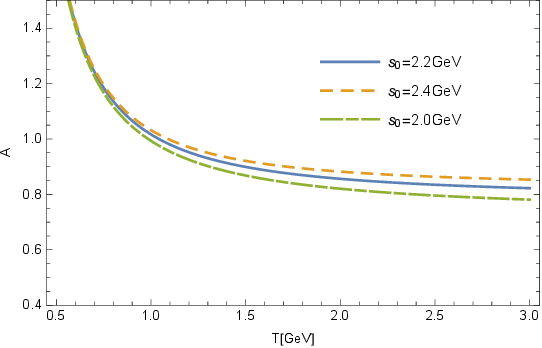}
\caption{$B_s \rightarrow \bar{K}_{1A}$ form factors as
functions of $T$ for different $s_0$ at $q^2=0$. } \label{Fig:BsK1AbT}
\end{figure}
\begin{figure}[H]
\centering
\includegraphics[width=3in]{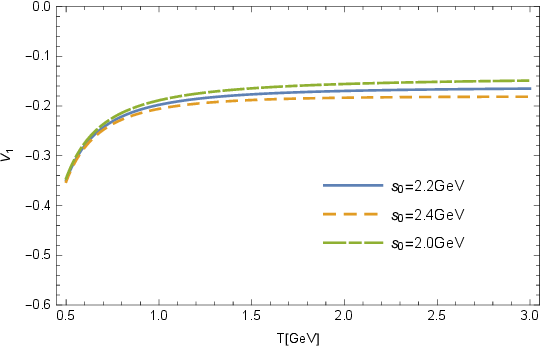} \hfill
\includegraphics[width=3in]{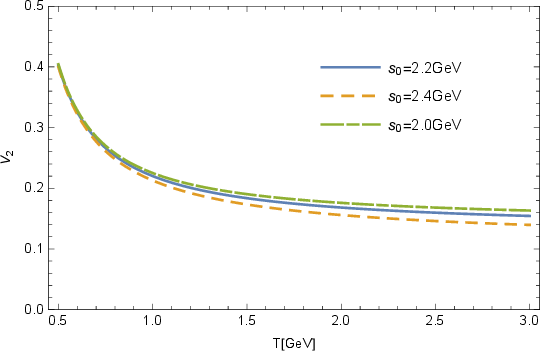} \\
\includegraphics[width=3in]{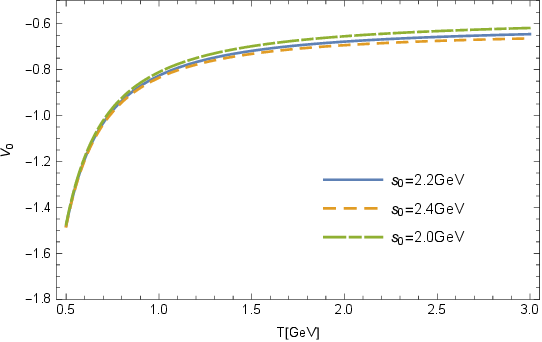}\hfill
\includegraphics[width=3in]{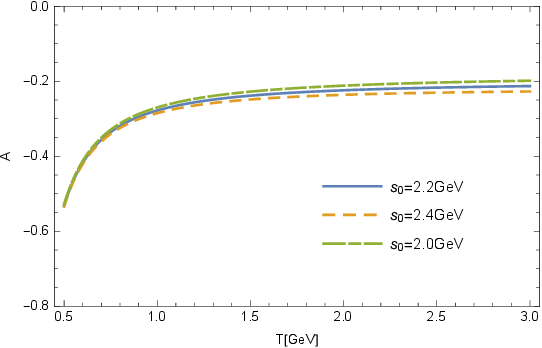}
\caption{$B_s \rightarrow \bar{K}_{1B}$ form factors as
functions of $T$ for different $s_0$ at $q^2=0$. } \label{Fig:BsK1BbT}
\end{figure}

\begin{figure}[H]
\centering
\includegraphics[width=3in]{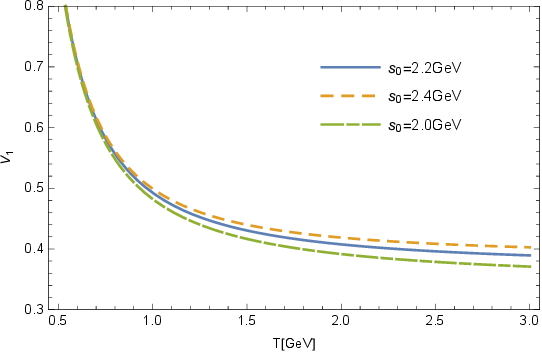} \hfill
\includegraphics[width=3in]{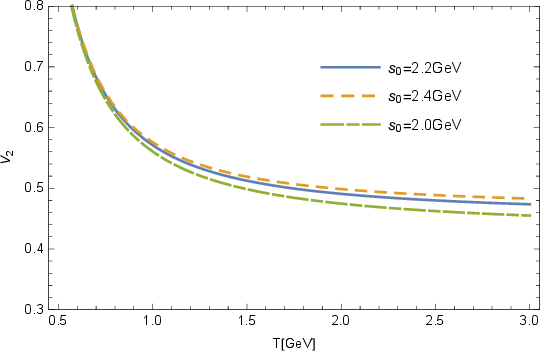} \\
\includegraphics[width=3in]{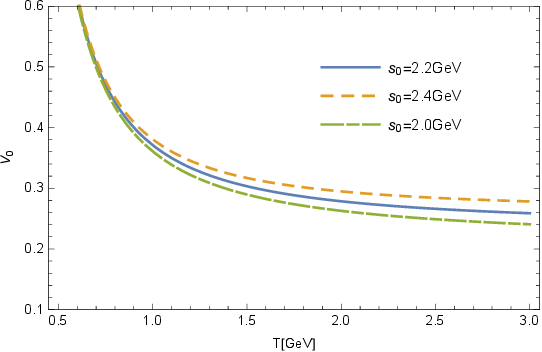}\hfill
\includegraphics[width=3in]{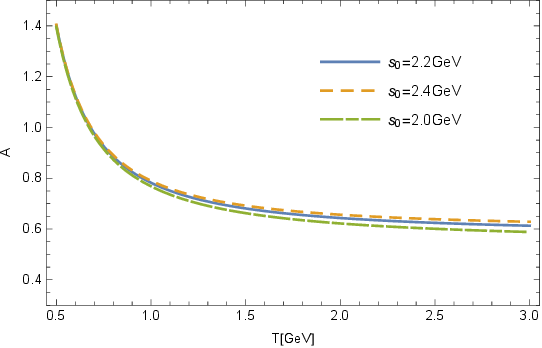}
\caption{$B_s \rightarrow K_{1A}$ form factors as
functions of $T$ for different $s_0$ at $q^2=0$. } \label{Fig:BsK1AT}
\end{figure}
\begin{figure}[H]
\centering
\includegraphics[width=3in]{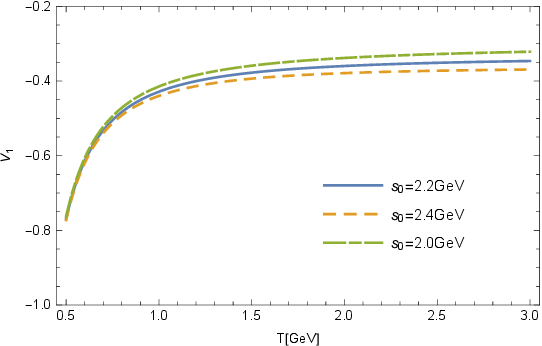} \hfill
\includegraphics[width=3in]{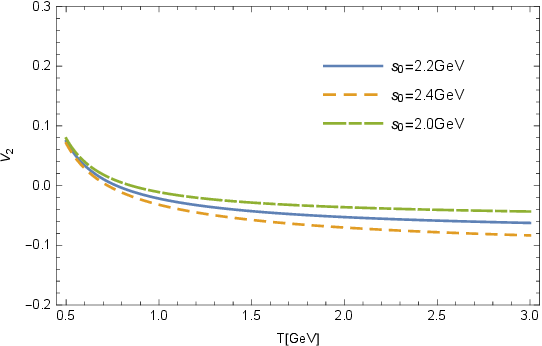} \\
\includegraphics[width=3in]{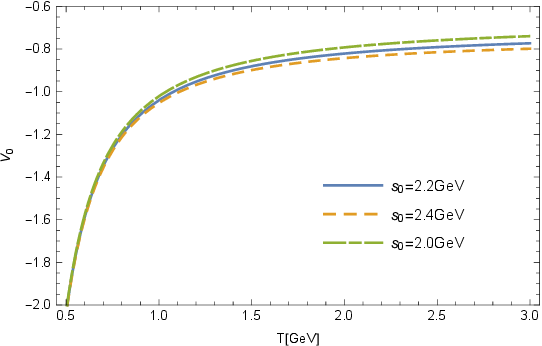}\hfill
\includegraphics[width=3in]{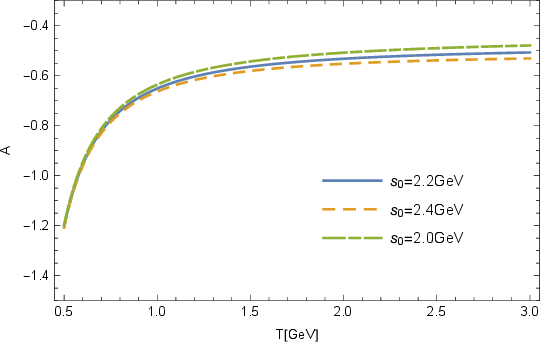}
\caption{$B_s \rightarrow K_{1B}$ form factors as
functions of $T$ for different $s_0$ at $q^2=0$. } \label{Fig:BsK1BT}
\end{figure}

\begin{figure}[H]
\centering
\includegraphics[width=3in]{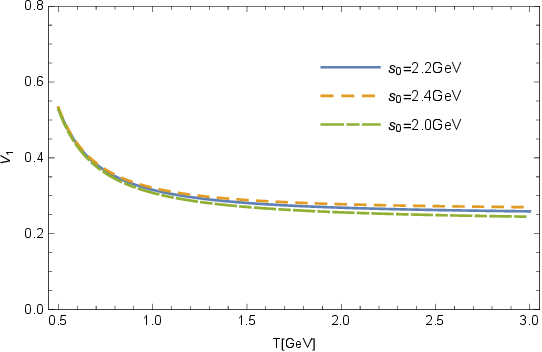} \hfill
\includegraphics[width=3in]{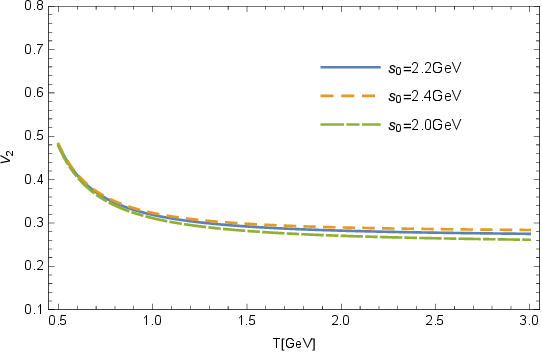} \\
\includegraphics[width=3in]{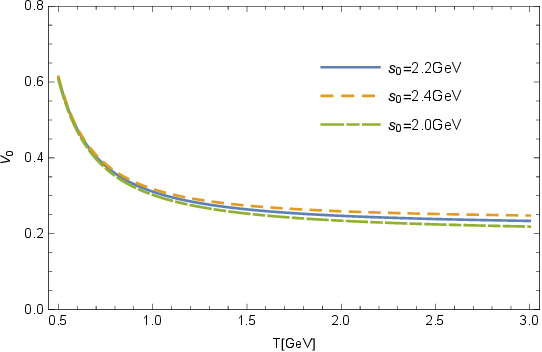}\hfill
\includegraphics[width=3in]{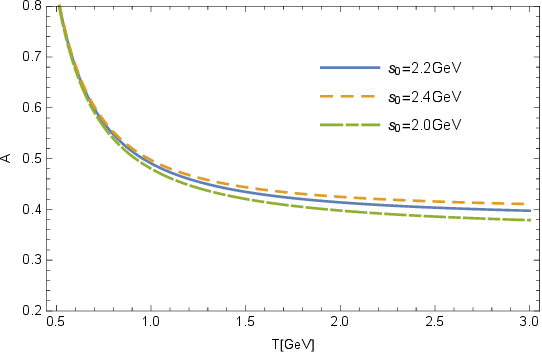}
\caption{$B_s \rightarrow f_1(1^3P_1)$ form factors as
functions of $T$ for different $s_0$ at $q^2=0$. } \label{Fig:Bsf1T}
\end{figure}
\begin{figure}[H]
\centering
\includegraphics[width=3in]{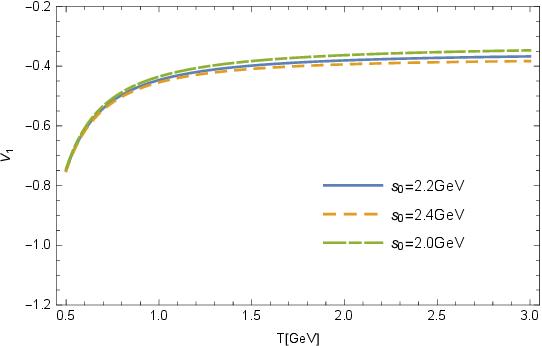} \hfill
\includegraphics[width=3in]{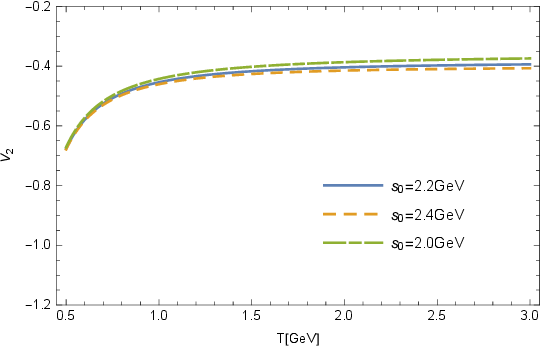} \\
\includegraphics[width=3in]{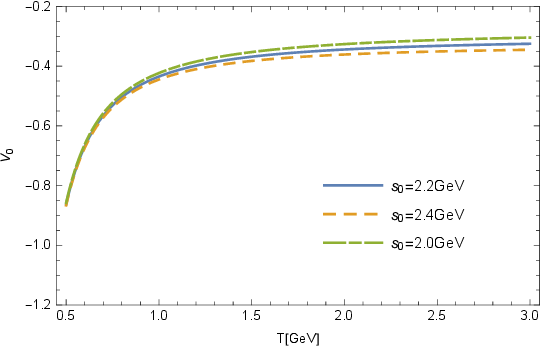}\hfill
\includegraphics[width=3in]{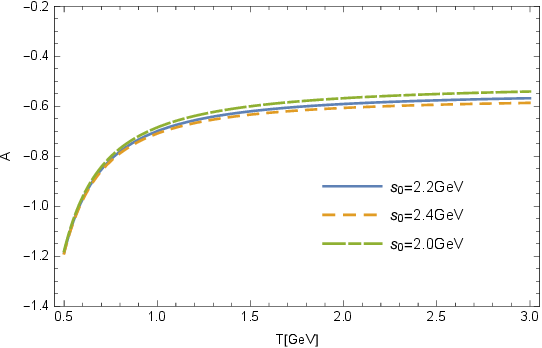}
\caption{$B_s \rightarrow f_8(1^3P_1)$ form factors as
functions of $T$ for different $s_0$ at $q^2=0$. } \label{Fig:Bsf8T}
\end{figure}

\begin{figure}[H]
\centering
\includegraphics[width=3in]{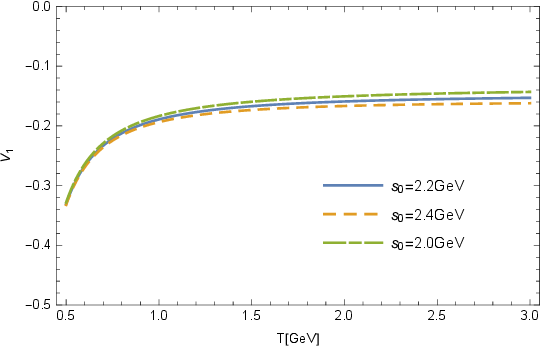} \hfill
\includegraphics[width=3in]{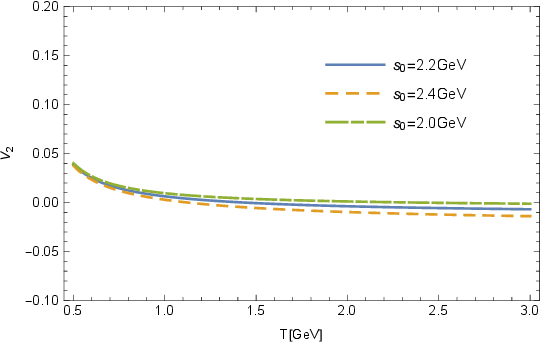} \\
\includegraphics[width=3in]{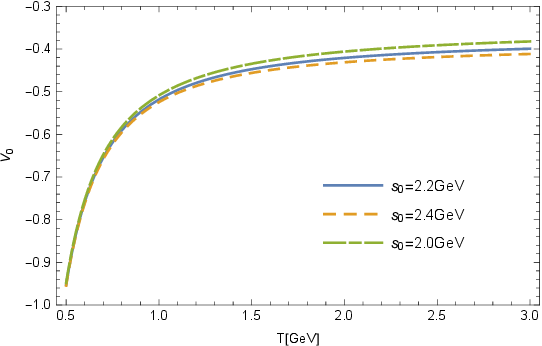}\hfill
\includegraphics[width=3in]{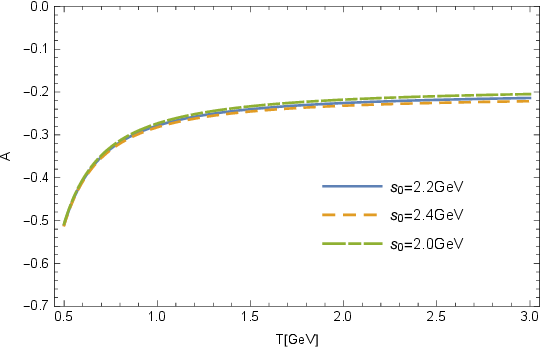}
\caption{$B_s \rightarrow h_1(1^1P_1)$ form factors as
functions of $T$ for different $s_0$ at $q^2=0$. } \label{Fig:Bsh1T}
\end{figure}
\begin{figure}[H]
\centering
\includegraphics[width=3in]{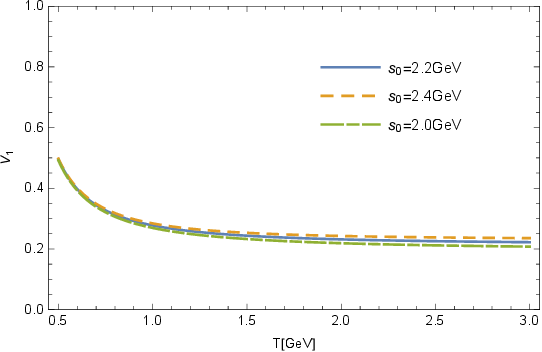} \hfill
\includegraphics[width=3in]{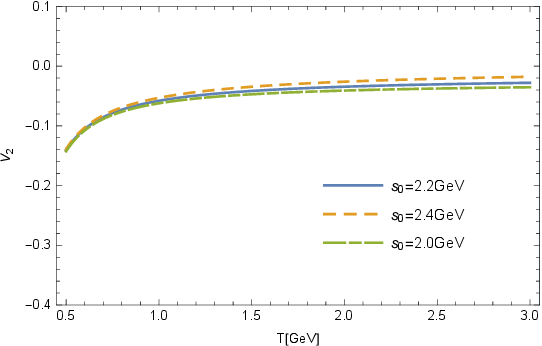} \\
\includegraphics[width=3in]{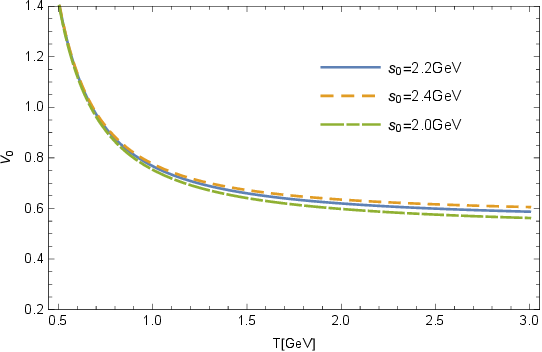}\hfill
\includegraphics[width=3in]{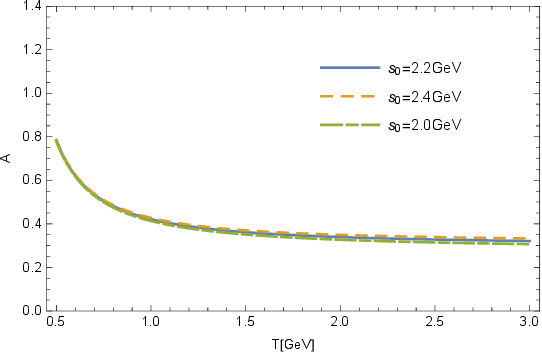}
\caption{$B_s \rightarrow h_8(1^1P_1)$ form factors as
functions of $T$ for different $s_0$ at $q^2=0$. } \label{Fig:Bsh8T}
\end{figure}

\section{The behaviors of form factors as functions of $q^2$}\label{FFfigq}

In this appendix, we give the behaviors of form factors as functions of $q^2$  for the $B_{(s)}$ decays into $K_{1A}$, $K_{1B}$, $f_1(1285)$, $f_1(1420)$, $h_1(1415)$, $h_1(1170)$.

\begin{figure}[H]
\centering
\includegraphics[width=3in]{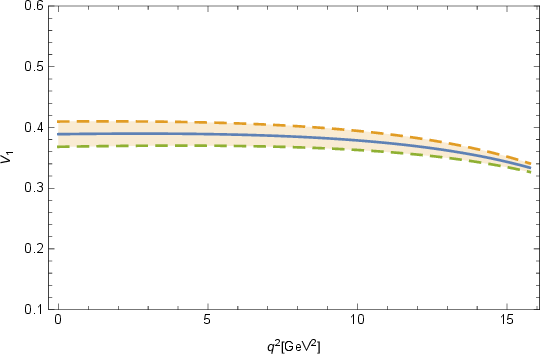} \hfill
\includegraphics[width=3in]{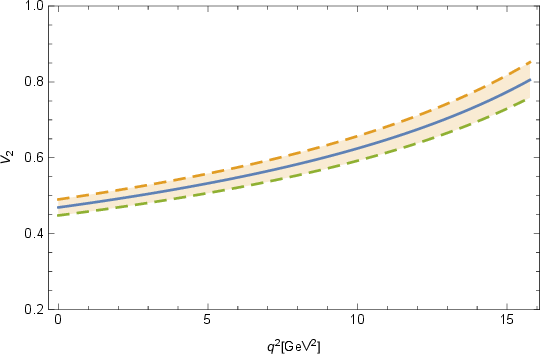} \\
\includegraphics[width=3in]{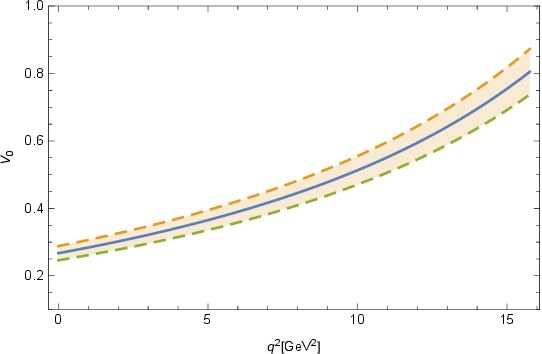}\hfill
\includegraphics[width=3in]{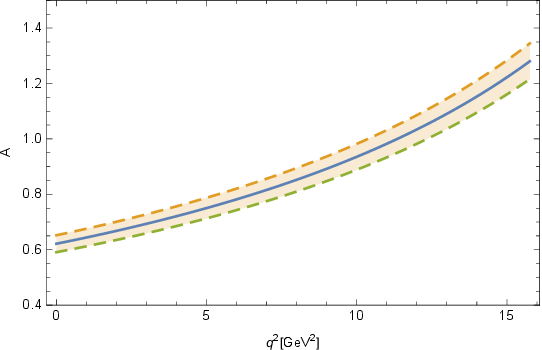}
\caption{$B \rightarrow K_{1A}$ form factors as
functions of $q^2$. } \label{Fig:BK1Aq}
\end{figure}
\begin{figure}[H]
\centering
\includegraphics[width=3in]{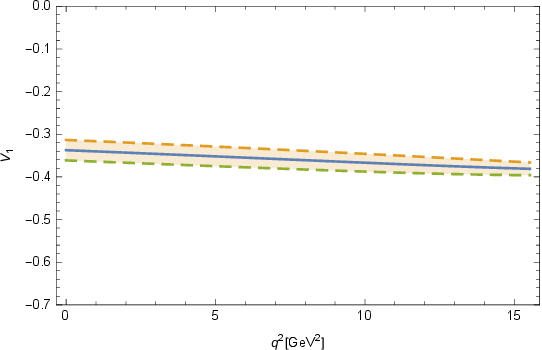} \hfill
\includegraphics[width=3in]{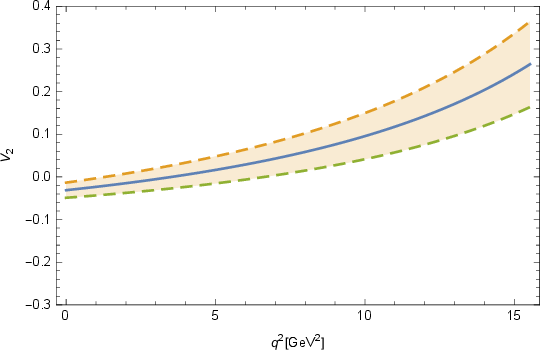} \\
\includegraphics[width=3in]{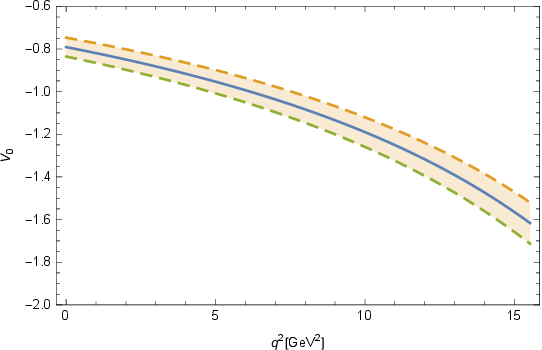}\hfill
\includegraphics[width=3in]{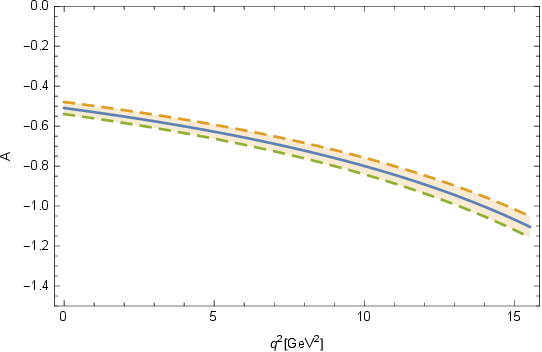}
\caption{$B \rightarrow K_{1B}$ form factors as
functions of $q^2$. } \label{Fig:BK1Bq}
\end{figure}

\begin{figure}[H]
\centering
\includegraphics[width=3in]{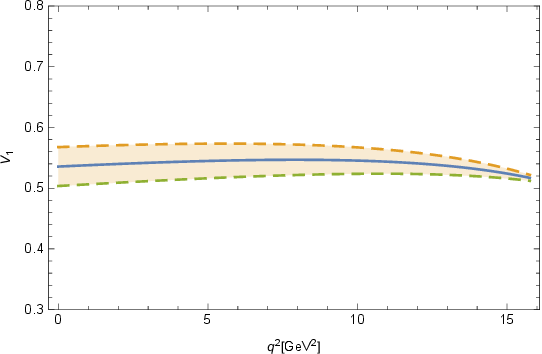} \hfill
\includegraphics[width=3in]{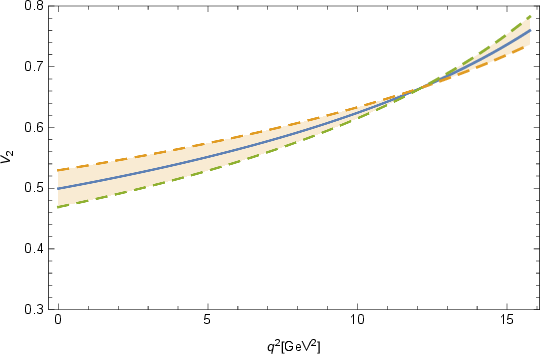} \\
\includegraphics[width=3in]{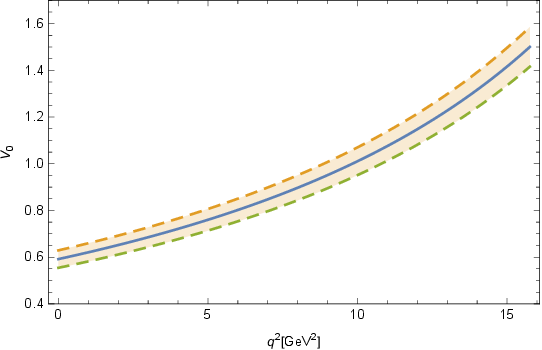}\hfill
\includegraphics[width=3in]{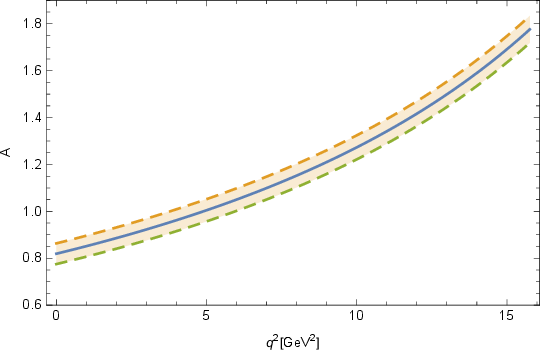}
\caption{$B \rightarrow \bar{K}_{1A}$ form factors as
functions of $q^2$. } \label{Fig:BK1Abq}
\end{figure}
\begin{figure}[H]
\centering
\includegraphics[width=3in]{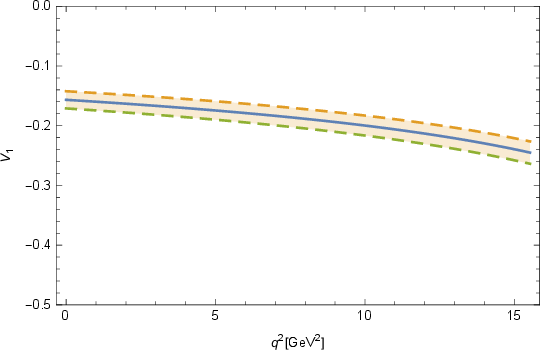} \hfill
\includegraphics[width=3in]{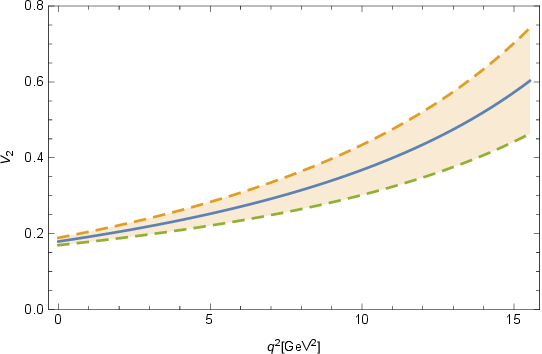} \\
\includegraphics[width=3in]{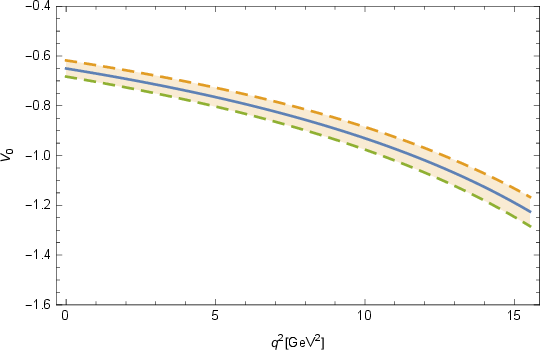}\hfill
\includegraphics[width=3in]{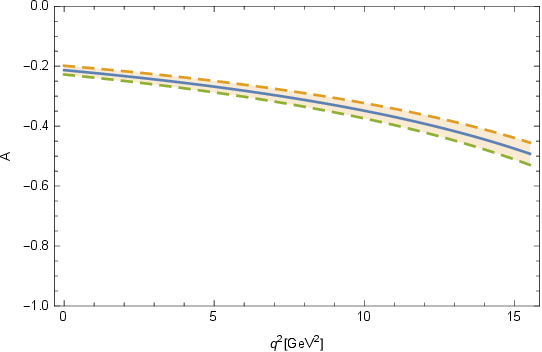}
\caption{$B \rightarrow \bar{K}_{1B}$ form factors as
functions of $q^2$. } \label{Fig:BK1Bbq}
\end{figure}

\begin{figure}[H]
\centering
\includegraphics[width=3in]{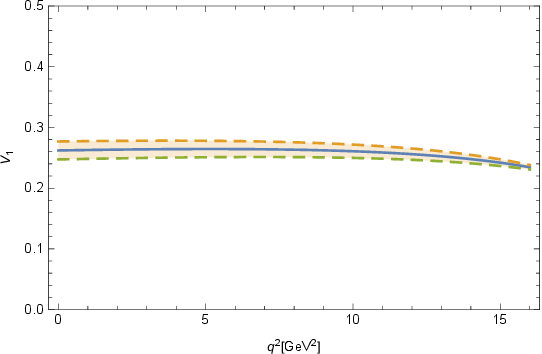} \hfill
\includegraphics[width=3in]{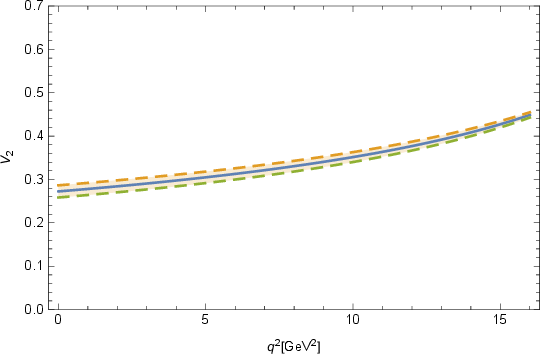} \\
\includegraphics[width=3in]{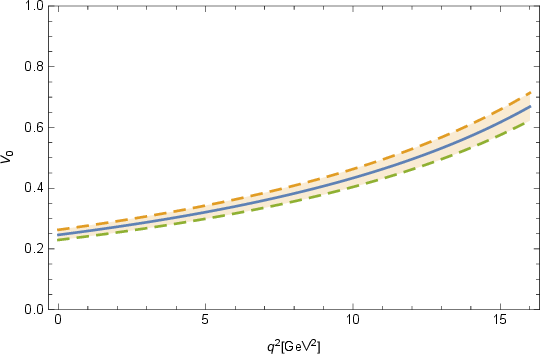}\hfill
\includegraphics[width=3in]{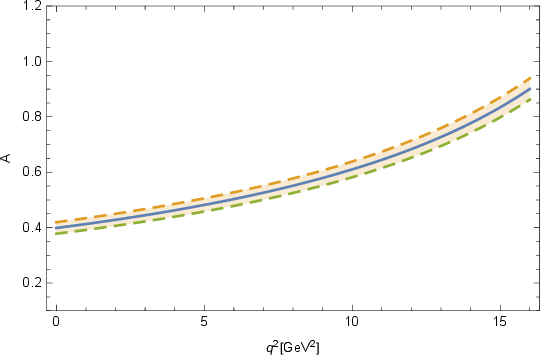}
\caption{$B \rightarrow f_1(1^3P_1)$ form factors as
functions of $q^2$. } \label{Fig:Bf1q}
\end{figure}
\begin{figure}[H]
\centering
\includegraphics[width=3in]{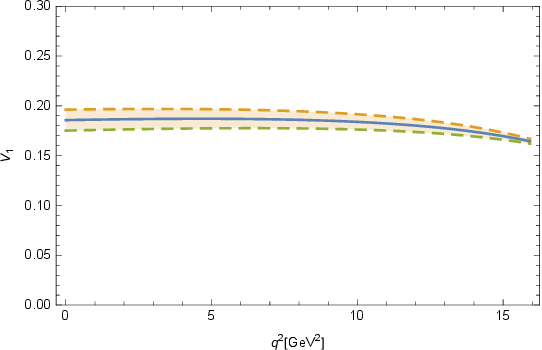} \hfill
\includegraphics[width=3in]{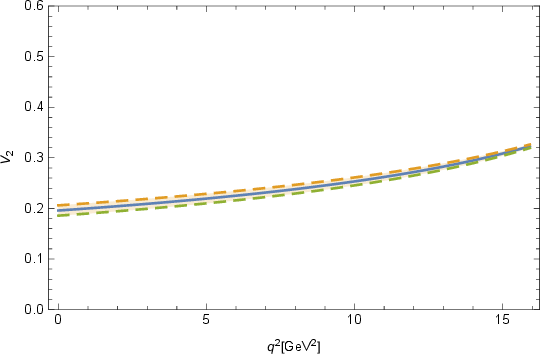} \\
\includegraphics[width=3in]{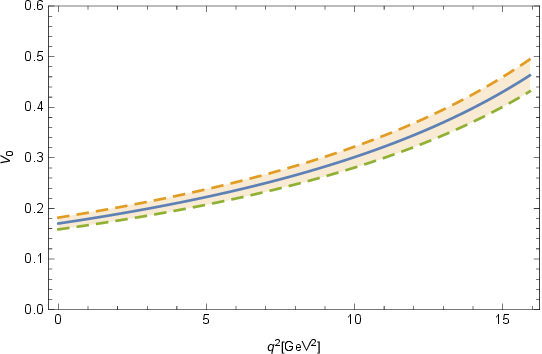}\hfill
\includegraphics[width=3in]{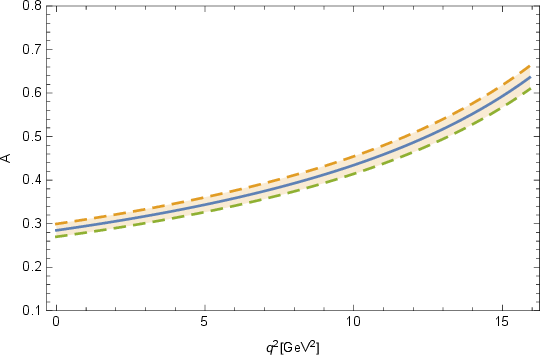}
\caption{$B \rightarrow f_8(1^3P_1)$ form factors as
functions of $q^2$. } \label{Fig:Bf8q}
\end{figure}

\begin{figure}[H]
\centering
\includegraphics[width=3in]{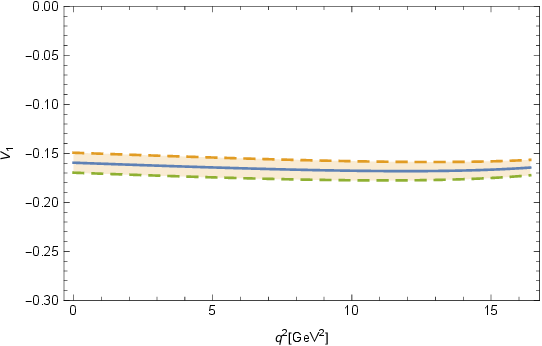} \hfill
\includegraphics[width=3in]{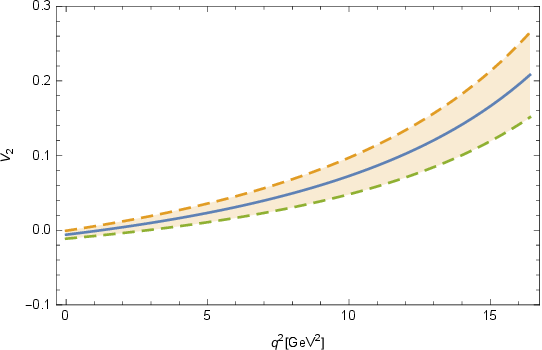} \\
\includegraphics[width=3in]{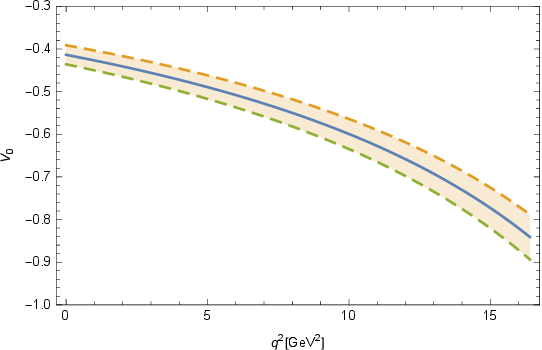}\hfill
\includegraphics[width=3in]{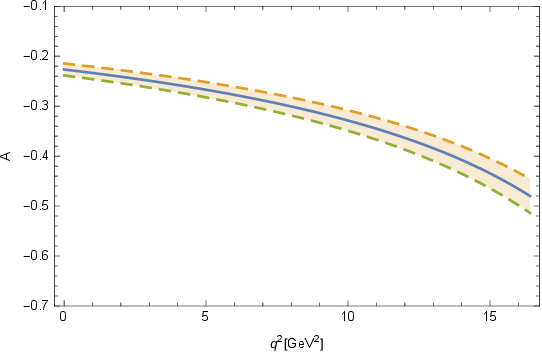}
\caption{$B \rightarrow h_1(1^1P_1)$ form factors as
functions of $q^2$. } \label{Fig:Bh1q}
\end{figure}
\begin{figure}[H]
\centering
\includegraphics[width=3in]{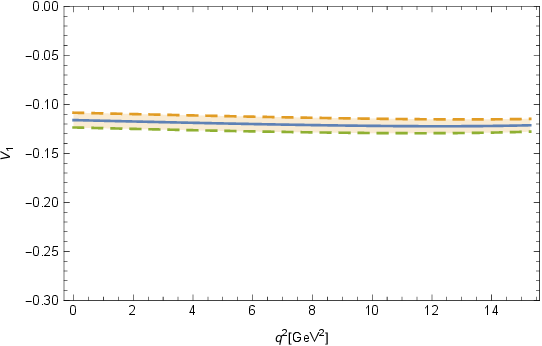} \hfill
\includegraphics[width=3in]{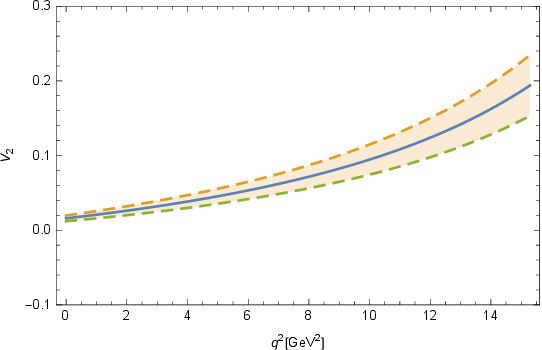} \\
\includegraphics[width=3in]{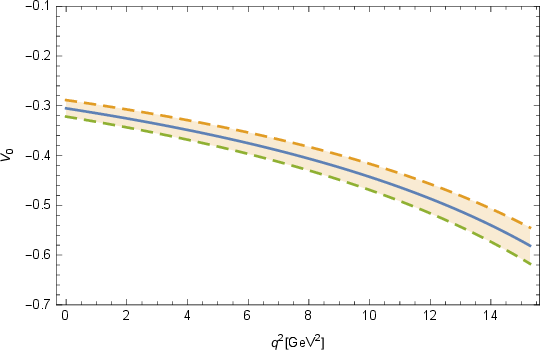}\hfill
\includegraphics[width=3in]{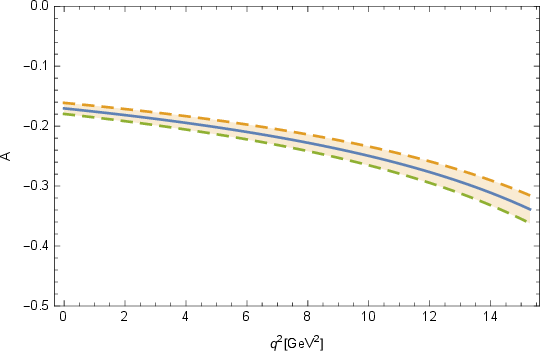}
\caption{$B \rightarrow h_8(1^1P_1)$ form factors as
functions of $q^2$. } \label{Fig:Bh8q}
\end{figure}

\begin{figure}[H]
\centering
\includegraphics[width=3in]{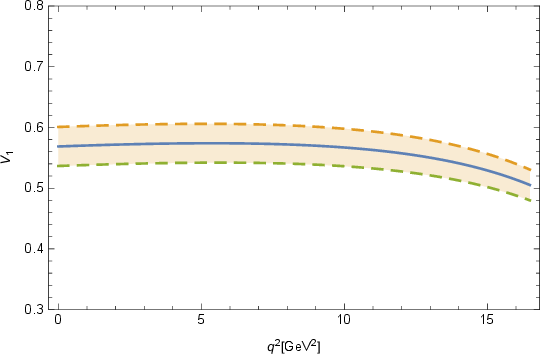} \hfill
\includegraphics[width=3in]{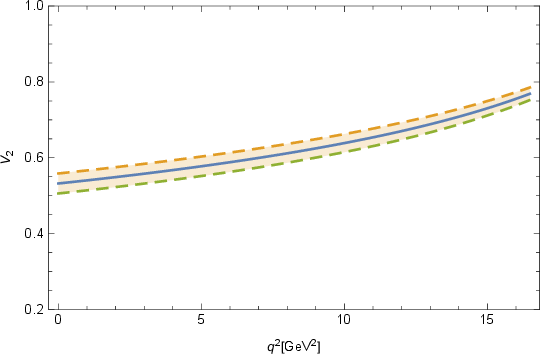} \\
\includegraphics[width=3in]{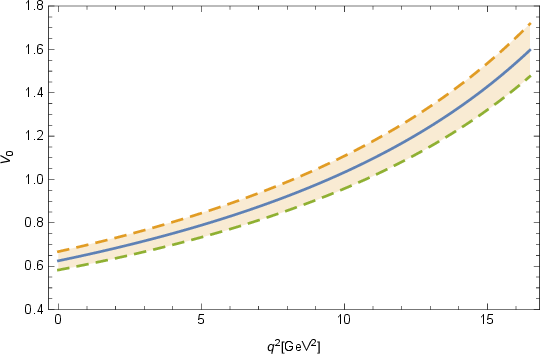}\hfill
\includegraphics[width=3in]{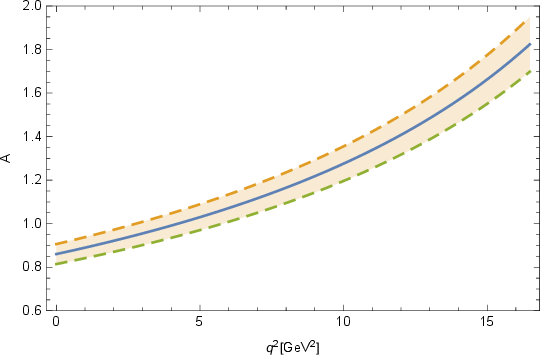}
\caption{$B_s \rightarrow \bar{K}_{1A}$ form factors as
functions of $q^2$. } \label{Fig:BsK1Abq}
\end{figure}
\begin{figure}[H]
\centering
\includegraphics[width=3in]{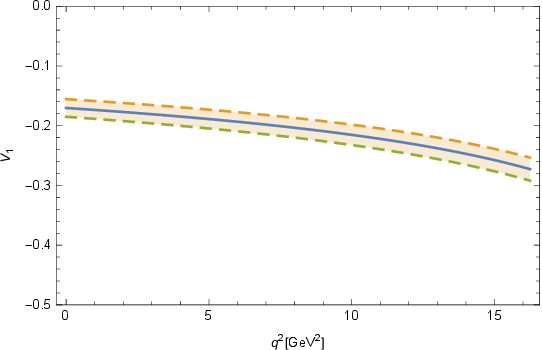} \hfill
\includegraphics[width=3in]{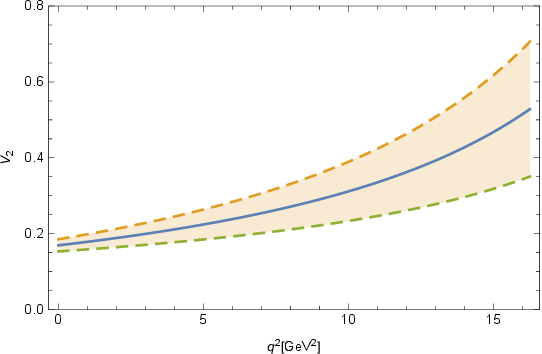} \\
\includegraphics[width=3in]{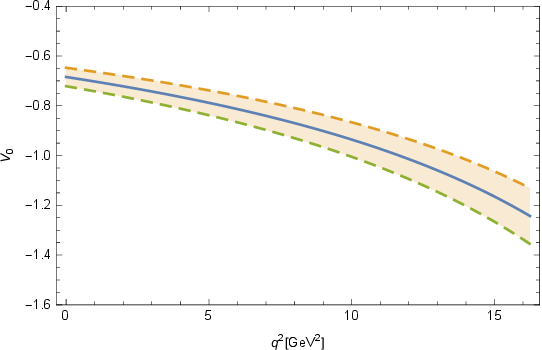}\hfill
\includegraphics[width=3in]{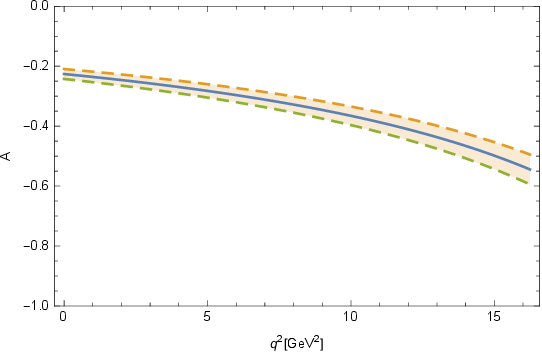}
\caption{$B_s \rightarrow \bar{K}_{1B}$ form factors as
functions of $q^2$. } \label{Fig:BsK1Bbq}
\end{figure}

\begin{figure}[H]
\centering
\includegraphics[width=3in]{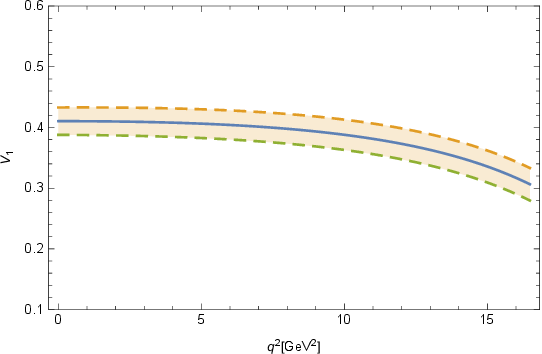} \hfill
\includegraphics[width=3in]{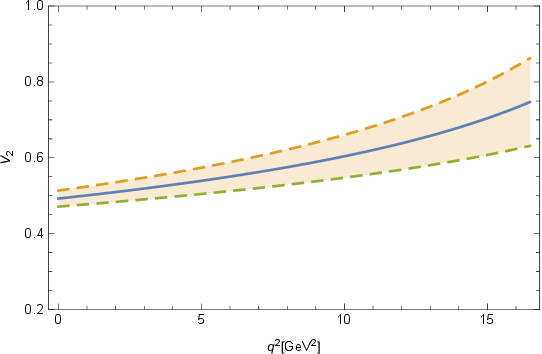} \\
\includegraphics[width=3in]{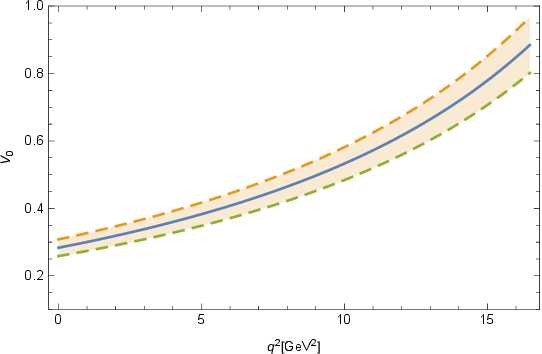}\hfill
\includegraphics[width=3in]{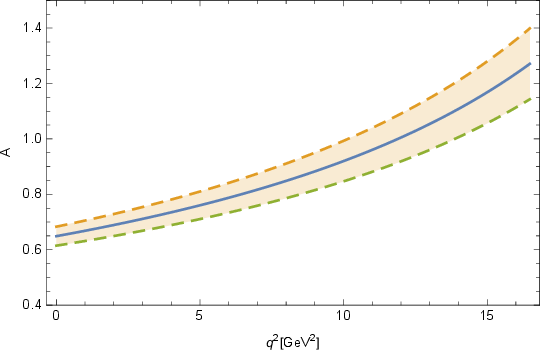}
\caption{$B_s \rightarrow K_{1A}$ form factors as
functions of $q^2$. } \label{Fig:BsK1Aq}
\end{figure}
\begin{figure}[H]
\centering
\includegraphics[width=3in]{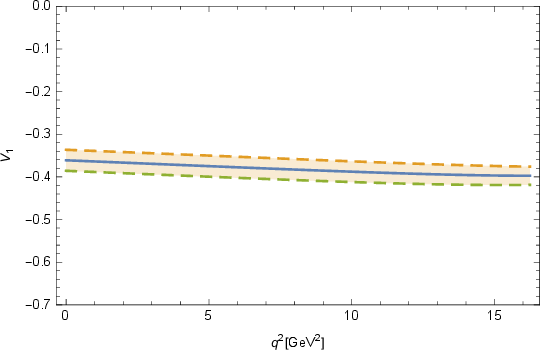} \hfill
\includegraphics[width=3in]{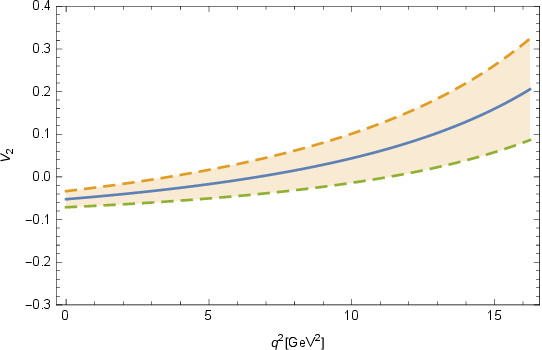} \\
\includegraphics[width=3in]{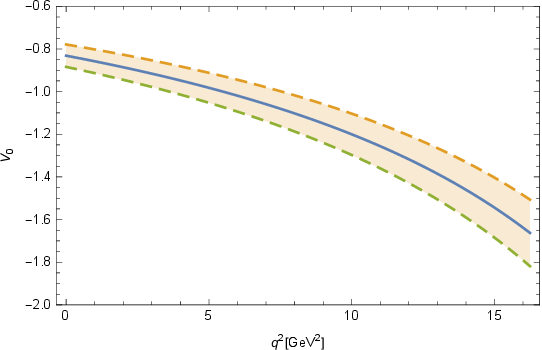}\hfill
\includegraphics[width=3in]{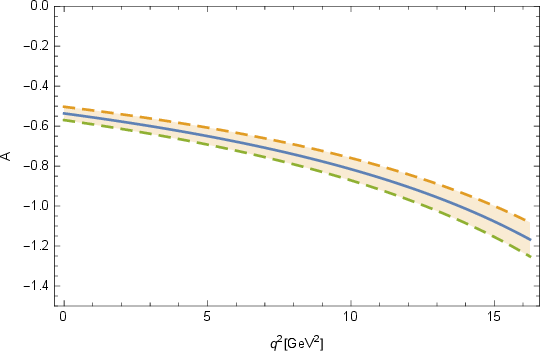}
\caption{$B_s \rightarrow K_{1B}$ form factors as
functions of $q^2$. } \label{Fig:BsK1Bq}
\end{figure}

\begin{figure}[H]
\centering
\includegraphics[width=3in]{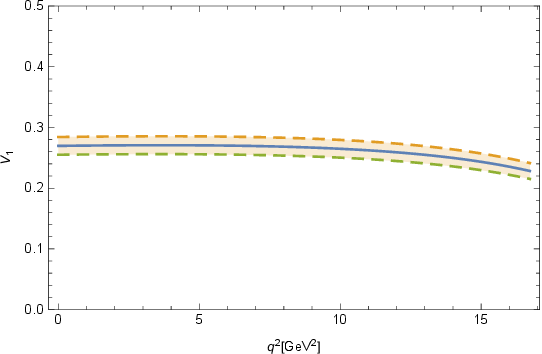} \hfill
\includegraphics[width=3in]{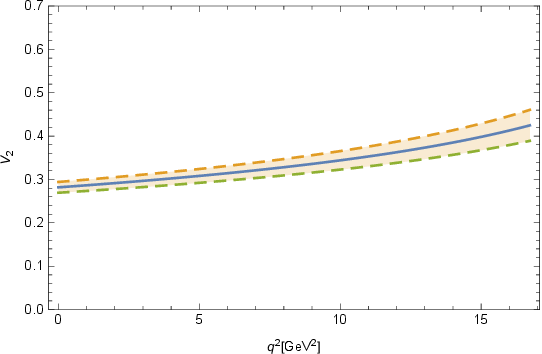} \\
\includegraphics[width=3in]{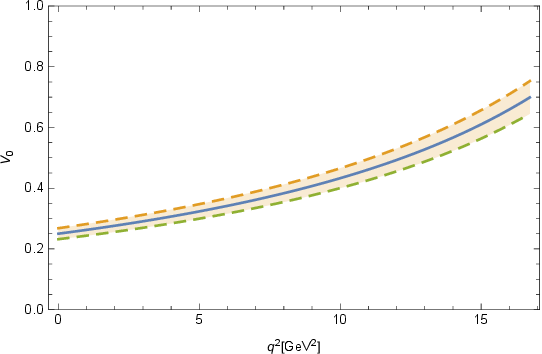}\hfill
\includegraphics[width=3in]{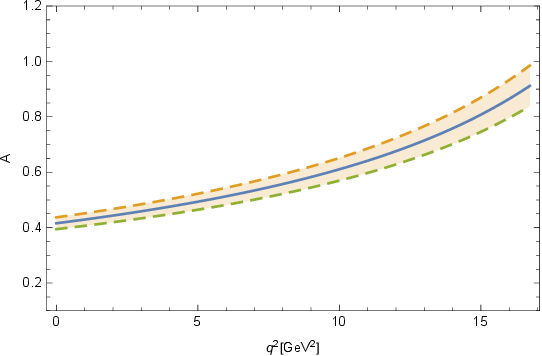}
\caption{$B_s \rightarrow f_1(1^3P_1)$ form factors as
functions of $q^2$. } \label{Fig:Bsf1q}
\end{figure}
\begin{figure}[H]
\centering
\includegraphics[width=3in]{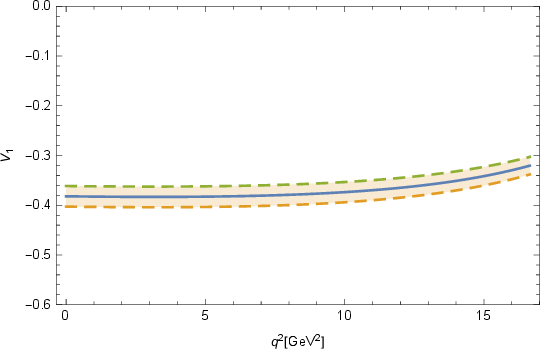} \hfill
\includegraphics[width=3in]{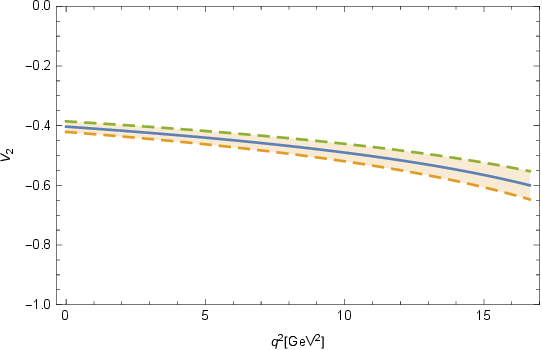} \\
\includegraphics[width=3in]{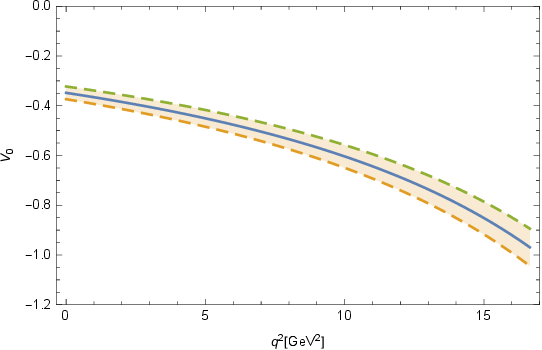}\hfill
\includegraphics[width=3in]{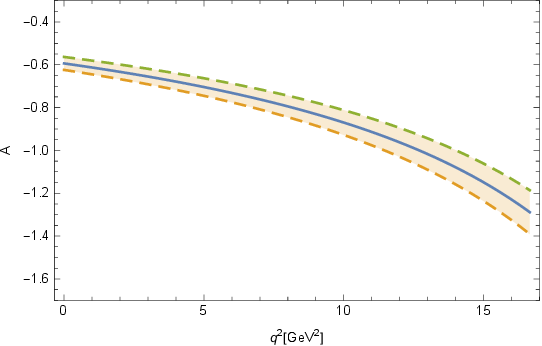}
\caption{$B_s \rightarrow f_8(1^3P_1)$ form factors as
functions of $q^2$. } \label{Fig:Bsf8q}
\end{figure}

\begin{figure}[H]
\centering
\includegraphics[width=3in]{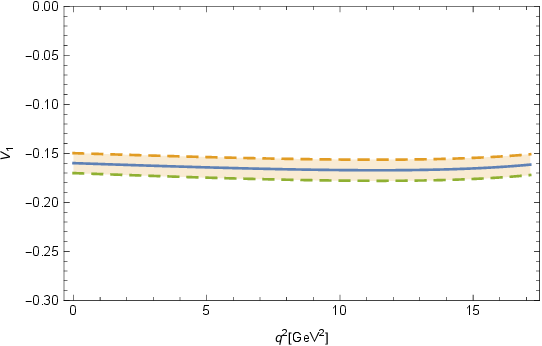} \hfill
\includegraphics[width=3in]{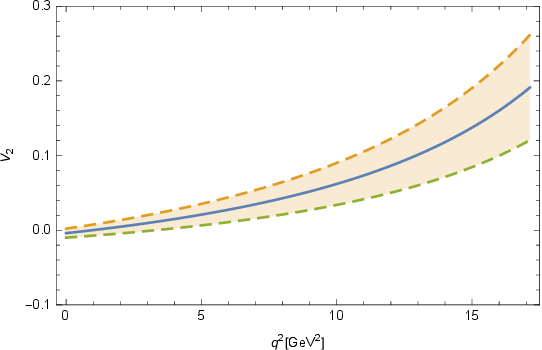} \\
\includegraphics[width=3in]{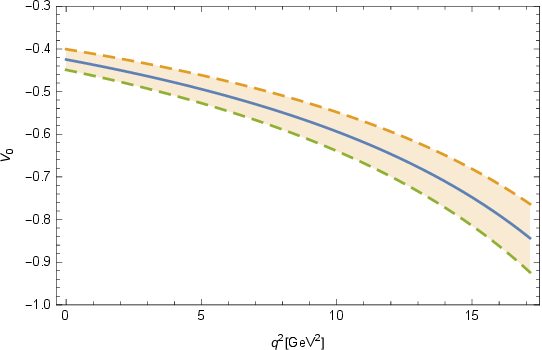}\hfill
\includegraphics[width=3in]{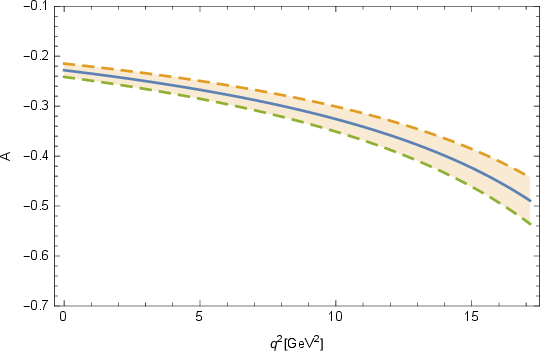}
\caption{$B_s \rightarrow h_1(1^1P_1)$ form factors as
functions of $q^2$. } \label{Fig:Bsh1q}
\end{figure}
\begin{figure}[H]
\centering
\includegraphics[width=3in]{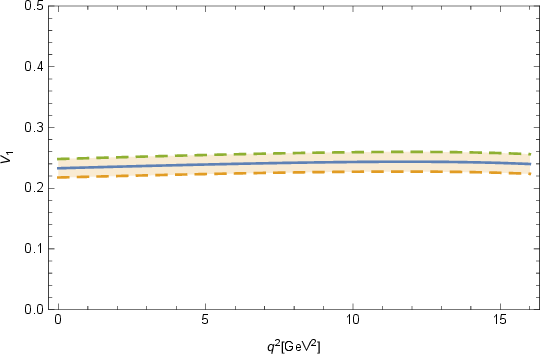} \hfill
\includegraphics[width=3in]{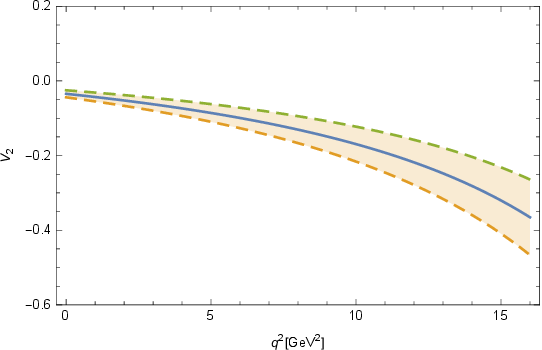} \\
\includegraphics[width=3in]{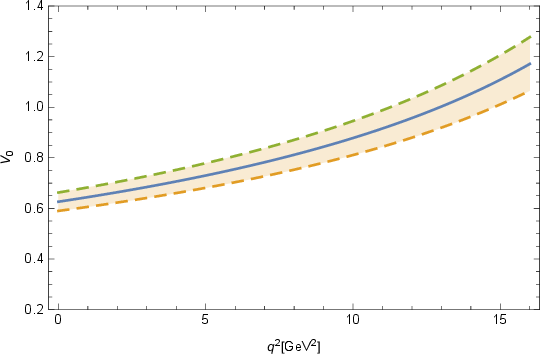}\hfill
\includegraphics[width=3in]{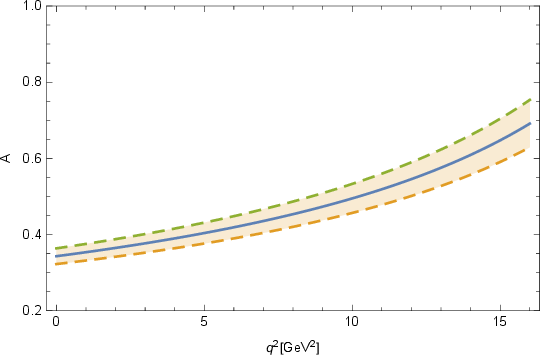}
\caption{$B_s \rightarrow h_8(1^1P_1)$ form factors as
functions of $q^2$. } \label{Fig:Bsh8q}
\end{figure}

\end{document}